# *Taxator-tk*: Fast and Precise Taxonomic Assignment of Metagenomes by Approximating Evolutionary Neighborhoods

J. Dröge[1,2], I. Gregor[1,2] and A. C. McHardy[1,2,3*]

[1] Department for Algorithmic Bioinformatics, Heinrich Heine University, Universitätsstraße 1, 40225 Düsseldorf, Germany

[2] Max-Planck Research Group for Computational Genomics and Epidemiology, Max-Planck Institute for Informatics, University Campus E1 4, 66123 Saarbrücken, Germany

[3] Computational Biology of Infection Research, Helmholtz Center for Infection Research, Inhoffenstraße 7, 38124 Braunschweig, Germany

*Correspondence to mchardy@hhu.de

## Abstract

Metagenomics characterizes microbial communities by random shotgun sequencing of DNA isolated directly from an environment of interest. An essential step in computational metagenome analysis is taxonomic sequence assignment, which allows us to identify the sequenced community members and to reconstruct taxonomic bins with sequence data for the individual taxa. We describe an algorithm and the accompanying software, *taxator-tk*, which performs taxonomic sequence assignments by fast approximate determination of evolutionary neighbors from sequence similarities. *Taxator-tk* was precise in its taxonomic assignment across all ranks and taxa for a range of evolutionary distances and for short sequences. In addition to the taxonomic binning of metagenomes, it is well suited for profiling microbial communities from metagenome samples becauseit identifies bacterial, archaeal and eukaryotic community members without being affected by varying primer binding strengths, as in marker gene amplification, or copy number variations of marker genes across different taxa. *Taxator-tk* has an efficient, parallelized implementation that allows the assignment of 6 Gb of sequence data per day on a standard multiprocessor system with ten CPU cores and microbial *RefSeq* as the genomic reference data.

## Introduction

Metagenomics allows us to study microbial communities from natural environments without the need to obtain pure cultures for the individual member species (Hugenholtz, 2002; Riesenfeld *et al.*, 2004). Metagenome sequencing of microbial community DNA with current shotgun techniques generates reads that range from less than 100 to several thousand nucleotides (Dröge and McHardy, 2012; Klumpp *et al.*, 2012). By computational analyses of metagenome samples, we can estimate the abundances of different taxa for the sampled communities, known as taxonomic profiling, characterize of the functional and metabolic potential based on the predicted proteins and resolve the contributions of individual taxa to the latter by reconstructing taxonomic 'bins' of unassembled or assembled sequences originating from a common taxon.

A taxonomic profile of a microbial community can be inferred by either targeted sequencing of taxonomic marker genes or from metagenome shotgun data-sets (Lindner and Renard, 2013; Sunagawa et al., 2013). Most metagenome profiling methods classify reads based on predefined taxon-specific (Segata *et al.*, 2012) or 'universal' marker genes (Darling *et al.*, 2014), or directly estimate a taxonomic profile for the underlying microbial community from their k-mer composition (Koslicki et al., 2013). Frequently used phylogenetic placement programs within such frameworks are *pplacer* (Matsen *et al.*, 2010) or *EPA/RAxML* (Berger *et al.*, 2011), which both operate in a probabilistic framework to position a query gene sequence in a pre-computed reference phylogeny of a particular gene family. If this gene tree is an approximate representation of the respective species tree – or reference taxonomy – this can be used to assign a taxonomic identifier (ID) to the query sequence (Stark *et al.*, 2010; Matsen and Gallagher, 2012). Taxon abundances are then derived from the individual read counts or gene frequencies within each taxonomic group.

Metagenome binning places arbitrary shotgun sequences of a metagenome into bins representing the different taxa of the sampled microbial community. If a bin represents a low-ranking taxon, such as the species, then the set of reads or contigs of an individual taxonomic bin serves as a draft-genome reconstruction for a community member (Pope et al., 2011). Binning methods are either based on clustering or classification. Clustering methods group sequences into bins without consideration of external reference sequences or taxonomic information. Instead, bins are inferred based on similarities in GC-content, oligomer frequencies, the abundance of genes or contig coverage within one or multiple samples (Baran and Halperin, 2012; Carr *et al.*, 2013; Albertsen *et al.,* 2013; Alneberg *et al.*, 2013), or by using a combination of these sequence features (Iverson *et al.*, 2012). This allows draft genome recovery from deep lineages for sequences of sufficient length. Taxonomic classification, like profiling, uses the resemblance of a sequence to known taxa in either global sequence composition or local sequence similarity to assign a taxonomic ID. For the human gut microbiome, extensive genome sequencing of isolate cultures allowed species-level taxonomic binning for a substantial portion (approx. 40%) of a metagenome sample (Schloissnig *et al.*, 2013) by mapping the reads to isolate genomes, which exist for the majority of abundant species (Sunagawa et al., 2013). However, this procedure is not suitable for environments in which most species are from deep-branching lineages without available reference genomes. Taxonomic binning of these requires more sophisticated similarity-based or composition-based taxonomic assignment methods (McHardy *et al.*, 2007; Huson *et al.*, 2011; Brady and Salzberg, 2011). Classification by sequence composition also allows draft genome recovery from deep-branching lineages, based on limited amounts of sequences for the individual taxa. Such programs also achieve linear classification times regarding metagenome sample size whereas similarity-based methods require considerably more computational resources due to homology searches in large reference sequence collections. Nevertheless, they are more accurate in the assignment of sequences shorter than approx. 1 kb (Patil *et al.*, 2011).

As described, a common problem of both taxonomic profiling and binning is the identification of known taxa. As 'taxonomic profiling' estimates only taxon abundances, it can be performed by analysis of smaller sets of marker genes alone, though one needs to account for variations in read counts among different clade-specific markers within taxa (Lindner and Renard, 2013). Taxonomic binning assigns taxonomic IDs to arbitrary sequences, to allow taxon-specific functional and metabolic analyses, where as a side-effect, a bin can also be used to quantify the corresponding taxon abundance based on individual read counts assigned to these bins. From a methodological standpoint, the differences between the mentioned phylogenetic-placement-based methods for profiling and alignment-score-based binning methods, such as *MEGAN4* (Huson *et al.*, 2011), *CARMA3* (Gerlach and Stoye, 2011) or *SOrT-ITEMS* (Monzoorul Haque *et al.*, 2009) are that the latter lack a well-motivated evolutionary framework. However, they have the advantages of being computationally lightweight and applicable to arbitrary genes. This is necessary for binning because pre-computing or even inferring *de-novo* trees for non-marker genes on a metagenome-wide scale is too computationally demanding – particularly with next-generation sequencing (NGS) data.

Our taxator toolkit (*taxator-tk*) is a software package for the taxonomic assignment of genomic sequences with application to metagenome profiling and binning. Conceptually, it lies between sequence-similarity-based programs which use local sequence alignment scores and those using trees. *Taxator-tk* extends the alignment-score-based approach by approximating phylogenetic gene trees and thereby provides more accurate taxonomic assignments, without assuming universal, rank or clade-specific gene conservation levels as parameters. We improve in terms of applicability to large data sets compared to phylogenetic methods by assigning arbitrary genomic sequences without the computationally demanding steps of *de-novo* multiple sequence alignment (MSA) and tree inference. *Taxator-tk* determines a subset of homologs, which represent the approximate evolutionary neighbors for a query sequence with a linear number of pairwise sequence comparisons with regard to the number of considered homologs and then assigns a taxonomic ID using a reference taxonomy based on the taxonomic identities of these neighbors. We have furthermore reduced the run-time by limiting the analysis to distinct homology-supported regions of the query sequence, which we termed query segmentation. Our method can be applied to arbitrary nucleotide sequences including large assembled or unassembled metagenome sequence samples. The software is released as an open-source package (GPLv3) and can be downloaded from http://algbio.cs.uni-duesseldorf.de/software/.

## Methods (Algorithm and Evaluation)

### *Taxator-tk*'s workflow for taxonomic assignment

The workflow for the taxonomic assignment of a query sequence comprises three stages (Fig. 1a–c). The first stage uses a (nucleotide) local sequence aligner to identify similar regions from a reference sequence collection, such as microbial *RefSeq* (*mRefSeq*) (Sayers *et al.*, 2009). The program supports different aligners and we used *NCBI BLAST* (Camacho *et al.*, 2009) and *LAST* (Frith *et al.*, 2010). Before applying the *taxator* algorithm in stage two, overlapping regions on the query, each defined by local alignment to a reference sequence, are merged into larger subsequences called segments (Supplementary Fig. 24). These query segments are flanked by regions without similarity to any reference data (Supplementary Fig. 16) and are not considered further. This step reduces the overall number of positions in the following alignment computations and improves the taxonomic assignment of queries that have undergone genome rearrangements, resulting in a different order of these segments. The reference sequence regions corresponding to the local alignments are extended at both sides by the missing number of nucleotides to match to the corresponding query segment with respect to its length. We refer to these as reference segments. Each independent set of homologous segments is the input to core algorithm in the program *taxator* in stage two (Fig. 1b), which calculates a taxonomic ID independently and in parallel for the corresponding query segment.

In the third stage (Fig. 1c), multiple segments belonging to the same query are considered and their IDs are combined in the program *binner*, to derive a consensus taxonomic ID. The corresponding algorithm weights the individual segment assignments by the number of positions matching the closest reference sequence and assigns to the entire query the taxonomic ID supported by the majority (default = 70%) of weighted assignments with a minimum number of matching positions (default = 50 bp) (Supplementary Methods, "Consensus binning algorithm"). *Binner* has optional parameters to specify additional constraints for binning (minimum sequence identity, minimum. sample abundance) but these were not applied in our analysis. If the taxonomic information is limited or contradictory, *taxator* and *binner* assign identifiers to higher ranking taxa in a conservative fashion to obtain maximum reliability for the resulting taxonomic assignments.

**The taxonomic assignment algorithm (*taxator*)**

The input to the algorithm is a segment *q* of the original query sequence from an (unknown) taxon Q and a set of homologous segments with known taxonomic IDs. The term "segment" refers to a gap-less subsequence of either thequery or a reference sequence. Given that for the set of homologs we know the correct underlying tree of taxa (Fig. 2a), we can see that for our query taxon Q, the closest evolutionary neighbors would be A, B and S. If we simply assign X, the parental taxon of A, B and S, as a taxonomic identifier, this would be inaccurate, as A, B and S are more closely related to each other than to Q. Instead, the correct taxonomic assignment would be a parent of X, Q and at least one additional outgroup taxon (O) in the reference tree, such that Q becomes a descendant of the identified parent (R in Fig. 2a). If we therefore can identify the taxa A, B, S and O in the reference tree, we can determine the taxonomic ID of R as the lowest common ancestor (LCA) of these taxa and assign it to Q (and *q*).

Assuming that the underlying gene tree for a set of homologs is similar to the species tree, a natural procedure to identify the segments corresponding to the leaf taxa within R among the homologs would be to construct a MSA for the segment and a phylogenetic tree with a corresponding subtree as in Fig. 2a. However, the computational effort for this approach is superlinear with respect to the number of homologs being compared and substantial for all the query segments in a large sample, even using fast techniques for MSA construction and tree inference. The *taxator* algorithm attempts to identify these segments with a linear number of pair-wise segment comparisons. Let us consider the evolutionary distances between pairs of segments within the underlying tree to be represented in an undirected graph with the nodes representing the segments (tree leaves) and the edges scaled by evolutionary distances between pairs of segments (Fig. 2b). In this graph, a monophyletic group in the species tree is a subgraph. For all pairs of subgraph nodes, the following inequality is true, given that the segments have evolved with a constant rate of evolution (i.e. the segment tree is ultrametric): The distance between any two subgraph nodes is smaller than that to any other node outside the subgraph. The relationship becomes clearer when thinking of the evolutionary distance between two nodes as the divergence time from their most recent ancestor. Members of a monophyletic group derive from a single common ancestor and thus there is a maximum distance for all possible pairs. If one member's distance to an outside node is smaller than this maximum, both must share a more recent common ancestor and the corresponding group is not monophyletic by definition. The stated inequality can be used to augment an incomplete group or corresponding subgraph iteratively by taking an internal distance, ideally close to the maximum, as a threshold and adding outside nodes to the group which have a smaller distance to some internal node.

In this manner, *taxator-tk* searches for the leaf node taxa of clade R among all segments based on a linear number of sequence comparisons between the input segments and adds them to an (initially empty) working set *M*.

0.  A ranking by alignment scores from the input local alignments is used at the beginning to identify the reference segment *s* that is most similar to the query *q*.

The working set *M* is then augmented in two passes:

1.  In the first pass, all segments are aligned to *s* using fast nucleotide alignment and the edit distance. All segment taxa with a distance less than or equal to the threshold, *distance(s,q)*, are added to *M* (Fig. 2c).

2.  The outgroup segment *o* is determined as the first segment for which *distance(s,o)* larger than *distance(s,q)*. In the second pass, all segments are then aligned to *o* and segment taxa with distances smaller than or equal to *distance(o,q)* are added to *M* as well (Fig. 2d).

This procedure requires approximately 2*n* alignments, where *n* is the number of reference segments.

3.  The resulting set *M* of taxa (implicit in the partially resolved tree in Fig. 2e) is used to determine the taxonomic ID for *q*, corresponding to the LCA of these taxa in a reference taxonomy, such as the NCBI taxonomy.

If no outgroup could be determined or if *M* is so diverse that the LCA corresponds to the taxonomy root, *q* is left unassigned. The algorithm requires at least two homologous segments (*s* and *o*) to determine a valid taxonomic ID. The taxa in *M* become more diverse if the alignment scores are inaccurate ultrametric distance estimates, if the species subtree's topology deviates from the respective part of the taxonomy or if the gene tree's topology deviates from the species tree, for instance due to varying rates of evolution or the inclusion of non-homologous segments in the analysis. The robustness of the algorithm in avoiding incorrect assignments under these circumstances depends on the number of taxa in *M* and the subsequent LCA operation. Further details relating to the robustness of the implementation are given in the Supplementary Methods, "Taxonomic assignment of sequence segments".

**Evaluation procedures**

Before evaluating all the methods, we removed the smallest predicted bins (1%) as likely errors. We used the macro-precision and macro-recall as measures of assignment performance (Supplementary Methods, "Performance measures"). The macro-precision specifies the fraction of correct assignments per predicted bin (precision), averaged over all such bins, while the macro-recall measures the fraction of correctly recovered sequence data per truly existing bin (recall), averaged over all such bins. To account for strong differences in bin size, we also pooled the species, genus and family assignments, and reported the overall precision for these ranks as the total fraction of correct assignments. We tested the assignment performance of different methods using three simulated short read datasets, a simulated 16S rRNA dataset, three simulated assembled metagenome samples and a cow rumen metagenome sample. For every simulated dataset, we performed seven cross-validation experiments (Supplementary Methods, "Cross-validation"). In each experiment, we created a partition for every query sequence considered, simulating a specific taxonomic distance between this query and the reference sequences. For the first scenario, all reference data, including the species genome from which the query had been sampled, were made available to the method for assigning a single query sequence as an idealized test case. In the other six scenarios, all reference data belonging to the species, genus, family, order, class or phylum of the query sequence, respectively, were made inaccessible for the method. We added the sequence assignmentsfrom these experiments to characterize a method's assignment performance across the entire range of taxonomic distances. For evaluation with the cow rumen metagenome sample, for which no true taxonomic labels were known, we divided the assembled contig sequences into multiple 'chunks' and characterized the consistency of taxonomic assignments for chunks originating from same contig (Supplementary Methods, "Consistency Analysis").

**Results**

**Evaluation with unassembled data**

We first evaluated the performance of *taxator-tk* for classification of the most widely used taxonomic marker in bacterial diversity studies, the 16S rRNA gene (Supplementary Material, Supplementary Fig. 1). This served as a proof of concept, as *taxator-tk* classifies arbitrary sequence regions including taxonomic marker genes. We did not expect it to perform better than sophisticated phylogenetic models for this task, but wanted to confirm a satisfactory performance. The macro-precision for the taxonomic assignment of 7176 16S rRNA genes was constantly above 92% (Fig. 3a) in the combined cross-validation (Methods), using the whole-genome reference sequences in *mRefSeq47* (Supplementary Fig. 19), not just the 16S genes. More precisely, the average error rate per bin (one minus precision) was 7.4% at the species level and 4.6% at the order level.

Next, we simulated 100,000 reads at 100, 500 and 1000 bp by subsampling randomly from 1729 species in *mRefSeq47* and evaluated *taxator-tk* with these three datasets using (combined) cross-validation. The performance was very similar for the different fragment sizes (Fig 2b, 2c and Supplementary Fig. 2–4a). Overall, *taxator-tk* showed high precision in simulated read assignment: The macro-precision for all short read lengths remained above 74% and was 82–99% for the genus to kingdom ranks, about 10% lower on average than for the 16S data. This was still good for the assignment of short sequence fragments from arbitrary genomic regions compared to marker genes. Longer reads had a slightly higher macro-recall than the shorter ones. At genus level, the macro-recall was 19–23 % (~33% genera recovered) if genomes of the same species as the query sequence were provided in the reference (Supplementary Fig. 2–4b) and as low as 5–7 % (~16% genera recovered) otherwise (Supplementary Fig. 2–4c). The macro-recall must decrease when removing reference data for cross-validation. For example, if all reference data at genus level are removed, then no correct assignments to the genus rank should be possible. The macro-recall scale also very much depends on the validation sample. Generally, it increases with higher genome coverage or lower organismal sample complexity. Here, the macro-recall was low due to the large number of sample taxa and their uneven sampling caused by the taxonomic bias in *mRefSeq47*. Since longer sequences yield better recall and because overlapping reads contain redundant information, leading to more alignment computations, we recommend applying *taxator-tk* to (partially) assembled data. We observed that on longer query sequences, we were more likely to find segments for processing and therefore assign a larger portion of the sample.

**Evaluation with simulated metagenome samples**

For the tests on assembled simulated samples, we compared *taxator-tk* to *CARMA3* and *MEGAN4* using the same taxonomy and the same nucleotide alignments against *mRefSeq54* (Supplementary Fig. 20). We used the recommended parameter settings (Supplementary Methods, "Program parameters") and cross-validation, as before (Methods, "Evaluation procedures"). Both *CARMA3* and *MEGAN4* allow the use of protein or nucleotide alignments. We exclusively used faster nucleotide alignments because of the large size of the metagenome datasets and because we did not observe an improvement in performance when using protein instead of nucleotide local alignments.

We used the SimMC/AMD and SimHC/soil simulated metagenome datasets of the FAMeS collection (Mavromatis *et al.,* 2007) for our evaluation. These metagenome datasets were generated by Sanger sequencing in the year 2006 and are several orders of magnitude smaller than datasets generated with the current NGS technologies (Dröge and McHardy, 2012). The medium complexity SimMC/AMD consists of ~17 Mb/7307 contigs and the high complexity SimHC/soil sample comprises ~1 Mb/578 contigs. We evaluated *MEGAN4* and *taxator-tk* with SimMC and SimHC (Supplementary Fig. 21, 22), but omitted *CARMA3* due to its long run-time for each of the seven cross-validation experiments (Methods, "Evaluation procedures"). Due to the small sample sizes and because the FAMeS data could have been used for the method development, we created an additional simulated NGS metagenome dataset (simArt49e, composition in Supplementary Fig. 23) for our evaluation. This sample included 49 equally abundant species (51 strains) and was created by Illumina paired read simulation with *pIRS* (Hu *et al*., 2012), followed by *SOAPdenovo* assembly (Luo *et al.,* 2012). Around 160 Mb or 267,178 contigs remained after removal of 0.03% chimeric sequences.

On the FAMeS datasets, *taxator-tk* produced fewer errors for all taxonomic ranks than *MEGAN4*, which was accompanied by a moderate reduction in macro-recall (Supplementary Fig. 5-8) throughout all individual experiments and in the combined cross-validation experiments: For SimMC, the macro-precision was three to four times as large as *MEGAN4*'s for species to order, with higher macro-recall (Supplementary Fig. 5-6). The species to family overall precision was ~91% for *taxator-tk* (~59% for *MEGAN4)* and *taxator-tk* estimated 54 species bins (*MEGAN4* 188) for the 47 actual species in SimMC. Similarly, for SimHC, *taxator-tk* achieved a higher macro-precision for all ranks, which was most pronounced for class and phylum (Supplementary Fig. 7–8). By contrast, the macro-recall was slightly reduced and both methods underestimated the 96 existing species in SimHC.

Our simulated metagenome sample simArt49e was difficult to assign for all methods when the sequences from the corresponding species and genus were removed from the reference (Supplementary Fig. 9–11d). In this case, all methods showed a reduced macro-precision for the assignment at the family rank in comparison to the FAMeS datasets. Still, *taxator-tk* was the most precise, though it had a lower recall than the other methods (*taxator-tk*: 56% family macro-precision, 60% overall precision for species to family, 10% family macro-recall; *CARMA3*: 13%, 27% and 20%; *MEGAN4*: 22%, 27% and 31%). Similar to this individual experiment, in the combined cross-validation (Supplementary Fig. 9–11a), most sequences were assigned to bacteria or archaea by all methods or, in the case of *CARMA3*, remained unassigned. For the data assigned to the species to family ranks, *taxator-tk* had 91% correct assignments compared to 52% for *CARMA3* and 59% for *MEGAN4*. The macro-precision was substantially higher for *taxator-tk* than for the other methods, e.g. 61% at the species level (*taxator-tk*), compared to 3% (*CARMA3)* and 5% (*MEGAN4)*. The low macro-precision observed for *CARMA3* and *MEGAN4* is largely due to the prediction of many small bins with many false assignments (Supplementary Methods, "Performance measures"). Likewise, consistent with the results for FAMeS, more species bins were predicted by *CARMA3* (1672) and *MEGAN4* (824) than by *taxator-tk* (65), with 49 species being actually present in the sample. When simulating novel families (Supplementary Fig. 9–11e), *MEGAN4* predicted 69 distinct orders to be present, *CARMA3* 81 and *taxator-tk* only 27, compared to the existing 32 in simArt49e (Fig. 4). Taxonomic assignments of *taxator-tk* were considerably rarer to false taxa at low ranks than with the other methods, and instead were to higher-ranking existing taxa. The other two methods assigned a substantial amount of sequence data incorrectly to bins at the family level or below, which can be a seriously misleading result, depending on its further use. The prediction of much fewer taxa which are not truly present in the sample makes *taxator-tk* more suitable as a tool for determining microbial community members, in addition to the plain recovery of taxonomic sequence bins from shotgun datasets.

To investigate the reason for the observed differences between overall and macro-precision, we plotted the per-bin precision at the family level in the combined cross-validation as a function of predicted size with a k-nearest-neighbor (kNN) estimate of macro-precision (Fig. 5; for all ranks, see Supplementary Fig. 17). Overall, the bins predicted by *taxator-tk* were smaller, more precise and much more likely to represent truly existing taxa than those predicted by the other two programs; larger bins tended to be more accurate for all methods. *CARMA3* and *MEGAN4* predicted a substantial number of mostly smaller-sized incorrect bins with zero precision. Even though the size-dependent kNN precision curve should be unaffected by these bins, for larger bin sizes, they never reached 70% (*CARMA3*) or 80% (*MEGAN4*), whereas the *taxator-tk* curve reached almost 100%. For the smallest bins, *taxator-tk*'s kNN precision was ~20% whereas bins below 500 kb for *CARMA3* and *MEGAN4* were practically indistinguishable from noise.

When comparing to the composition-based program *PhyloPythiaS* (Patil *et al.*, 2011), we could not apply cross-validation due to the computational effort of training a large number of models. Therefore we evaluated this using the published evaluation scenario with SimMC (Patil *et al.*, 2011), in which all genome sequences of the SimMC genera were removed from the reference sequenced genomes. All programs were provided with the remaining sequenced genomes and an additional 100 kb of reference data for each of the three dominant strains. The latter could be used by *PhyloPythiaS* to infer a corresponding species model but were less helpful for the similarity-based classifiers. We generated assignments with *taxator-tk*, *CARMA3* and *MEGAN4* under equivalent conditions and compared them to the published *PhyloPythiaS* assignments (Supplementary Fig. 12). The results for the similarity-based programs were consistent with the previous evaluations with SimMC concerning the error distributions. The *PhyloPythiaS* results showed that composition-based classification with additional training data correctly assigned most data at the genus and family levels (species assignments were not given in the original publication), which were either rarely assigned (*taxator-tk*) or mostly incorrectly assigned (*MEGAN4*, *CARMA3*) by the other programs. However, *PhyloPythiaS* predicted fewer families (6), compared to 29 underlying families versus 14 (*taxator-tk)*, 50 (*CARMA3)* and 17 (*MEGAN4)*. Apart from an increased macro-recall with *PhyloPythiaS*, the macro-precision (~50% at genus to order level) was also higher than it was with *MEGAN4* (~9–30%) or *CARMA3* (~7–24%) but less than with *taxator-tk* (~50–68%). However, unlike the other programs, for *PhyloPythiaS* the modeled taxa should be specified *a priori* to achieve optimal performance. It is therefore best applied when the taxonomic composition of a microbial community has already been determined and sufficient training data are available for the identified taxa.

**Evaluation with a real metagenome sample**

For microbial communities in many environments, only distantly related reference genomes are available. We analyzed a medium complex metagenome sample of such a microbial community from a cow rumen (Hess *et al.*, 2011) with *taxator-tk*, *CARMA3*, *MEGAN4* and *PhyloPythiaS* (the general model with the 100 most abundant species among sequenced prokaryotes). For this particular sample, we considered scaffolds to be unreliable compared to contigs, which we reconstructed by splitting the available scaffolds at gaps of more than 200 positions (A. Sczyrba, personal communication). We subsequently divided contigs longer than 10 kb into sequence 'chunks' of 2 kb (minimum 5), resulting in a 319 Mb dataset, which we used to assess the assignment consistency for chunks originating from the same contig. The chunk sequences were assigned independently with *taxator-tk, CARMA3* and *MEGAN4* (identical alignments) and with *PhyloPythiaS*. As the standard of truth for each contig, we determined the taxon minimizing the inconsistency between all corresponding chunk assignments (Gregor *et al.,* unpublished). A chunk assignment was considered consistent if it was to the same taxon as the one for entire contig, and was inconsistent otherwise. The consistency of a taxonomic bin is the fraction of chunk data with matching contig assignments and the macro-consistency is the consistency averaged over all predicted taxa, comparable to the macro-precision.

In agreement with the tests on the simulated metagenome datasets, the *taxator-tk* results were more consistent than those of the other methods (Supplementary Fig. 13): 76–78% macro-consistency at species to order level, in comparison to *MEGAN4* (34–40%) and *PhyloPythiaS* (56–65%). The overall consistency (analogous to overall precision) for species to family levels was 97% with *taxator-tk*, 39% with *CARMA3*, 62% with *MEGAN4* and 82% with *PhyloPythiaS*. Likewise, *taxator-tk* assigned less data at species to family level, with a total of 12.8 Mb being consistent compared to *CARMA3* (8kb), *MEGAN4* (46.9 Mb) or *PhyloPythiaS* (13.8 Mb). Different methods again identified different numbers of taxa to be present: CARMA3 identified 572 genera with a macro-consistency of 53%, *MEGAN4* 264 (34%), *PhyloPythiaS* 33 genera (63%) and *taxator-tk* found 110 genera (76%). These results are in agreement with the results for simulated samples and suggest that *taxator-tk* is a precise taxonomic classifier for metagenomic shotgun sequences.

**Run-time analyses**

The run-time for taxonomic assignment of a metagenome sample consists of the time to find homologs and to process them for assigning taxonomic IDs to all sequences. We evaluated the run-times of all methods using the same set of alignments generated with either *BLAST* or *LAST*, so the run-time for the initial similarity search was identical with all methods. We determined the time for the taxonomic assignment of simArt49e for all methods when performing a cross-validation with the data of the families present in the test dataset removed from the reference data. This took less than one hour for *MEGAN4* (interactive mode), less than 6 hours for *taxator-tk* (~10 CPU cores) and almost a week for *CARMA3* (~20 CPU cores). On our system, the parallelization of *taxator-tk* led to a linear parallel speedup (with the number of CPU cores) for up to 15 cores and deteriorated with 20 cores (Supplementary Fig. 14). To provide a more specific estimate of the throughput of *taxator-tk*, we aligned ~1 Gb of cow rumen sequence data with *BLAST* against *mRefSeq54* and assigned the data with *taxator-tk* on 10 CPU cores (AMD Opteron 6386 SE). We measured an average throughput of 5.9 Gb per day for the combined alignment and taxonomic assignment steps with this dataset. We also determined how our implementation scaled for increasing input sequence lengths and reference exclusion scenarios (Supplementary Fig. 15a). The run-time scaled approximately linearly except when the same or very similar species were among the reference genomes. In general, the greater the number of similar sequences in the reference data, the longer *taxator-tk's* run-time was for the alignment of longer sequence stretches with more homologs. Simultaneously, we investigated the impact of the query segmentation on *taxator-tk's* run-time (Supplementary Fig. 15b) and found that it reduced the total run-time by up to 30%.

**Discussion**

Here, we have described *taxator-tk*, a taxonomic assignment software package which generates very precise taxonomic assignments (i.e. assignments with few errors) for metagenome shotgun sequencing data. To provide a fair comparison, we invested extensive effort into ensuring that we evaluated all methods under identical conditions with the same reference sequences, test datasets and background taxonomies, using their recommended settings. We evaluated *taxator-tk* on 16S gene sequences, simulated short reads, and with a simulated and a cow rumen assembled metagenome sample across a wide range of evolutionary distances between the query and reference sequences by using cross-validation. For comparisons to other methods, we used older and new larger NGS samples. *Taxator-tk* was the most precise of all tested methods with the most realistic number of identified taxa overall. This property was very pronounced for lower taxonomic ranks from species to family level. However, this also means that *taxator-tk* assigned fewer data overall than other methods from species to family. For the small assembled SimMC dataset, it assigned fewer data, particularly in comparison to the composition-based classifier *PhyloPythiaS* when 100 kb of data were provided for individual community members to train species-level models. For the real cow rumen dataset, *taxator-tk* was the most consistent in terms of classifying multiple pieces of one contig. All results consistently indicate that *taxator-tk*'s strength is its high precision of assignments, which allows us to confidently assign a core of sample sequences and thereby to infer the taxonomic composition of the community. In comparison to assignments based on marker genes, it has the advantages that it makes assignments across all domains of life and that corresponding abundance estimates from shotgun sequences are less affected by copy number variations of individual genes. Such shotgun estimates are also unaffected by PCR primer amplification biases, unlike marker gene sequencing techniques, and do not require high-quality reference gene phylogenies of individual marker genes. The amount of recoverable low ranking taxonomic assignments depends on the available reference data, as it does for all taxonomic classification methods. To target draft genome reconstructions, the data assigned to individual taxonomic bins by *taxator-tk* can be used as training data for complementary approaches, such as composition-based methods, or as independent information in combination with recently proposed clustering methods using the abundance of genes or contigs across multiple samples.

From a methodological point of view, we have introduced a method for the fast approximation of the evolutionary neighborhood of a query sequence with a run-time that increases linearly with the number of homologs. In *de-novo* phylogenetic inference methods, the run-time increases at least log-linearly with the number of homologs or they rely on time-consuming optimizations of parameter-rich phylogenetic models, which generate excessive computational requirements for the analysis of Gb-sized NGS samples. Our software, therefore, provides an easy to use alternative to taxonomic classification of marker genes that is applicable to any gene or gene fragment in a scalable manner. Unlike similarity-based taxonomic classifiers for shotgun data, our algorithm handles different degrees of sequence conservation without preset or user-specified parameters and without being restricted to the analysis of a number of high-quality homologs with a minimal length. At the same time, the inferred evolutionary neighborhood is extended by the identification of an outgroup, leading to more precise taxonomic assignments. Importantly, genes are evaluated with regard to their taxonomic information content in the process of assignment, which discards information in conflict with the taxonomy. We post-process independent taxonomic assignments of query segments to infer an assignment for the entire query and do this using a majority vote algorithm with a few robust default parameters. This computationally lightweight step is quickly repeated with other values for the majority and minimum support parameters, if required. In addition to the algorithmic considerations and other run-time optimizations, we implemented query sequence segmentation and program parallelization, which allow large-scale data analysis with a throughput of several Gb per day on a standard multiprocessor system.

We have applied *taxator-tk* to characterize the taxonomic composition of Bacteria, Archaea and Eukaryotes in multiple samples of the barley rhizosphere, which correlated with results from 16S rRNA profiling and showed the most notable deviations for taxa known to be affected by primer biases or having multiple copies of the 16S rRNA gene (results not shown). The program's scope is also not limited to the taxonomic assignment of metagenomes: It can be applied to arbitrary DNA or RNA sequences. A successful in-house application was the detection of contamination in isolate sequencing data. Furthermore, the program *taxator* within *taxator-tk* provides taxonomic information for individual query segments (Supplementary Fig. 16), which could be used to identify assembly errors or regions acquired by lateral gene transfer.

# Main Figures

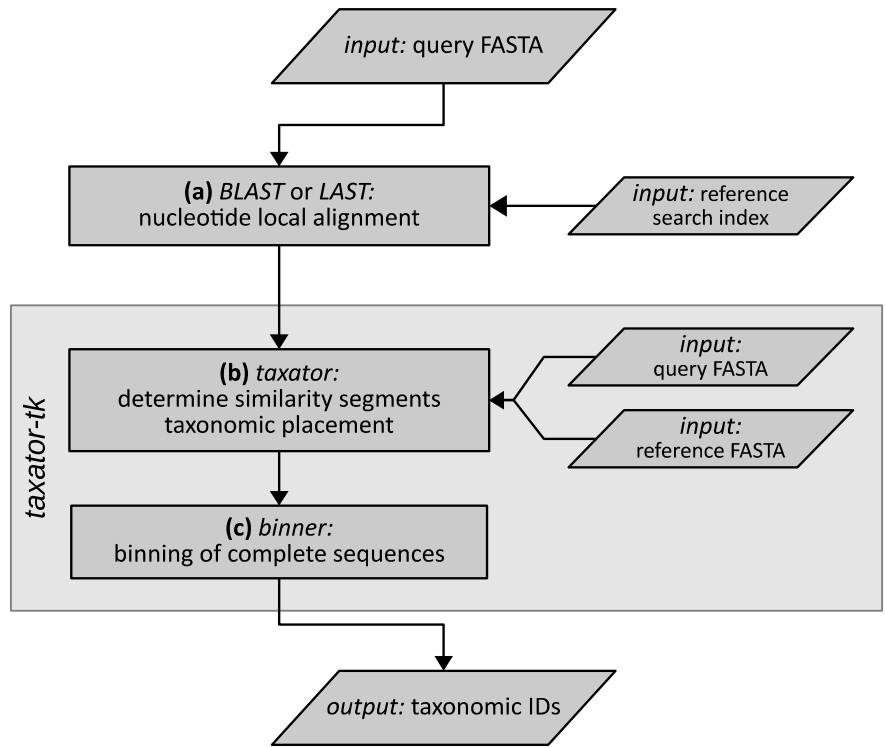

**Figure 1: Workflow diagram for the taxonomic assignment of a query sequence with *taxator-tk*.**

Taxonomic assignment with *taxator-tk* includes three steps. In step (a), local alignments between a query sequence and sequences from a reference collection are identified using e.g. NCBI's *BLAST* or *LAST*. In step (b), the program *taxator* separates the query sequence into distinct segments with homologs and determines a taxonomic ID for each segment. In step (c), the program *binner* determines a taxonomic ID for the entire query sequence based on the taxonomic assignments of the individual segments.

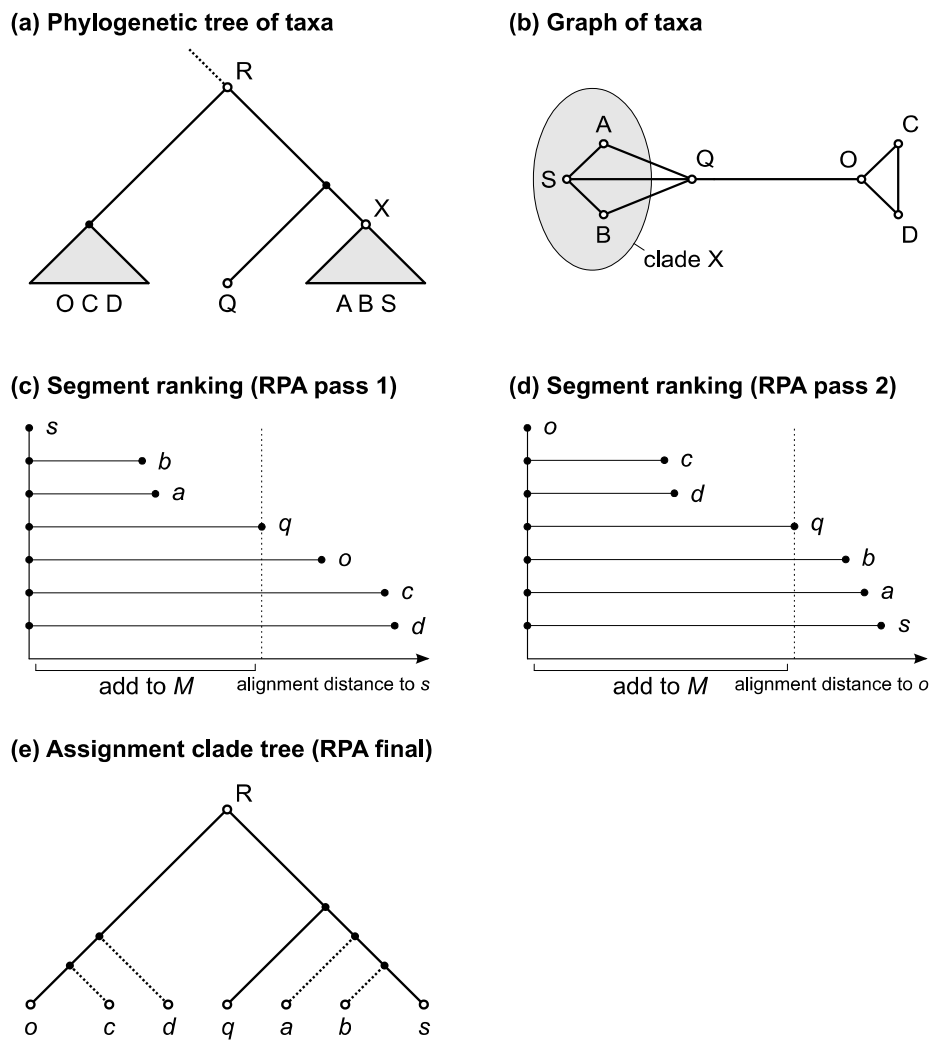

**Figure 2: Algorithm for taxonomic labeling of query segments (realignment placement algorithm)**

Realignment placement algorithm (RPA) for assigning a taxon ID to a query segment $q$. Panel (a): Exemplary phylogenetic tree with query taxon Q and reference taxa A, B, C, D, O and S. Panel (b): Approximate graph representation of the tree metric (pair-wise distances between taxa in the tree). The subgraph corresponding to the clade X is highlighted. Panels (c and d) show the two augmentation passes of the RPA in which segment taxa are added to an (empty) set $M$. (c) In the first pass, all segments are aligned to $s$, which is the segment with the best initial local alignment score (program input). The resulting scores are ordered and the taxa are added to $M$ if they have an equal or smaller score than the threshold, *distance(s,q)*. The outgroup segment $o$ is the segment with the shortest distance but which is larger than the threshold. (d) In the second pass, all segments are aligned to $o$ and the resulting scores are ranked. All taxa of segments with a score smaller than *distance(o,q)* are again added to $M$. Finally, $M$ includes all the nearest evolutionary neighbors for the query segment $q$; in this example, they are the taxa corresponding to segments $a, b, c, d, o$ and $s$. The taxonomic ID then assigned to $q$ is the lowest common ancestor in a reference taxonomy such as the NCBI taxonomy of all taxa in $M$. Panel (e) gives an example of the constructed subtree at node R from the pair-wise segment alignments in (d and e), where the exact position of segments $a, b, c$ and $d$ is left unresolved (dashed lines) by the RPA.

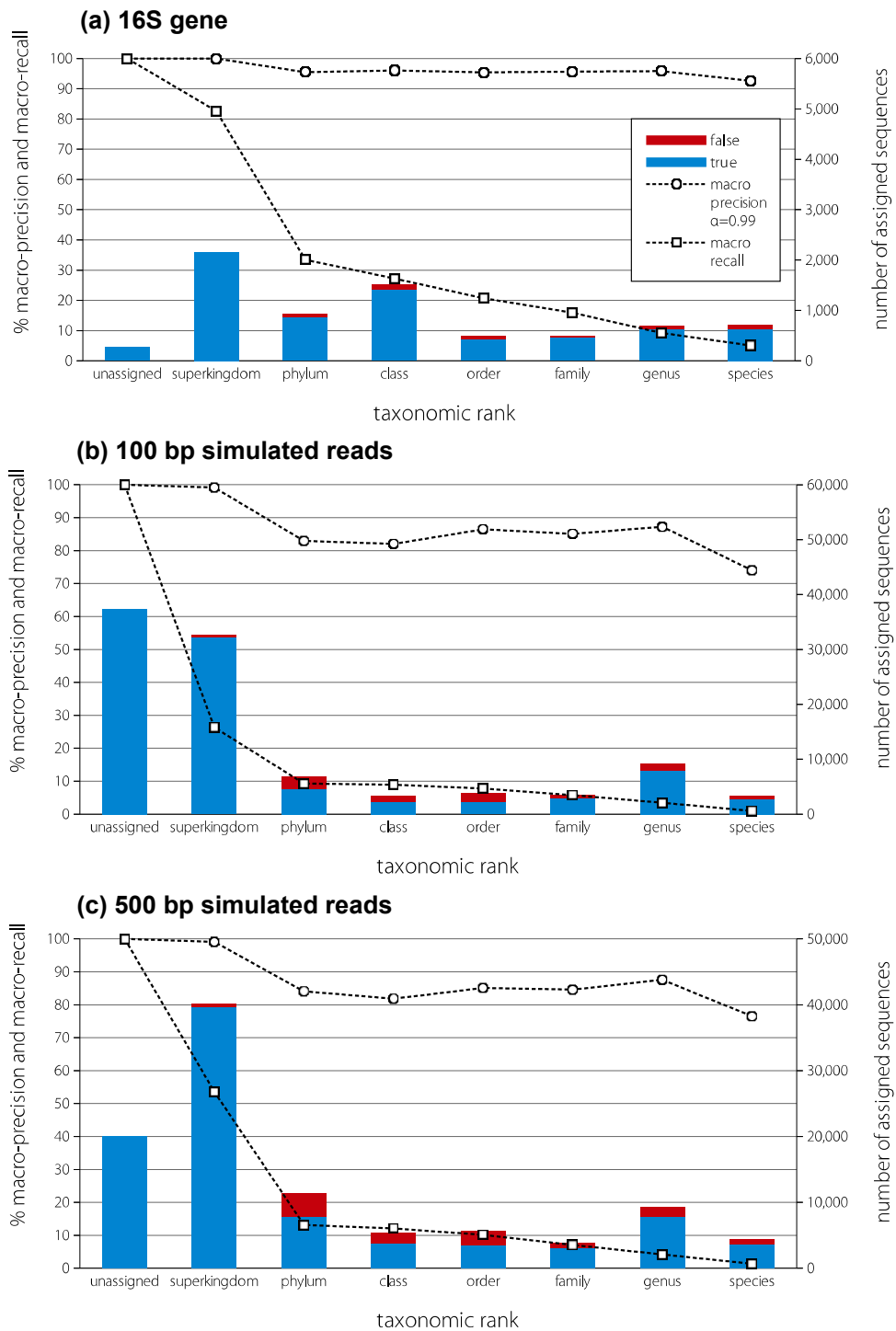

**Figure 3: *Taxator-tk* combined cross-validation performance for unassembled sequence data.**
Taxonomic assignment performance over seven cross-validation experiments simulating taxonomic assignments across a range of evolutionary distances. The bars show the absolute number of correct and false assignments for the corresponding rank (x axis). The macro-precision and macro-recall shown are cumulative from low to high taxonomic ranks, such that, for instance, the family level macro-precision includes the species, genus and family assignments. Assignment performance is shown for 16S rRNA genes in *mRefSeq47* with a minimum length of (a) 1000 bp and simulated reads with a length of (b) 100 bp and (c) 500 bp.

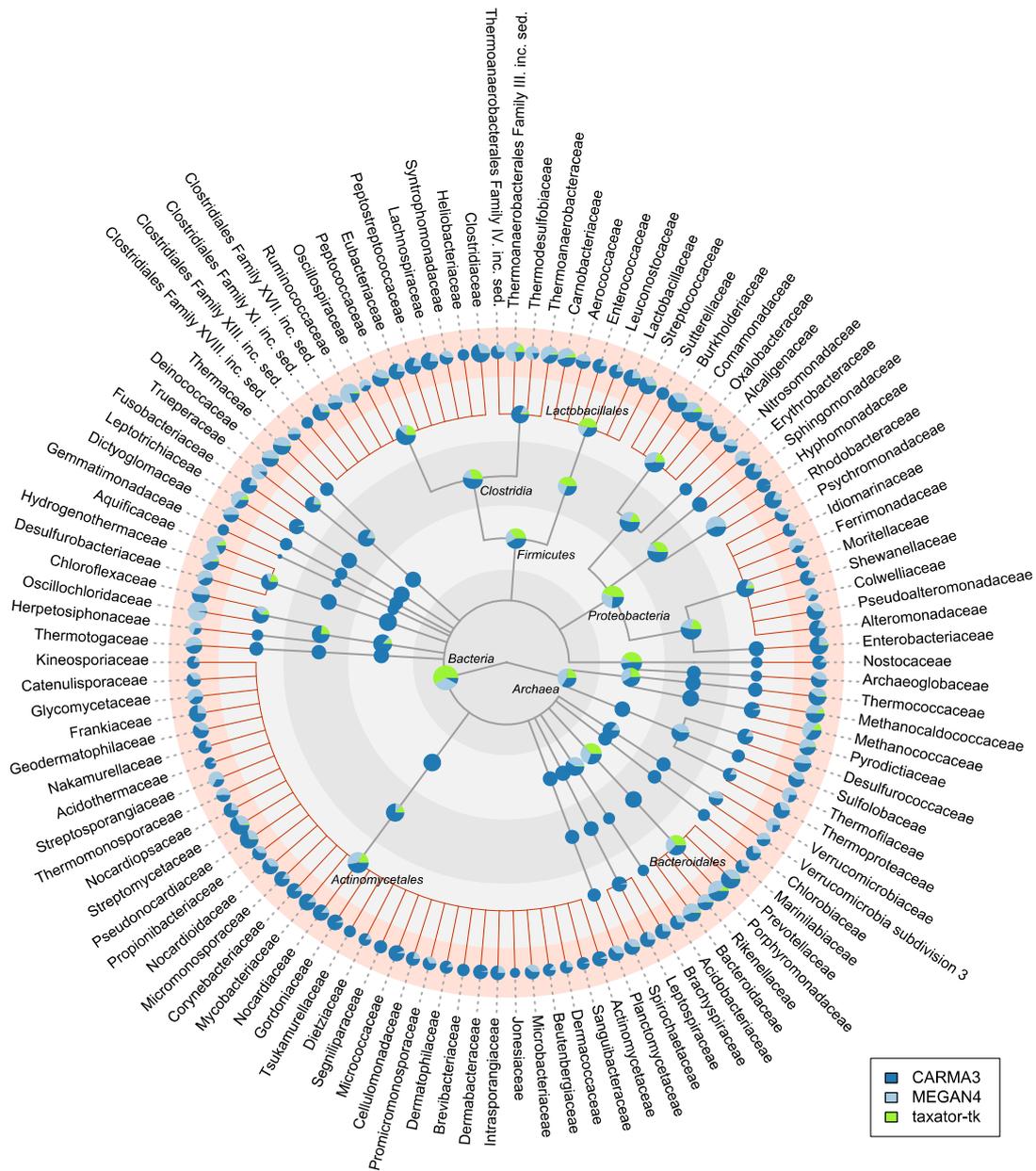

**Figure 4: Comparison of taxonomic assignments for sequences from novel families with three different classifiers on the simulated metagenome sample (simArt49e) with 49 species**

Taxonomic assignments of *CARMA3* (dark blue), *MEGAN4* (light blue) and *taxator-tk* (green) for simArt49e in the cross-validation experiment, for which all sequences of the same families were removed from the reference sequence data (Supplementary Fig. 9–11e). Only assignments at the order level or above can therefore be correctly assigned, and all family level assignments (shaded in light red) are incorrect. For further details on the precision and recall of the methods, see Supplementary Fig. 9–11e and Fig. 5. Due to the large number of taxa additionally predicted by CARMA3 and MEGAN4, only the 32 existing order level taxa of simArt49e are shown, as well as the predicted families of these orders, where -assignments below family level were included in counts at the family level.

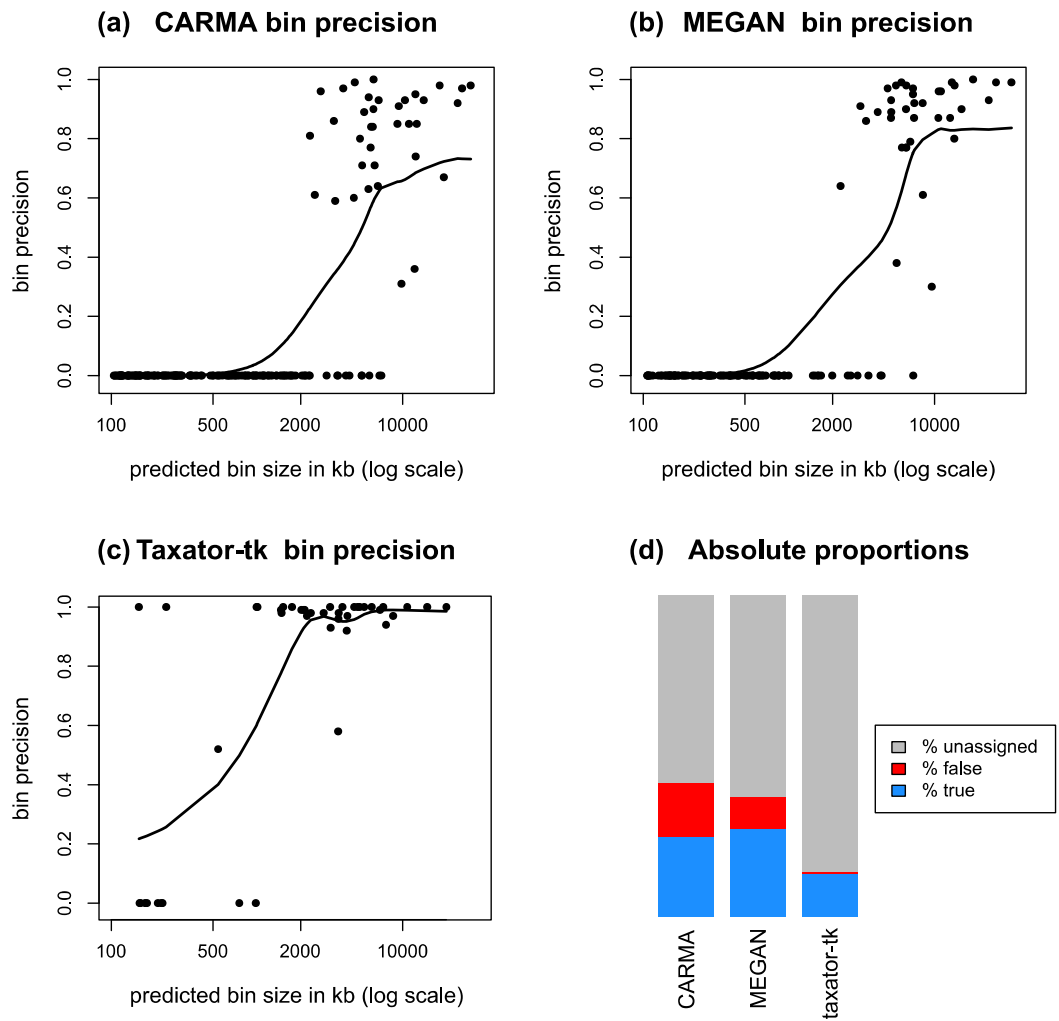

**Figure 5: Family-level bin precision for the simulated metagenome sample with 49 species (simArt49e)**

Comparison of the assignment precision to bin sizes (logarithmic scale) for individual family-level bins of simArt49e sample. We show the precision for the seven cross-validation experiments assessing taxonomic assignment performance combined across a range of evolutionary distances between the query sequence and the reference sequences. Correct assignments at ranks below family were also considered correct at the family level. (a–c) show the corresponding data, excluding the 1% (in total bp) of least abundant bins for (a) *CARMA3,* (b) *MEGAN4* and (c) *taxator-tk*. We added a smoothed k-nearest-neighbor estimate of the mean precision using the R function *wapply* (width=0.3) followed by *smooth.spline* (df=10). *CARMA3* and *MEGAN4* identified substantially more small taxonomic bins which were not present in the analyzed dataset than *taxator-tk*. Panel (d) gives the number of correct, false and undetermined family-level assignments for the different classifiers on the dataset.

# Supplementary Methods
# for
# 'Taxator-tk: Fast and Reliable Taxonomic Assignment of Metagenomes by Approximating Evolutionary Neighborhoods'

**Taxonomic Assignment of Sequence Segments**

   Here we describe in detail the individual steps and the run-time properties of the algorithm which is implemented in the program *taxator*, the second stage of the overall binning workflow using *taxator-tk* (Fig. 2b). We propose the re-alignment placement algorithm (RPA) for the taxonomic assignment of a query segment $q$, which can be any subsequence of the full query sequence (i.e. read, contig or scaffold). The algorithm constitutes **two pair-wise alignment passes** and in each, $q$ is aligned to segments of nucleotide reference sequences. It aims at identifying as many as possible taxa of the prediction clade (node R in Fig. 2a) in a set $M$ without explicitly resolving its phylogenetic structure.

1. Among the given set of homologous segments constructed from overlapping alignments before application of the RPA, we define $s$ to be the most similar segment to $q$, i.e. the one with the best local alignment score of all reference segments. In the first pass, all segments are aligned against $s$ ($n$ alignments). The resulting pair-wise scores, our implementation uses the **edit distance** (mismatches + gaps), define an ordering among all segments or their corresponding taxa. The distinction between segments and associated taxa will be neglected in the following for better readability. All taxa which are less distant to $s$ than $q$, including $s$ itself, are added to $M$ which holds all identified taxa of the prediction clade. The first more distant taxon than $q$ is defined to be the outgroup segment $o$ (Fig. 1) and used as the alignment target in the following second and last pass in which similar taxa to $o$ are added $M$.

2. We align all segments, including $q$, against $o$ and rank the resulting scores. Then we add all taxa to $M$ which have a lower score than $q$. With some fine-tuning, we chose to also add taxa with a higher score than $q$, within a small range accounting for erroneous scores, because $o$ and $q$ can be very distant homologs with noisy alignment. The width of this **error band** is determined on a per-segment basis as a linear score function of the taxonomic disorder in the alignment scores and not a universal or configurable run-time parameter. We interpret a rank disorder (e.g. a known family member of $o$ being more similar to *o* than a corresponding species member segment) as a discordance between gene tree and taxonomy and proportionally scale the effective score of *q* to enlarge *M* by taxa which are slightly more distant to *o* than *q*. This second pass requires $n-1$ new alignments, or less if some segments are identical to either *q or s*.

If multiple best references (*s)* or outgroup segments (*o)* were present in these two passes with identical alignment scores, the calculations are repeated for every such segment in order to produce stable output. We reduced the additional computational effort in our implementation by detecting frequent identical segments and uninformative homologs. The final assignment taxon ID of *q* is the LCA of the taxa in *M*, or none if no outgroup had been found. The theoretical run-time of the segment assignment algorithm is in $O(n)$ and about $2n$. Segments with an excessive number of homologs, most often short segments of abundant and uninformative regions, have a negative impact on the program run-time. Therefore we currently limit the number of homologs per query to the **top-scoring 50** by default (configurable run-time parameter in program *alignments-filter* or directly in the local alignment search program), before passing them to *taxator*. Other tested **values gave similar results** and the parameter, if changed, should be chosen based on hardware limitations. If this parameter is set lower, then the number of reference segments drops below a critical value such that no outgroup can be determined for some *q* and which therefore remain unassigned (but without impacting the taxon ID of other segments).

**Consensus Binning Algorithm**

Due to sparse segments and taxonomic assignment thereof with *taxator* in stage two of the workflow (Fig. 2b), a final processing step (Fig. 2c) is required to determine a taxon ID for the entire query sequence. Therefore we have implemented a simplistic, weighted consensus assignment scheme in the program *binner,* which optionally permits to apply custom constraints, e.g. the minimum percentage identity (PID) for classification at the species level or the removal of taxa with low counts in the whole sample. However, there are currently only **two mandatory run-time parameters** to control the actual post-processing consensus algorithm. First we define the support of a query segment to be the number of total matching positions to the best reference segment. The first run-time parameter specifies the **minimum combined support** at any rank (50 positions by default) and serves to ignore false predictions caused by short and often noisy segments. The other parameter specifies the **minimum percentage of the summed support** (70% by default) to allow a majority taxon to outvote a contradicting minority. Inconsistent taxa below this support are resolved by the LCA operation until the threshold is reached. Probably due to the conservative nature of the RPA, we found those two parameters to have minimal impact on the binning results in practice. The actual output of *taxator* is very detailed, allowing diverse information to be taken into account. We provide Python language bindings for processing with other applications.

**Taxonomy and Phylogeny**

*Taxator-tk* assumes that the NCBI taxonomy used for the assignment correctly captures the evolutionary process of speciation, although we know that the categorization of some taxa might be inconsistent with their evolution. If the phylogenetic information inferred from similarity scores disagrees with the taxonomic structure, assignments are made to a consistent higher rank. For instance horizontal gene transfer and upstream sequence misassembly can cause multiple similar copies of a sequence to be distributed across unrelated taxa. In case a query sequence cannot be traced by the algorithm to have evolved with either copy, it is usually assigned to the LCA of these clades. However, if the donor clade is unknown, the query may also be assigned to the recipient clade and the horizontal transfer or missassembly can go undetected. Thus assignment errors caused by the evolution of genes, upstream technical errors or taxonomy cannot always be eliminated in this framework. It remains to be assessed whether the use of an alternative microbial taxonomy such as the GreenGenes or the SILVA taxonomy would improve on the taxonomic assignment.

**Comparison and Innovations**

*Taxator-tk* shares some ideas with previous programs: Starting with *MEGAN*[1], which uses local alignments scores to derive a "neighborhood of related sequences" and the taxonomic estimate is the LCA of the corresponding taxa. This neighborhood threshold is a percentage of the local alignment score and can be interpreted to reflect the rate of evolution within a taxonomic group. Its value is empirical and lacks stronger justification. The threshold selection has been improved in *taxator-tk* and other programs. To our knowledge, SOrt-ITEMS[2] was the first algorithm to use the logic of re-alignment to the best reference (termed reciprocal similarity) for read assignment but is restricted to protein level alignment and is implemented as a wrapper around NCBI BLAST[3]. Protein alignment in general triples the run-time of the local alignment step (translation into three frame shifts) and cannot make use of faster nucleotide aligners. SOrt-ITEMS also uses fixed similarity thresholds in terms of percentage identity to define universal levels of conservation within taxonomic groups assuming the same rate of evolution for different genetic regions and clades. Furthermore SOrt-ITEMS was primarily designed for reads and if it performs well for longer sequences, its run-time increases proportionally with input sequence lengths. Both, *taxator-tk* and *CARMA3*[4], adopt the logic of reciprocal alignment and extends it and remove the assumption of universal conservation levels. *CARMA3* accounts for a heterogenous rate of evolution for different genetic regions. The initial identification of similar sequences in the reference can be based on nucleotide or protein BLAST search or profile Hidden Markov Models with HMMER[5]. In BLAST mode, *CARMA3*, like SOrt-ITEMS, uses one reciprocal alignment search and then extra or interpolates alignment scores to select a taxonomic rank for prediction. It therefore assumes a parametric model for the conservation level at a taxonomic rank, a linear function which is fitted to local alignment scores.

With *taxator-tk*, we use a non-parametric score ranking algorithm, instead. Also, to our knowledge, we provide the first algorithm to determine an outgroup and to sparsify the input data being able to assign distinct regions on the query sequence to possibly different taxonomic groups. Also, we at most assume segment-wise constant rates of evolution (equally long branches from a common ancestor). This makes the major algorithmic component parameter-less and robust in itself, independent of the segment size. Through the sparsification procedure it can deal with structural rearrangements among distant relatives and scales better with the length of the input sequences. The individual segment assignments allow for a robust consensus voting scheme for the assignment of entire sequence fragments. The segment-specific classification could also be used for the detection of horizontal gene transfers events (HGTs) and assembly errors. Different from most previous approaches, *taxator-tk* was developed and tested using fast nucleotide sequence local alignments instead of protein sequence alignments, although for the local alignments in stage 1 of the workflow both can be used, but with our data we did not find advantages in using protein alignments as input. Thus, taxonomic binning of a metagenome sample with *taxator-tk* requires no more than specification of reference sequences, their taxonomic affiliations and an aligner like *BLAST* or *LAST*[6]. On the implementation side, all workflow steps for taxonomic assignment with *taxator-tk* are designed in a modular way making it easy to save, compress, reuse or recompute results. The computation intensive classification of segments in *taxator* can be run in parallel on many CPU cores in which we make use of the open source C++ algorithm library SeqAn[7] for fast alignment.

**Performance Measures**

As metagenome data sets can have varying taxonomic composition in terms of which taxa are present and their relative abundances, this needs to be taken into consideration in evaluating taxonomic assignment methods. If an algorithm performs better for some clades than for others at a given rank we call it taxonomically biased. Oftentimes a classifier is biased, if it uses parameters that fit one clade better than another. This can be the case if the parameters were chosen to give good overall assignment accuracy (low total number of false predictions) on training data with uneven taxonomic composition. Such a method will not generalize well when applied to a sample of different taxonomic structure and abundances. To account for uneven taxonomic composition in evaluation data sets and to obtain comparable performance estimates across data sets of different taxonomic composition, we used as the primary evaluation measure the bin-averaged **precision** (or **positive predictive value**), also known as **macro-precision**.

$$\text{macro-precision} = \frac{1}{N_p} \sum_{i=1}^{N_p} \text{precision}_i \quad \text{(Equation 1)}$$

where $N_p$ is number of all predicted bins with a single bin precision

$$\text{precision}_i = \frac{\text{TP}_i}{\text{TP}_i + \text{FP}_i} \quad \text{(Equation 2)}$$

True positives ($\text{TP}_i$) are the correct assignments to the $i^{\text{th}}$ bin and false positives ($\text{FP}_i$) the incorrect assignments to the same bin.

The macro-precision is the fraction of correct sequence assignments over all assignments to a given taxonomic bin, averaged over all predicted bins for a given rank. For falsely predicted bins which do not occur in the data, the precision is therefore zero. This value reflects how trustworthy the bin assignments are on average from a user's perspective, as it is averaged overall predicted bins.

In addition to the macro-precision, we report the raw numbers of true and false predictions for every cross-validation, as well as a quick **overall precision** for pooled ranks.

This overall precision is most informative for species+genus+family and reports the fraction of true classifications among the predictions for all these ranks in a single pooled bin.

$$\text{overall-precision} = \frac{\text{TP}}{\text{TP} + \text{FP}} \qquad \text{(Equation 3)}$$

We measure the taxonomic bias of a method in terms of the standard deviation over all individual bin precisions.

$$\text{sd}_{prec} = \sqrt{\frac{1}{N_p} \sum_{i=1}^{N_p} \left(\text{precision}_i - \overline{\text{precision}}\right)^2} \qquad \text{(Equation 4)}$$

where

$$\overline{\text{precision}} = \frac{1}{N_p} \sum_{i=1}^{N_p} \text{precision}_i \qquad \text{(Equation 5)}$$

The standard deviation is small if all predicted bins have a similar precision. A universally good method should have a high macro-precision with a low taxonomic bias.

The **recall** (or **sensitivity**) is a measure of completeness of a predicted bin and, analogously, the **macro-recall** is the fraction of correctly assigned sequences of all sequences belonging to a certain bin, averaged over all existing bins in the test data[8].

$$\text{macro-recall} = \frac{1}{N_r} \sum_{i=1}^{N_r} \text{recall}_i \qquad \text{(Equation 6)}$$

where $C_r$ is a set of all actually existing bins in the test data and $c_i$ is a single such bin.

$$\text{recall}_i = \frac{\text{TP}_i}{\text{TP}_i + \text{FN}_i} \qquad \text{(Equation 7)}$$

False negatives ($\text{FN}_i$) are the assignments belonging to the $i^{\text{th}}$ bin but which where classified to another bin or left unassigned.

The macro-recall reflects how well the classifier works more from a developer's perspective than from the user perspective, as it is usually not known which predicted bins correspond to existing ones and which do not.

**Low-abundance Filtering**

The number of predicted bins at each rank can be quite large, at most the number of known taxa in the taxonomy and reference sequence data. When noise is considered to occur evenly distributed across this large output space, bins with few assigned sequences are more likely to be falsely identified, than larger bins (the chance to independently classify the same bin by chance *n* times is $(1/\text{number of possible bins})^n$. Since the macro precision is an average over all predicted bins, it is heavily affected by bins with few sequences assigned. As a result, classifiers that predict clades present at low frequencies in the sample score badly under this measure. To correct for this effect, we define a truncated average precision ignoring the least abundant predicted bins and consider only the **largest predicted bins constituting a minimum fraction $\alpha$ of the total assignments** (equal size bins are also included). This modification acts as a noise filter and accounts for different behavior of classifiers without explicitly considering the size of the model space or the number of existing species in the actual sample. We set $\alpha$ to 0.99 for our evaluations.

**Cross-validation**

Despite the limitations of simulated metagenomes, which incorporate assumptions about sequencing error rates or species abundance distributions, it is very informative to evaluate taxonomic assignment methods on simulated sequence data as real metagenome samples lack taxon IDs for evaluation. Our canonical way of evaluating a method on simulated data is a version of **leave-one-out cross-validation**: Each query sequence is classified by removing all identical or related sequences up to a given rank from the reference collection: For example, to assess the performance in assigning query sequences from a new species, all sequences belonging to this species are removed from the reference sequence collection for the classifier. Performance measures (macro-recall, macro-precision), along with other statistics (true/false/unassigned data, overall precision, bin counts) which are available in the coupled tables, were normally calculated for number of assigned basepairs or for the number of assigned sequences, if these had comparable lengths. These values were calculated for all ranks (species, genus, family, order, class, phylum, domain/superkingdom) for seven simulations: either all reference data was used (per query) or all data from the query species, genus, family, order, class or phylum was removed from the reference data prior to classification. The assignments of these seven cross-validation experiments were averaged for a combined performance summary with standard measures.

**Consistency Analysis**

In order to evaluate the predictions of *taxator-tk* for real metagenome samples where no underlying correct taxon IDs are known for the sequences, we assigned sequences linked by assembly and calculate a measure of assignment consistency. We split long contigs into multiple pieces and classified each piece independently. Assuming that the sequence assembly was correct in the first place, contradicting assignments of pieces that originate from the same contig represent false assignments. This unveils part of the errors made by a particular method but some, if not the majority, will go undetected because the actual ID stays unknown and the assignments for a contig can be consistently wrong. Hence these results are generally more difficult to interpret than those from simulated data.

**Nucleotide Alignment**

In the course of evaluation we created many local alignments as input to the taxonomic assignment programs *CARMA3*, *MEGAN4* and *taxator-tk*. These were usually generated using the fast alignment program *LAST* because it ran faster without noticeable differences in the output alignments than *BLAST* nucleotide alignments. For short sequence length evaluation (Supplementary Fig. 2-4), evaluation of a published SimMC scenario (Supplementary Fig. 12) and evaluation of a simulated metagenome sample with 49 species (Fig. 3, Supplementary Fig. 9-11), standard *BLAST* search was chosen. We used the standard alignment parameters and scoring schemes with each aligner. The generated alignments were converted into BLAST tabular format to work with *CARMA3* and *MEGAN4*.

**Program Parameters**

For taxonomic assignment with *MEGAN4* we used minscore=2, toppercent=20, minsupport=5 and mincomplexity=0.44 parameters. In *CARMA3*, we used the standard parameters in the contained configuration file. *Taxator-tk* was restricted the 50 best scoring local alignments to avoid long run-times for some of the query sequences. This was purely a convenient filter at the current state of development and is meant to be replaced by an adaptive per-segment heuristic.

**16S Evaluation**

We evaluated the performance of *taxator-tk* in classifying the most widely used taxonomic marker gene in studies of microbial diversity, the 16S rRNA gene, as a proof of concept. For our evaluation, we extracted 7,175 annotated 16S rRNA genes each with a minimum length of 1 kb from *mRefSeq47* (Suppl. Fig. 19, 20). The sequences were assigned with *taxator-tk* using the entire mRefSeq as reference, not just 16S genes. The cross-validation assesses the performance of 16S gene assignment in a wide range of situations. The performance statistics were calculated based on the number of assigned sequences, as all have comparable length. When using the complete reference sequences, 87% of sequences were assigned to the ranks of species, genus and family with 100% accuracy (Supplementary Fig. 1b), the remaining 13% were correctly assigned at higher ranks. This is an ideal situation showing the rank depth baseline on our data set. In more realistic simulations, when we tested assignment of genes from novel species or novel higher-level clades, assignments were accordingly made to higher ranks in most cases. For instance, when simulation novels species, 2678 contigs were assigned to the correct genera, while 491 erroneous species and genus assignments were made. The macro-precision in the combined cross-validation (Fig. 2) was always above 92%, with standard deviations from 10 to 25%, which demonstrates a good and even performance of *taxator-tk* for all clades in the case of 16S rRNA data.

**Supplementary Files**

The submission includes the files which are necessary to reproduce the results which are shown in the article. A more complete benchmark data-set can be downloaded from the software download web page.

Supplementary Figure 1: 16S gene assignment with taxator-tk **(a) summary scenario**

| rank | depth | true | false | unknown | macro precision α=0.99 | stdev | pred. bins | macro recall | stdev | real bins | sum true | sum false | overall prec. | description |
|---|---|---|---|---|---|---|---|---|---|---|---|---|---|---|
| unassigned | 0 | 274.4 | 0.0 | 0 | 100.0 | 0.0 | 1 | 100.0 | 0.0 | 1 | | | | root+superkingdom |
| superkingdom | 1 | 2159.6 | 0.0 | 0 | 100.0 | 0.0 | 2 | 82.7 | 14.2 | 2 | 4593.6 | 0.0 | 100.0 | |
| phylum | 2 | 869.1 | 66.6 | 0 | 95.5 | 13.8 | 13 | 33.5 | 23.6 | 32 | | | | phylum+class+order |
| class | 3 | 1417.9 | 92.6 | 0 | 96.1 | 10.7 | 25 | 27.3 | 18.4 | 52 | 2707.1 | 228.6 | 92.2 | |
| order | 4 | 420.1 | 69.4 | 0 | 95.4 | 12.6 | 62 | 20.7 | 14.2 | 109 | | | | |
| family | 5 | 471.6 | 26.1 | 0 | 95.7 | 13.5 | 148 | 16.0 | 11.9 | 235 | | | | family+genus+species |
| genus | 6 | 636.7 | 65.0 | 0 | 95.8 | 16.1 | 342 | 9.2 | 8.9 | 615 | 1732.1 | 174.1 | 90.9 | |
| species | 7 | 623.8 | 83.0 | 0 | 92.6 | 24.5 | 570 | 5.1 | 6.7 | 1416 | | | | |
| avg/sum | 2.6 | 6598.8 | 402.7 | 0 | 95.9 | 13.0 | 166.0 | 27.8 | 14.0 | 351.6 | | | 94.2 | all but unassigned |
| avg/sum | 2.6 | 6873.3 | 402.7 | 0 | 96.4 | 11.4 | 145.4 | 36.8 | 12.2 | 307.8 | | | 94.5 | all with unassigned |

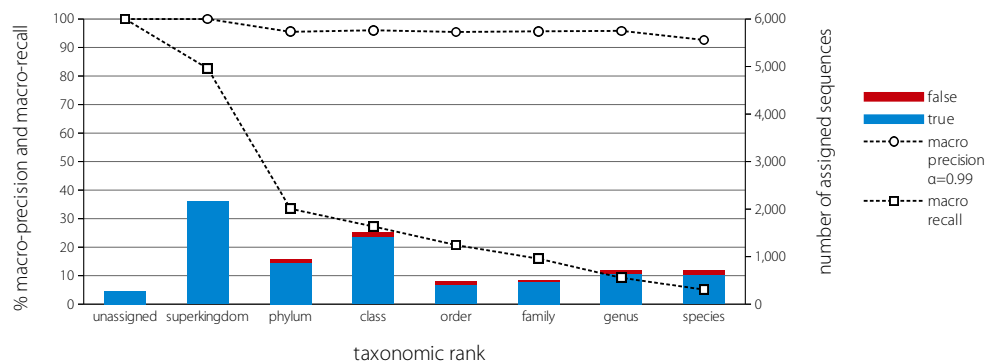

Supplementary Figure 1: 16S gene assignment with taxator-tk **(b) all reference scenario**

| rank | depth | true | false | unknown | macro precision α=0.99 | stdev | pred. bins | macro recall | stdev | real bins | sum true | sum false | overall prec. | description |
|---|---|---|---|---|---|---|---|---|---|---|---|---|---|---|
| unassigned | 0 | 10 | 0 | 0 | 100.0 | 0.0 | 1 | 100.0 | 0.0 | 1 | | | | root+superkingdom |
| superkingdom | 1 | 80 | 0 | 0 | 100.0 | 0.0 | 2 | 97.6 | 2.4 | 2 | 170 | 0 | 100.0 | |
| phylum | 2 | 113 | 0 | 0 | 100.0 | 0.0 | 16 | 78.3 | 38.3 | 32 | | | | phylum+class+order |
| class | 3 | 428 | 0 | 0 | 100.0 | 0.0 | 29 | 77.6 | 37.2 | 52 | 813 | 0 | 100.0 | |
| order | 4 | 272 | 0 | 0 | 100.0 | 0.0 | 67 | 72.4 | 39.6 | 109 | | | | |
| family | 5 | 750 | 0 | 0 | 100.0 | 0.0 | 158 | 71.0 | 40.4 | 235 | | | | family+genus+species |
| genus | 6 | 1779 | 0 | 0 | 100.0 | 0.0 | 337 | 53.9 | 48.0 | 615 | 6272 | 0 | 100.0 | |
| species | 7 | 3743 | 0 | 0 | 100.0 | 0.0 | 504 | 35.8 | 46.8 | 1416 | | | | |
| avg/sum | 5.0 | 7165 | 0 | 0 | 100.0 | 0.0 | 159.0 | 69.5 | 36.1 | 351.6 | | | 100.0 | all but unassigned |
| avg/sum | 5.0 | 7175 | 0 | 0 | 100.0 | 0.0 | 139.3 | 73.3 | 31.6 | 307.8 | | | 100.0 | all with unassigned |

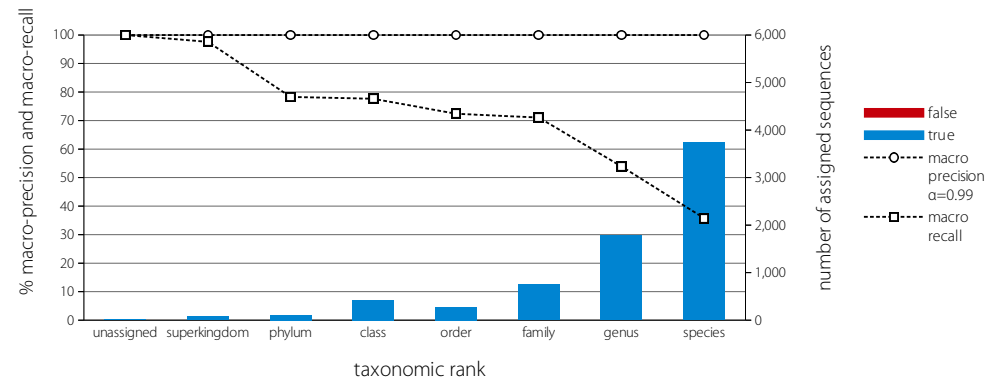

Supplementary Figure 1: 16S gene assignment with taxator-tk **(c) new species scenario**

| rank | depth | true | false | unknown | macro precision α=0.99 | stdev | pred. bins | macro recall | stdev | real bins | sum true | sum false | overall prec. | description |
|---|---|---|---|---|---|---|---|---|---|---|---|---|---|---|
| unassigned | 0 | 22 | 0 | 0 | 100.0 | 0.0 | 1 | 100.0 | 0.0 | 1 | | | | root+superkingdom |
| superkingdom | 1 | 313 | 0 | 0 | 100.0 | 0.0 | 2 | 96.7 | 3.2 | 2 | 648 | 0 | 100.0 | |
| phylum | 2 | 347 | 2 | 0 | 100.0 | 0.0 | 14 | 54.8 | 39.5 | 32 | | | | phylum+class+order |
| class | 3 | 989 | 8 | 0 | 98.9 | 3.9 | 26 | 53.9 | 41.6 | 52 | 2284 | 19 | 99.2 | |
| order | 4 | 948 | 9 | 0 | 98.8 | 7.9 | 54 | 44.2 | 39.6 | 109 | | | | |
| family | 5 | 1350 | 18 | 0 | 98.2 | 7.3 | 91 | 28.8 | 37.5 | 235 | | | | family+genus+species |
| genus | 6 | 2678 | 64 | 0 | 95.2 | 18.9 | 88 | 10.8 | 26.3 | 615 | 4028 | 509 | 88.8 | |
| species | 7 | 0 | 427 | 0 | 0.0 | 0.0 | 54 | 0.0 | 0.0 | 1416 | | | | |
| avg/sum | 4.6 | 6625 | 528 | 0 | 84.4 | 5.4 | 47.0 | 41.3 | 26.8 | 351.6 | | | 92.6 | all but unassigned |
| avg/sum | 4.6 | 6647 | 528 | 0 | 86.4 | 4.7 | 41.3 | 48.6 | 23.5 | 307.8 | | | 92.6 | all with unassigned |

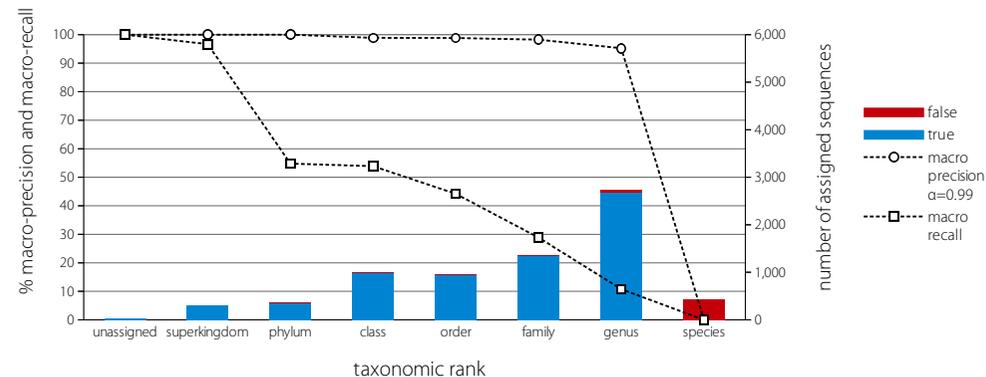

Supplementary Figure 1: 16S gene assignment with taxator-tk **(d) new genus scenario**

| rank | depth | true | false | unknown | macro precision α=0.99 | stdev | pred. bins | macro recall | stdev | real bins | sum true | sum false | overall prec. | description |
|---|---|---|---|---|---|---|---|---|---|---|---|---|---|---|
| unassigned | 0 | 48 | 0 | 0 | 100.0 | 0.0 | 1 | 100.0 | 0.0 | 1 | | | | root+superkingdom |
| superkingdom | 1 | 804 | 0 | 0 | 100.0 | 0.0 | 2 | 96.5 | 3.0 | 2 | 1656 | 0 | 100.0 | |
| phylum | 2 | 1098 | 2 | 0 | 100.0 | 0.0 | 12 | 46.8 | 38.9 | 32 | | | | phylum+class+order |
| class | 3 | 2392 | 8 | 0 | 98.5 | 4.7 | 22 | 36.3 | 35.7 | 52 | 4680 | 19 | 99.6 | |
| order | 4 | 1190 | 9 | 0 | 95.4 | 17.8 | 48 | 25.2 | 32.9 | 109 | | | | |
| family | 5 | 1201 | 36 | 0 | 78.8 | 39.2 | 59 | 11.7 | 26.7 | 235 | | | | family+genus+species |
| genus | 6 | 0 | 344 | 0 | 0.0 | 0.0 | 34 | 0.0 | 0.0 | 615 | 1201 | 423 | 74.0 | |
| species | 7 | 0 | 43 | 0 | 0.0 | 0.0 | 8 | 0.0 | 0.0 | 1416 | | | | |
| avg/sum | 3.3 | 6685 | 442 | 0 | 67.5 | 8.8 | 26.4 | 30.9 | 19.6 | 351.6 | | | 93.8 | all but unassigned |
| avg/sum | 3.3 | 6733 | 442 | 0 | 71.6 | 7.7 | 23.5 | 39.6 | 17.2 | 307.8 | | | 93.8 | all with unassigned |

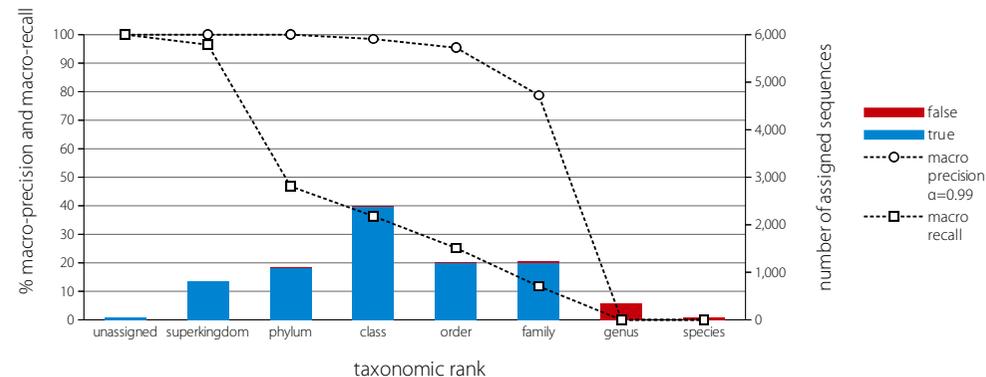

| rank | depth | true | false | unknown | macro precision α=0.99 | stdev | pred. bins | macro recall | stdev | real bins | sum true | sum false | overall prec. | description |
|---|---|---|---|---|---|---|---|---|---|---|---|---|---|---|
| unassigned | 0 | 299 | 0 | 0 | 100.0 | 0.0 | 1 | 100.0 | 0.0 | 1 | | | | root+superkingdom |
| superkingdom | 1 | 1321 | 0 | 0 | 100.0 | 0.0 | 2 | 82.8 | 13.7 | 2 | 2941 | 0 | 100.0 | |
| phylum | 2 | 1442 | 2 | 0 | 100.0 | 0.0 | 7 | 25.4 | 35.2 | 32 | | | | |
| class | 3 | 3485 | 11 | 0 | 97.7 | 7.2 | 13 | 15.0 | 26.2 | 52 | 5458 | 28 | 99.5 | phylum+class+order |
| order | 4 | 531 | 15 | 0 | 69.6 | 42.6 | 28 | 3.4 | 12.1 | 109 | | | | |
| family | 5 | 0 | 38 | 0 | 0.0 | 0.0 | 24 | 0.0 | 0.0 | 235 | | | | |
| genus | 6 | 0 | 17 | 0 | 0.0 | 0.0 | 9 | 0.0 | 0.0 | 615 | 0 | 69 | 0.0 | family+genus+species |
| species | 7 | 0 | 14 | 0 | 0.0 | 0.0 | 3 | 0.0 | 0.0 | 1416 | | | | |
| avg/sum | 2.4 | 6779 | 97 | 0 | 52.5 | 7.1 | 12.3 | 18.1 | 12.5 | 351.6 | | | 98.6 | all but unassigned |
| avg/sum | 2.4 | 7078 | 97 | 0 | 58.4 | 6.2 | 10.9 | 28.3 | 10.9 | 307.8 | | | 98.6 | all with unassigned |

### taxator-tk on RefSeq 16S genes

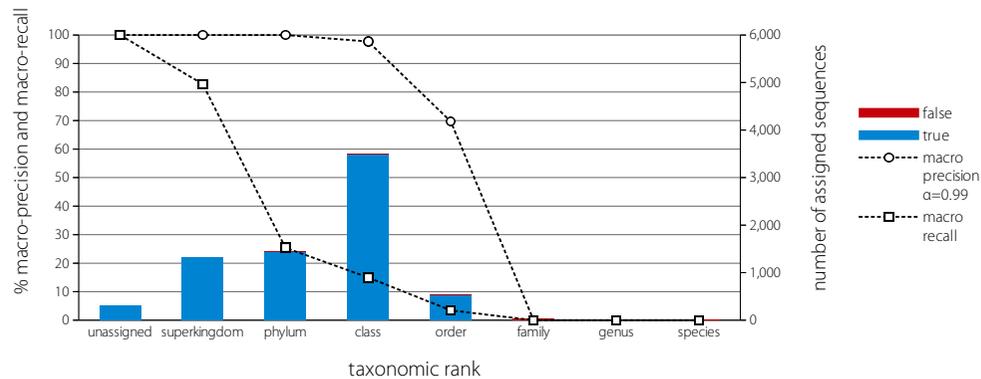

| rank | depth | true | false | unknown | macro precision α=0.99 | stdev | pred. bins | macro recall | stdev | real bins | sum true | sum false | overall prec. | description |
|---|---|---|---|---|---|---|---|---|---|---|---|---|---|---|
| unassigned | 0 | 424 | 0 | 0 | 100.0 | 0.0 | 1 | 100.0 | 0.0 | 1 | | | | root+superkingdom |
| superkingdom | 1 | 1920 | 0 | 0 | 100.0 | 0.0 | 2 | 72.9 | 22.3 | 2 | 4264 | 0 | 100.0 | |
| phylum | 2 | 1665 | 2 | 0 | 100.0 | 0.1 | 6 | 19.1 | 31.6 | 32 | | | | |
| class | 3 | 2631 | 12 | 0 | 99.6 | 1.1 | 8 | 7.9 | 20.8 | 52 | 4296 | 448 | 90.6 | phylum+class+order |
| order | 4 | 0 | 434 | 0 | 0.0 | 0.0 | 17 | 0.0 | 0.0 | 109 | | | | |
| family | 5 | 0 | 74 | 0 | 0.0 | 0.0 | 11 | 0.0 | 0.0 | 235 | | | | |
| genus | 6 | 0 | 2 | 0 | 0.0 | 0.0 | 3 | 0.0 | 0.0 | 615 | 0 | 87 | 0.0 | family+genus+species |
| species | 7 | 0 | 11 | 0 | 0.0 | 0.0 | 1 | 0.0 | 0.0 | 1416 | | | | |
| avg/sum | 2.1 | 6216 | 535 | 0 | 42.8 | 0.2 | 6.9 | 14.3 | 10.7 | 351.6 | | | 92.1 | all but unassigned |
| avg/sum | 2.1 | 6640 | 535 | 0 | 49.9 | 0.1 | 6.1 | 25.0 | 9.3 | 307.8 | | | 92.5 | all with unassigned |

### taxator-tk on RefSeq 16S genes

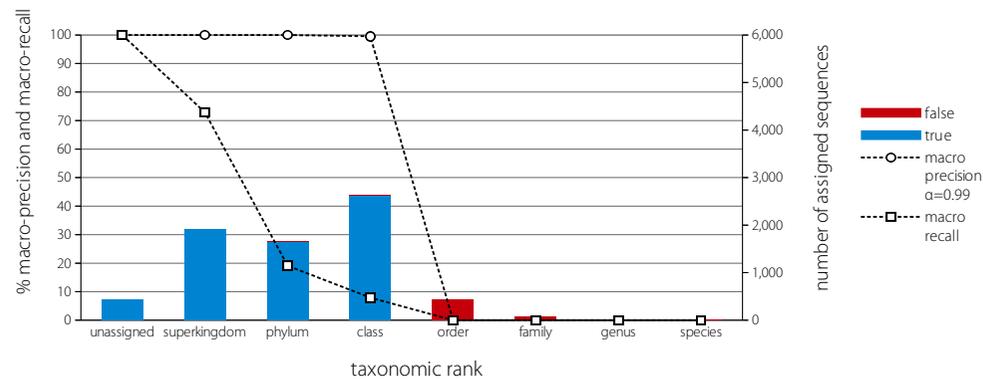

| rank | depth | true | false | unknown | macro precision α=0.99 | stdev | pred. bins | macro recall | stdev | real bins | sum true | sum false | overall prec. | description |
|---|---|---|---|---|---|---|---|---|---|---|---|---|---|---|
| unassigned | 0 | 549 | 0 | 0 | 100.0 | 0.0 | 1 | 100.0 | 0.0 | 1 | | | | root+superkingdom |
| superkingdom | 1 | 4734 | 0 | 0 | 100.0 | 0.0 | 2 | 73.7 | 19.6 | 2 | 10017 | 0 | 100.0 | |
| phylum | 2 | 1419 | 67 | 0 | 90.3 | 15.4 | 4 | 9.8 | 23.9 | 32 | | | | |
| class | 3 | 0 | 390 | 0 | 0.0 | 0.0 | 8 | 0.0 | 0.0 | 52 | 1419 | 460 | 75.5 | phylum+class+order |
| order | 4 | 0 | 3 | 0 | 0.0 | 0.0 | 8 | 0.0 | 0.0 | 109 | | | | |
| family | 5 | 0 | 9 | 0 | 0.0 | 0.0 | 6 | 0.0 | 0.0 | 235 | | | | |
| genus | 6 | 0 | 1 | 0 | 0.0 | 0.0 | 2 | 0.0 | 0.0 | 615 | 0 | 13 | 0.0 | family+genus+species |
| species | 7 | 0 | 3 | 0 | 0.0 | 0.0 | 1 | 0.0 | 0.0 | 1416 | | | | |
| avg/sum | 1.2 | 6153 | 473 | 0 | 27.2 | 2.2 | 4.4 | 11.9 | 6.2 | 351.6 | | | 92.9 | all but unassigned |
| avg/sum | 1.2 | 6702 | 473 | 0 | 36.3 | 1.9 | 4.0 | 22.9 | 5.4 | 307.8 | | | 93.4 | all with unassigned |

### taxator-tk on RefSeq 16S genes

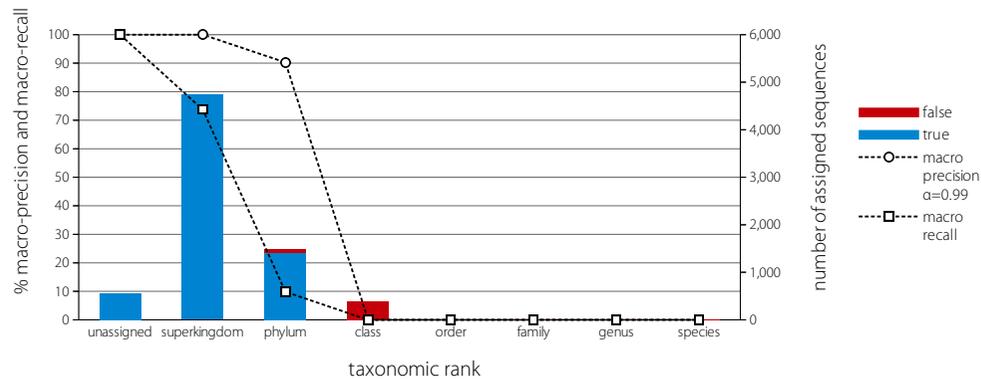

| rank | depth | true | false | unknown | macro precision α=0.99 | stdev | pred. bins | macro recall | stdev | real bins | sum true | sum false | overall prec. | description |
|---|---|---|---|---|---|---|---|---|---|---|---|---|---|---|
| unassigned | 0 | 569 | 0 | 0 | 100.0 | 0.0 | 1 | 100.0 | 0.0 | 1 | | | | root+superkingdom |
| superkingdom | 1 | 5945 | 0 | 0 | 100.0 | 0.0 | 2 | 58.5 | 35.3 | 2 | 12459 | 0 | 100.0 | |
| phylum | 2 | 0 | 391 | 0 | 0.0 | 0.0 | 5 | 0.0 | 0.0 | 32 | | | | |
| class | 3 | 0 | 219 | 0 | 0.0 | 0.0 | 8 | 0.0 | 0.0 | 52 | 0 | 626 | 0.0 | phylum+class+order |
| order | 4 | 0 | 16 | 0 | 0.0 | 0.0 | 7 | 0.0 | 0.0 | 109 | | | | |
| family | 5 | 0 | 8 | 0 | 0.0 | 0.0 | 5 | 0.0 | 0.0 | 235 | | | | |
| genus | 6 | 0 | 27 | 0 | 0.0 | 0.0 | 2 | 0.0 | 0.0 | 615 | 0 | 35 | 0.0 | family+genus+species |
| species | 7 | 0 | 0 | 0 | nan | nan | 0 | 0.0 | 0.0 | 1416 | | | | |
| avg/sum | 1.1 | 5945 | 661 | 0 | 16.7 | 0.0 | 4.0 | 8.4 | 5.0 | 351.6 | | | 90.0 | all but unassigned |
| avg/sum | 1.1 | 6514 | 661 | 0 | 28.6 | 0.0 | 3.6 | 19.8 | 4.4 | 307.8 | | | 90.8 | all with unassigned |

### taxator-tk on RefSeq 16S genes

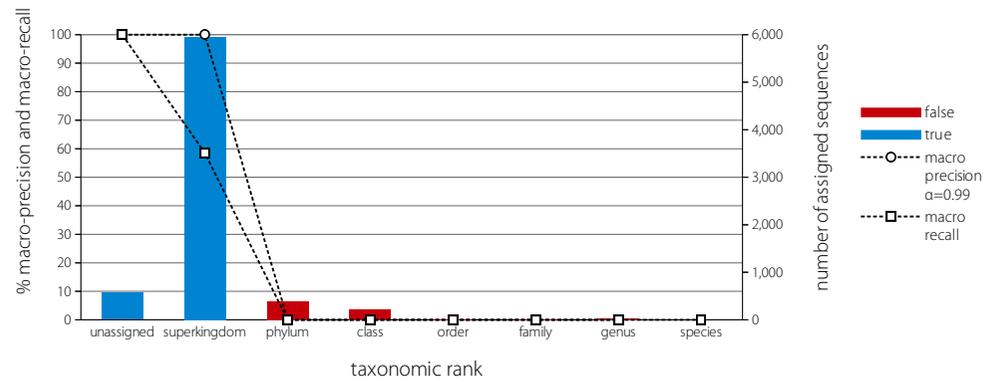

*Supplementary Figure 2: Simulated 100 bp sequence assignment with taxator-tk* **(a) summary scenario**

| rank | depth | true | false | unknown | macro precision α=0.99 | stdev | pred. bins | macro recall | stdev | real bins | sum true | sum false | overall prec. | description |
|---|---|---|---|---|---|---|---|---|---|---|---|---|---|---|
| unassigned | 0 | 37391.6 | 0.0 | 0 | 100.0 | 0.0 | 1 | 100.0 | 0.0 | 1 | 101937.3 | 427.3 | 99.6 | root+superkingdom |
| superkingdom | 1 | 32272.9 | 427.3 | 0 | 99.2 | 0.0 | 1 | 26.4 | 26.7 | 3 | | | | |
| phylum | 2 | 4563.7 | 2340.3 | 0 | 83.0 | 10.2 | 11 | 9.3 | 8.5 | 32 | 8977.7 | 4995.4 | 64.2 | phylum+class+order |
| class | 3 | 2164.1 | 1120.1 | 0 | 82.0 | 13.1 | 23 | 8.9 | 7.6 | 52 | | | | |
| order | 4 | 2249.9 | 1535.0 | 0 | 86.5 | 11.1 | 52 | 7.8 | 7.2 | 110 | | | | |
| family | 5 | 2859.3 | 591.9 | 0 | 85.1 | 14.7 | 98 | 5.8 | 6.8 | 240 | 13520.1 | 2415.0 | 84.8 | family+genus+species |
| genus | 6 | 7852.3 | 1275.7 | 0 | 87.3 | 17.6 | 202 | 3.5 | 5.6 | 656 | | | | |
| species | 7 | 2808.6 | 547.4 | 0 | 74.0 | 34.8 | 431 | 1.0 | 2.6 | 1697 | | | | |
| avg/sum | 2.4 | 54770.7 | 7837.7 | 0 | 85.3 | 14.5 | 116.9 | 8.9 | 9.3 | 398.6 | | | 87.5 | all but unassigned |
| avg/sum | 1.5 | 92162.3 | 7837.7 | 0 | 87.1 | 12.7 | 102.4 | 20.3 | 8.1 | 348.9 | | | 92.2 | all with unassigned |

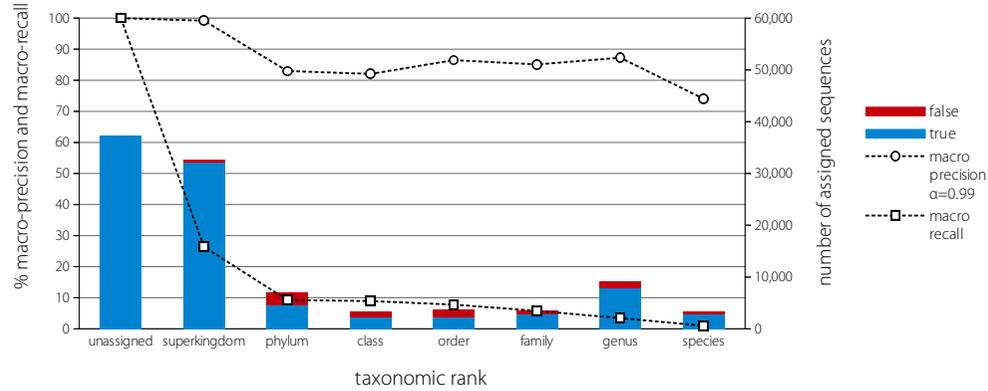

*Supplementary Figure 2: Simulated 100 bp sequence assignment with taxator-tk* **(b) all reference scenario**

| rank | depth | true | false | unknown | macro precision α=0.99 | stdev | pred. bins | macro recall | stdev | real bins | sum true | sum false | overall prec. | description |
|---|---|---|---|---|---|---|---|---|---|---|---|---|---|---|
| unassigned | 0 | 10662 | 0 | 0 | 100.0 | 0.0 | 1 | 100.0 | 0.0 | 1 | 47620 | 0 | 100.0 | root+superkingdom |
| superkingdom | 1 | 18479 | 0 | 0 | 100.0 | 0.0 | 2 | 48.0 | 37.0 | 3 | | | | |
| phylum | 2 | 4362 | 0 | 0 | 100.0 | 0.0 | 12 | 35.2 | 28.1 | 32 | 11598 | 0 | 100.0 | phylum+class+order |
| class | 3 | 2607 | 0 | 0 | 100.0 | 0.0 | 24 | 35.7 | 27.1 | 52 | | | | |
| order | 4 | 4629 | 0 | 0 | 100.0 | 0.0 | 54 | 33.9 | 28.2 | 110 | | | | |
| family | 5 | 8015 | 0 | 0 | 100.0 | 0.0 | 104 | 27.8 | 29.2 | 240 | 59261 | 0 | 100.0 | family+genus+species |
| genus | 6 | 31586 | 0 | 0 | 100.0 | 0.0 | 211 | 19.2 | 28.2 | 656 | | | | |
| species | 7 | 19660 | 0 | 0 | 100.0 | 0.0 | 365 | 6.9 | 18.2 | 1697 | | | | |
| avg/sum | 4.1 | 89338 | 0 | 0 | 100.0 | 0.0 | 110.3 | 29.5 | 28.0 | 398.6 | | | 100.0 | all but unassigned |
| avg/sum | 3.5 | 100000 | 0 | 0 | 100.0 | 0.0 | 96.6 | 38.4 | 24.5 | 348.9 | | | 100.0 | all with unassigned |

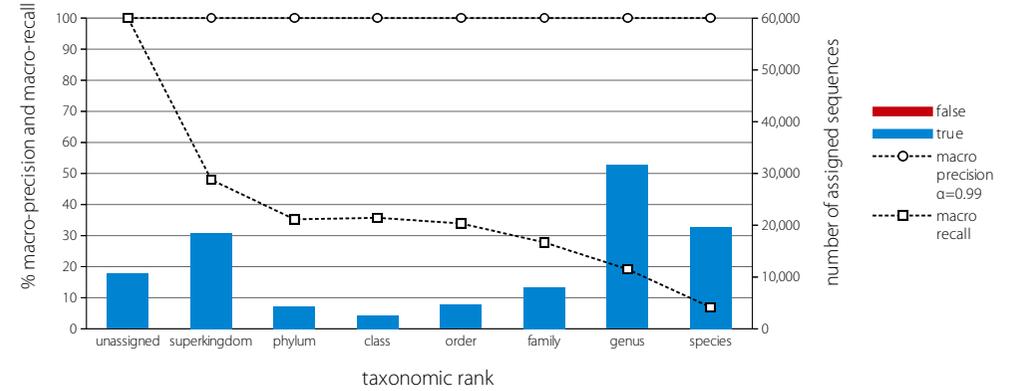

*Supplementary Figure 2: Simulated 100 bp sequence assignment with taxator-tk* **(c) new species scenario**

| rank | depth | true | false | unknown | macro precision α=0.99 | stdev | pred. bins | macro recall | stdev | real bins | sum true | sum false | overall prec. | description |
|---|---|---|---|---|---|---|---|---|---|---|---|---|---|---|
| unassigned | 0 | 22319 | 0 | 0 | 100.0 | 0.0 | 1 | 100.0 | 0.0 | 1 | 76901 | 252 | 99.7 | root+superkingdom |
| superkingdom | 1 | 27291 | 252 | 0 | 99.6 | 0.0 | 1 | 35.1 | 32.5 | 3 | | | | |
| phylum | 2 | 5362 | 746 | 0 | 97.2 | 1.8 | 10 | 17.5 | 18.3 | 32 | 13213 | 1541 | 89.6 | phylum+class+order |
| class | 3 | 3240 | 327 | 0 | 97.2 | 2.9 | 22 | 18.1 | 19.1 | 52 | | | | |
| order | 4 | 4611 | 468 | 0 | 97.3 | 3.3 | 45 | 15.4 | 18.5 | 110 | | | | |
| family | 5 | 7973 | 255 | 0 | 96.1 | 6.7 | 75 | 10.9 | 18.3 | 240 | 31353 | 4031 | 88.6 | family+genus+species |
| genus | 6 | 23380 | 776 | 0 | 90.4 | 21.6 | 100 | 5.0 | 14.2 | 656 | | | | |
| species | 7 | 0 | 3000 | 0 | 0.0 | 0.0 | 217 | 0.0 | 0.0 | 1697 | | | | |
| avg/sum | 3.4 | 71857 | 5824 | 0 | 82.5 | 5.2 | 67.1 | 14.6 | 17.3 | 398.6 | | | 92.5 | all but unassigned |
| avg/sum | 2.6 | 94176 | 5824 | 0 | 84.7 | 4.5 | 58.9 | 25.3 | 15.1 | 348.9 | | | 94.2 | all with unassigned |

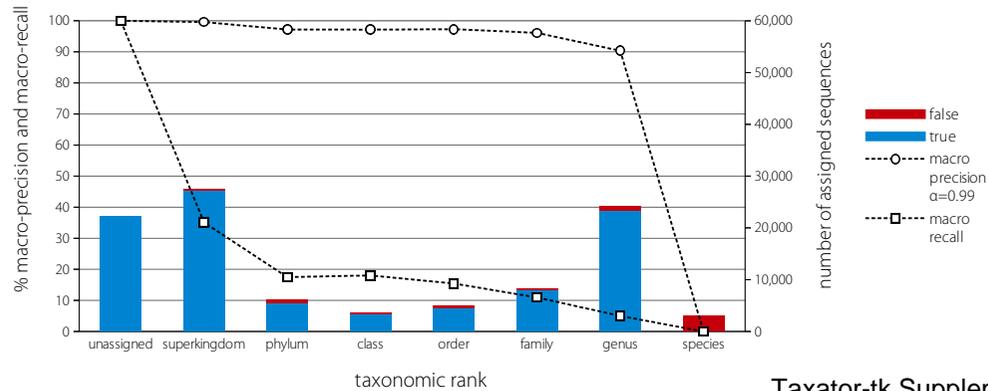

*Supplementary Figure 2: Simulated 100 bp sequence assignment with taxator-tk* **(d) new genus scenario**

| rank | depth | true | false | unknown | macro precision α=0.99 | stdev | pred. bins | macro recall | stdev | real bins | sum true | sum false | overall prec. | description |
|---|---|---|---|---|---|---|---|---|---|---|---|---|---|---|
| unassigned | 0 | 34909 | 0 | 0 | 100.0 | 0.0 | 1 | 100.0 | 0.0 | 1 | 110459 | 343 | 99.7 | root+superkingdom |
| superkingdom | 1 | 37775 | 343 | 0 | 99.3 | 0.0 | 1 | 27.2 | 28.0 | 3 | | | | |
| phylum | 2 | 6814 | 1406 | 0 | 82.6 | 22.0 | 8 | 6.9 | 9.4 | 32 | 14566 | 3057 | 82.7 | phylum+class+order |
| class | 3 | 3906 | 689 | 0 | 82.0 | 17.8 | 19 | 6.1 | 7.9 | 52 | | | | |
| order | 4 | 3846 | 962 | 0 | 80.6 | 17.2 | 44 | 4.4 | 7.5 | 110 | | | | |
| family | 5 | 4027 | 657 | 0 | 49.4 | 39.3 | 77 | 1.7 | 5.3 | 240 | 4027 | 5323 | 43.1 | family+genus+species |
| genus | 6 | 0 | 4422 | 0 | 0.0 | 0.0 | 193 | 0.0 | 0.0 | 656 | | | | |
| species | 7 | 0 | 244 | 0 | 0.0 | 0.0 | 103 | 0.0 | 0.0 | 1697 | | | | |
| avg/sum | 2.1 | 56368 | 8723 | 0 | 56.3 | 13.8 | 63.6 | 6.6 | 8.3 | 398.6 | | | 86.6 | all but unassigned |
| avg/sum | 1.4 | 91277 | 8723 | 0 | 61.7 | 12.0 | 55.8 | 18.3 | 7.3 | 348.9 | | | 91.3 | all with unassigned |

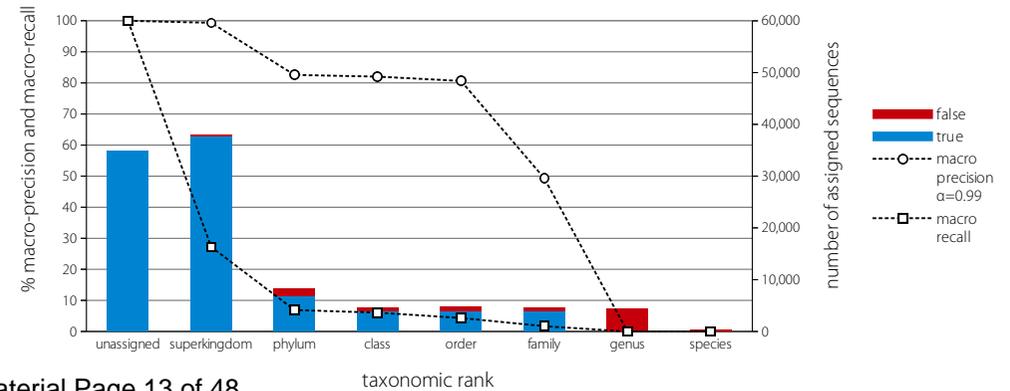

*Supplementary Figure 2: Simulated 100 bp sequence assignment with taxator-tk*  **(e) new family scenario**

| rank | depth | true | false | unknown | macro precision α=0.99 | stdev | pred. bins | macro recall | stdev | real bins | sum true | sum false | overall prec. | description |
|---|---|---|---|---|---|---|---|---|---|---|---|---|---|---|
| unassigned | 0 | 40215 | 0 | 0 | 100.0 | 0.0 | 1 | 100.0 | 0.0 | 1 | 120367 | 525 | 99.6 | root+superkingdom |
| superkingdom | 1 | 40076 | 525 | 0 | 98.9 | 0.0 | 1 | 22.2 | 27.0 | 3 | | | | |
| phylum | 2 | 6632 | 1627 | 0 | 80.7 | 7.2 | 6 | 2.9 | 5.6 | 32 | | | | |
| class | 3 | 3425 | 904 | 0 | 59.6 | 27.6 | 14 | 2.2 | 3.9 | 52 | 12720 | 3840 | 76.8 | phylum+class+order |
| order | 4 | 2663 | 1309 | 0 | 31.9 | 33.3 | 43 | 0.9 | 2.5 | 110 | | | | |
| family | 5 | 0 | 1045 | 0 | 0.0 | 0.0 | 120 | 0.0 | 0.0 | 240 | | | | |
| genus | 6 | 0 | 1413 | 0 | 0.0 | 0.0 | 133 | 0.0 | 0.0 | 656 | 0 | 2624 | 0.0 | family+genus+species |
| species | 7 | 0 | 166 | 0 | 0.0 | 0.0 | 80 | 0.0 | 0.0 | 1697 | | | | |
| avg/sum | 1.7 | 52796 | 6989 | 0 | 38.7 | 9.7 | 56.7 | 4.0 | 5.6 | 398.6 | | | 88.3 | all but unassigned |
| avg/sum | 1.0 | 93011 | 6989 | 0 | 46.4 | 8.5 | 49.8 | 16.0 | 4.9 | 348.9 | | | 93.0 | all with unassigned |

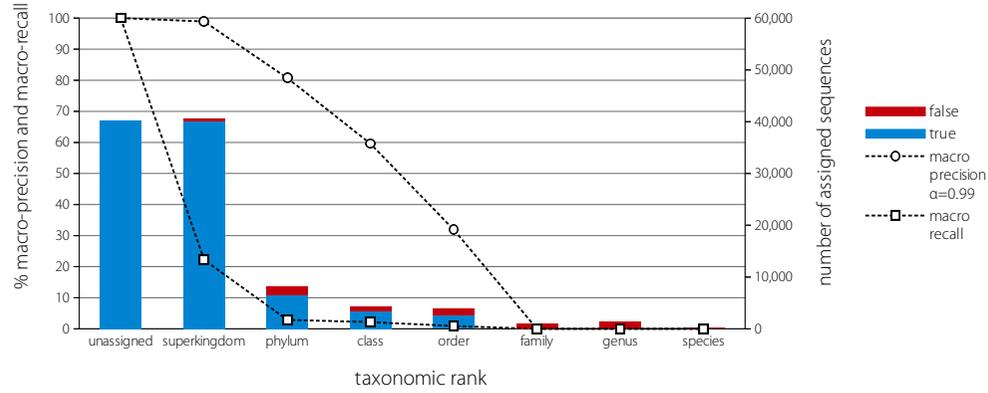

*Supplementary Figure 2: Simulated 100 bp sequence assignment with taxator-tk*  **(f) new order scenario**

| rank | depth | true | false | unknown | macro precision α=0.99 | stdev | pred. bins | macro recall | stdev | real bins | sum true | sum false | overall prec. | description |
|---|---|---|---|---|---|---|---|---|---|---|---|---|---|---|
| unassigned | 0 | 44037 | 0 | 0 | 100.0 | 0.0 | 1 | 100.0 | 0.0 | 1 | 121439 | 563 | 99.5 | root+superkingdom |
| superkingdom | 1 | 38701 | 563 | 0 | 98.8 | 0.0 | 1 | 20.3 | 25.5 | 3 | | | | |
| phylum | 2 | 5817 | 3314 | 0 | 60.2 | 21.5 | 6 | 1.8 | 4.0 | 32 | | | | |
| class | 3 | 1971 | 1414 | 0 | 23.3 | 23.8 | 18 | 0.6 | 1.5 | 52 | 7788 | 6917 | 53.0 | phylum+class+order |
| order | 4 | 0 | 2189 | 0 | 0.0 | 0.0 | 49 | 0.0 | 0.0 | 110 | | | | |
| family | 5 | 0 | 961 | 0 | 0.0 | 0.0 | 106 | 0.0 | 0.0 | 240 | | | | |
| genus | 6 | 0 | 873 | 0 | 0.0 | 0.0 | 118 | 0.0 | 0.0 | 656 | 0 | 1994 | 0.0 | family+genus+species |
| species | 7 | 0 | 160 | 0 | 0.0 | 0.0 | 72 | 0.0 | 0.0 | 1697 | | | | |
| avg/sum | 1.5 | 46489 | 9474 | 0 | 26.0 | 6.5 | 52.9 | 3.2 | 4.4 | 398.6 | | | 83.1 | all but unassigned |
| avg/sum | 0.9 | 90526 | 9474 | 0 | 35.3 | 5.7 | 46.4 | 15.3 | 3.9 | 348.9 | | | 90.5 | all with unassigned |

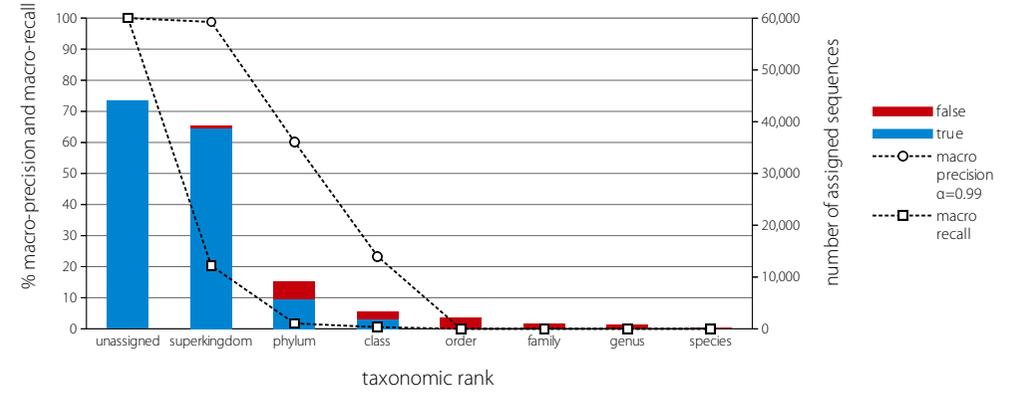

*Supplementary Figure 2: Simulated 100 bp sequence assignment with taxator-tk*  **(g) new class scenario**

| rank | depth | true | false | unknown | macro precision α=0.99 | stdev | pred. bins | macro recall | stdev | real bins | sum true | sum false | overall prec. | description |
|---|---|---|---|---|---|---|---|---|---|---|---|---|---|---|
| unassigned | 0 | 50150 | 0 | 0 | 100.0 | 0.0 | 1 | 100.0 | 0.0 | 1 | 122726 | 579 | 99.5 | root+superkingdom |
| superkingdom | 1 | 36288 | 579 | 0 | 98.6 | 0.0 | 1 | 17.9 | 22.8 | 3 | | | | |
| phylum | 2 | 2959 | 4203 | 0 | 27.7 | 24.4 | 7 | 0.7 | 2.0 | 32 | | | | |
| class | 3 | 0 | 2365 | 0 | 0.0 | 0.0 | 20 | 0.0 | 0.0 | 52 | 2959 | 8462 | 25.9 | phylum+class+order |
| order | 4 | 0 | 1894 | 0 | 0.0 | 0.0 | 49 | 0.0 | 0.0 | 110 | | | | |
| family | 5 | 0 | 692 | 0 | 0.0 | 0.0 | 100 | 0.0 | 0.0 | 240 | | | | |
| genus | 6 | 0 | 742 | 0 | 0.0 | 0.0 | 109 | 0.0 | 0.0 | 656 | 0 | 1562 | 0.0 | family+genus+species |
| species | 7 | 0 | 128 | 0 | 0.0 | 0.0 | 63 | 0.0 | 0.0 | 1697 | | | | |
| avg/sum | 1.5 | 39247 | 10603 | 0 | 18.0 | 3.5 | 49.9 | 2.7 | 3.5 | 398.6 | | | 78.7 | all but unassigned |
| avg/sum | 0.7 | 89397 | 10603 | 0 | 28.3 | 3.0 | 43.8 | 14.8 | 3.1 | 348.9 | | | 89.4 | all with unassigned |

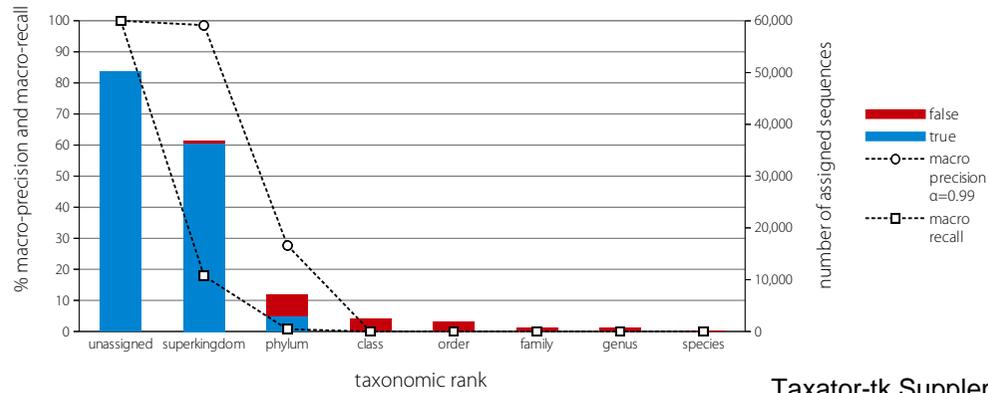

*Supplementary Figure 2: Simulated 100 bp sequence assignment with taxator-tk*  **(h) new phylum scenario**

| rank | depth | true | false | unknown | macro precision α=0.99 | stdev | pred. bins | macro recall | stdev | real bins | sum true | sum false | overall prec. | description |
|---|---|---|---|---|---|---|---|---|---|---|---|---|---|---|
| unassigned | 0 | 59449 | 0 | 0 | 100.0 | 0.0 | 1 | 100.0 | 0.0 | 1 | 114049 | 729 | 99.4 | root+superkingdom |
| superkingdom | 1 | 27300 | 729 | 0 | 97.9 | 0.0 | 1 | 14.0 | 18.6 | 3 | | | | |
| phylum | 2 | 0 | 5086 | 0 | 0.0 | 0.0 | 11 | 0.0 | 0.0 | 32 | | | | |
| class | 3 | 0 | 2142 | 0 | 0.0 | 0.0 | 20 | 0.0 | 0.0 | 52 | 0 | 11151 | 0.0 | phylum+class+order |
| order | 4 | 0 | 3923 | 0 | 0.0 | 0.0 | 44 | 0.0 | 0.0 | 110 | | | | |
| family | 5 | 0 | 533 | 0 | 0.0 | 0.0 | 98 | 0.0 | 0.0 | 240 | | | | |
| genus | 6 | 0 | 704 | 0 | 0.0 | 0.0 | 105 | 0.0 | 0.0 | 656 | 0 | 1371 | 0.0 | family+genus+species |
| species | 7 | 0 | 134 | 0 | 0.0 | 0.0 | 60 | 0.0 | 0.0 | 1697 | | | | |
| avg/sum | 1.7 | 27300 | 13251 | 0 | 14.0 | 0.0 | 48.4 | 2.0 | 2.7 | 398.6 | | | 67.3 | all but unassigned |
| avg/sum | 0.7 | 86749 | 13251 | 0 | 24.7 | 0.0 | 42.5 | 14.3 | 2.3 | 348.9 | | | 86.7 | all with unassigned |

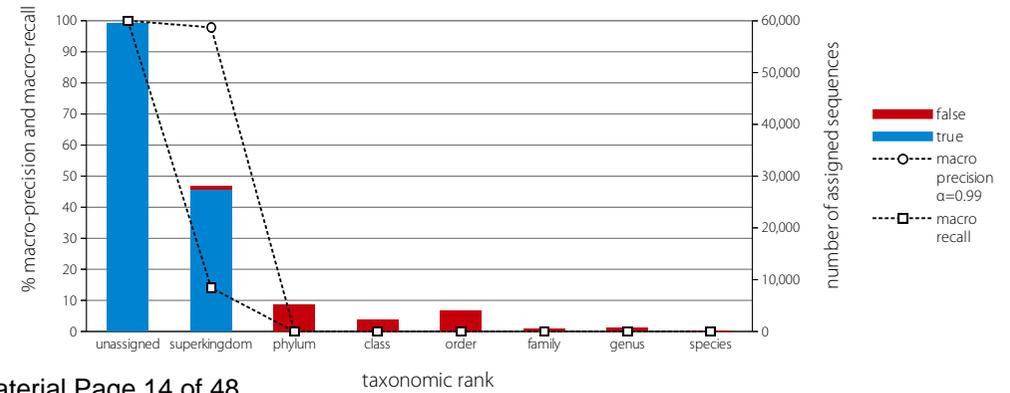

*Supplementary Figure 3: Simulated 500 bp sequence assignment with taxator-tk* **(a) summary scenario**

| rank | depth | true | false | unknown | macro precision α=0.99 | stdev | pred. bins | macro recall | stdev | real bins | sum true | sum false | overall prec. | description |
|---|---|---|---|---|---|---|---|---|---|---|---|---|---|---|
| unassigned | 0 | 20001.1 | 0.0 | 0 | 100.0 | 0.0 | 1 | 100.0 | 0.0 | 1 | 99200.0 | 582.0 | 99.4 | root+superkingdom |
| superkingdom | 1 | 39599.4 | 582.0 | 0 | 99.1 | 0.0 | 1 | 53.6 | 26.8 | 2 | | | | |
| phylum | 2 | 7862.7 | 3532.7 | 0 | 84.1 | 12.4 | 12 | 13.2 | 11.4 | 32 | | | | |
| class | 3 | 3756.1 | 1555.3 | 0 | 81.8 | 14.8 | 24 | 12.1 | 9.4 | 52 | 15065.4 | 7226.3 | 67.6 | phylum+class+order |
| order | 4 | 3446.6 | 2138.3 | 0 | 85.1 | 13.1 | 56 | 10.2 | 8.6 | 110 | | | | |
| family | 5 | 3162.4 | 702.6 | 0 | 84.6 | 17.2 | 104 | 7.1 | 7.8 | 240 | | | | |
| genus | 6 | 7880.9 | 1428.4 | 0 | 87.6 | 19.3 | 212 | 4.2 | 6.3 | 656 | 14666.0 | 2859.7 | 83.7 | family+genus+species |
| species | 7 | 3622.7 | 728.7 | 0 | 76.5 | 34.0 | 480 | 1.4 | 3.4 | 1693 | | | | |
| avg/sum | 2.3 | 69330.9 | 10668.0 | 0 | 85.6 | 15.8 | 127.0 | 14.5 | 10.5 | 397.9 | | | 86.7 | all but unassigned |
| avg/sum | 1.8 | 89332.0 | 10668.0 | 0 | 87.4 | 13.8 | 111.3 | 25.2 | 9.2 | 348.3 | | | 89.3 | all with unassigned |

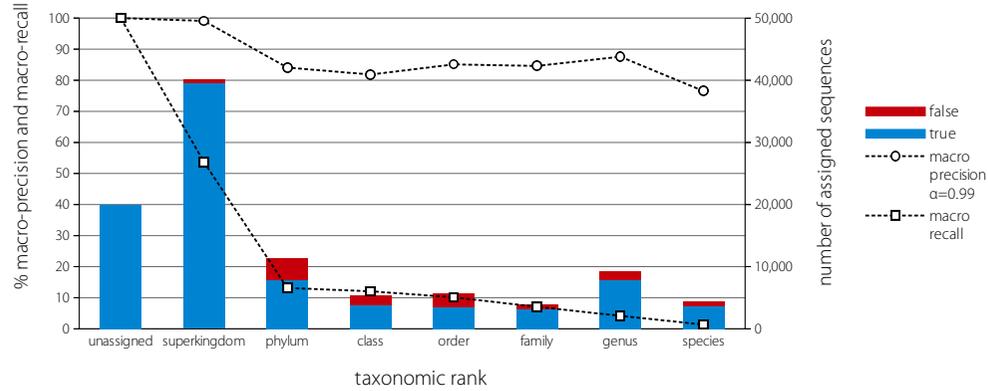

*Supplementary Figure 3: Simulated 500 bp sequence assignment with taxator-tk* **(b) all reference scenario**

| rank | depth | true | false | unknown | macro precision α=0.99 | stdev | pred. bins | macro recall | stdev | real bins | sum true | sum false | overall prec. | description |
|---|---|---|---|---|---|---|---|---|---|---|---|---|---|---|
| unassigned | 0 | 7999 | 0 | 0 | 100.0 | 0.0 | 1 | 100.0 | 0.0 | 1 | 42883 | 0 | 100.0 | root+superkingdom |
| superkingdom | 1 | 17442 | 0 | 0 | 100.0 | 0.0 | 2 | 81.8 | 10.7 | 2 | | | | |
| phylum | 2 | 4415 | 0 | 0 | 100.0 | 0.0 | 14 | 43.0 | 30.7 | 32 | | | | |
| class | 3 | 2699 | 0 | 0 | 100.0 | 0.0 | 27 | 43.7 | 29.3 | 52 | 11750 | 0 | 100.0 | phylum+class+order |
| order | 4 | 4636 | 0 | 0 | 100.0 | 0.0 | 59 | 40.6 | 30.9 | 110 | | | | |
| family | 5 | 7889 | 0 | 0 | 100.0 | 0.0 | 109 | 32.6 | 31.8 | 240 | | | | |
| genus | 6 | 29561 | 0 | 0 | 100.0 | 0.0 | 221 | 23.0 | 32.0 | 656 | 62809 | 0 | 100.0 | family+genus+species |
| species | 7 | 25359 | 0 | 0 | 100.0 | 0.0 | 408 | 9.5 | 23.5 | 1693 | | | | |
| avg/sum | 4.0 | 92001 | 0 | 0 | 100.0 | 0.0 | 120.0 | 39.2 | 27.0 | 397.9 | | | 100.0 | all but unassigned |
| avg/sum | 3.6 | 100000 | 0 | 0 | 100.0 | 0.0 | 105.1 | 46.8 | 23.6 | 348.3 | | | 100.0 | all with unassigned |

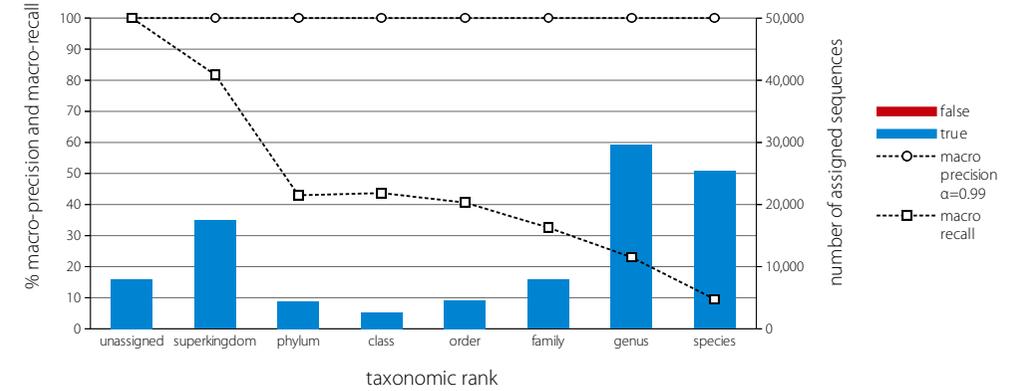

*Supplementary Figure 3: Simulated 500 bp sequence assignment with taxator-tk* **(c) new species scenario**

| rank | depth | true | false | unknown | macro precision α=0.99 | stdev | pred. bins | macro recall | stdev | real bins | sum true | sum false | overall prec. | description |
|---|---|---|---|---|---|---|---|---|---|---|---|---|---|---|
| unassigned | 0 | 10520 | 0 | 0 | 100.0 | 0.0 | 1 | 100.0 | 0.0 | 1 | 67276 | 224 | 99.7 | root+superkingdom |
| superkingdom | 1 | 28378 | 224 | 0 | 99.2 | 0.5 | 2 | 68.6 | 21.6 | 2 | | | | |
| phylum | 2 | 8027 | 773 | 0 | 98.4 | 1.2 | 11 | 24.7 | 23.6 | 32 | | | | |
| class | 3 | 4991 | 337 | 0 | 97.8 | 2.6 | 23 | 24.9 | 23.3 | 52 | 19494 | 1591 | 92.5 | phylum+class+order |
| order | 4 | 6476 | 481 | 0 | 97.7 | 2.8 | 50 | 21.0 | 22.3 | 110 | | | | |
| family | 5 | 9009 | 253 | 0 | 96.6 | 6.6 | 79 | 14.2 | 21.7 | 240 | | | | |
| genus | 6 | 25605 | 910 | 0 | 90.8 | 21.1 | 107 | 6.3 | 16.4 | 656 | 34614 | 5179 | 87.0 | family+genus+species |
| species | 7 | 0 | 4016 | 0 | 0.0 | 0.0 | 237 | 0.0 | 0.0 | 1693 | | | | |
| avg/sum | 3.5 | 82486 | 6994 | 0 | 82.9 | 5.0 | 72.7 | 22.8 | 18.4 | 397.9 | | | 92.2 | all but unassigned |
| avg/sum | 3.1 | 93006 | 6994 | 0 | 85.1 | 4.3 | 63.8 | 32.5 | 16.1 | 348.3 | | | 93.0 | all with unassigned |

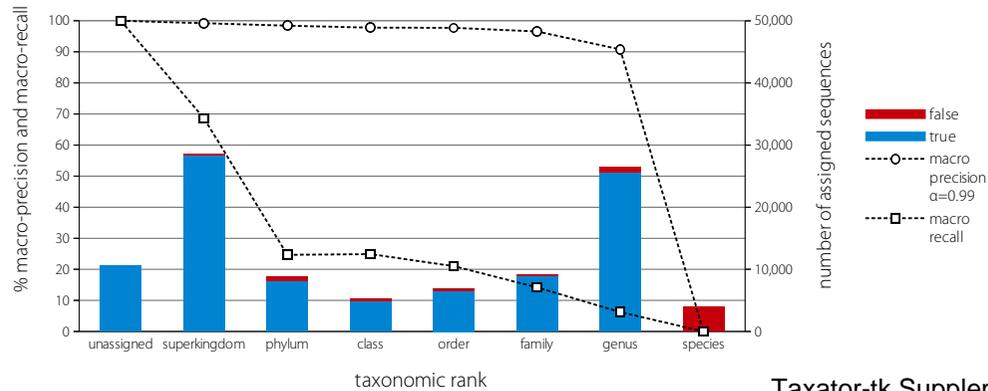

*Supplementary Figure 3: Simulated 500 bp sequence assignment with taxator-tk* **(d) new genus scenario**

| rank | depth | true | false | unknown | macro precision α=0.99 | stdev | pred. bins | macro recall | stdev | real bins | sum true | sum false | overall prec. | description |
|---|---|---|---|---|---|---|---|---|---|---|---|---|---|---|
| unassigned | 0 | 13884 | 0 | 0 | 100.0 | 0.0 | 1 | 100.0 | 0.0 | 1 | 100784 | 357 | 99.6 | root+superkingdom |
| superkingdom | 1 | 43450 | 357 | 0 | 99.5 | 0.0 | 1 | 60.2 | 26.6 | 2 | | | | |
| phylum | 2 | 12278 | 1548 | 0 | 95.9 | 2.9 | 9 | 13.4 | 15.2 | 32 | | | | |
| class | 3 | 7723 | 760 | 0 | 87.2 | 13.6 | 21 | 10.9 | 12.4 | 52 | 27611 | 3340 | 89.2 | phylum+class+order |
| order | 4 | 7610 | 1032 | 0 | 86.3 | 12.4 | 47 | 7.8 | 11.2 | 110 | | | | |
| family | 5 | 5239 | 761 | 0 | 52.7 | 40.4 | 81 | 2.8 | 7.9 | 240 | | | | |
| genus | 6 | 0 | 5064 | 0 | 0.0 | 0.0 | 136 | 0.0 | 0.0 | 656 | 5239 | 6119 | 46.1 | family+genus+species |
| species | 7 | 0 | 294 | 0 | 0.0 | 0.0 | 105 | 0.0 | 0.0 | 1693 | | | | |
| avg/sum | 2.2 | 76300 | 9816 | 0 | 60.2 | 9.9 | 57.1 | 13.6 | 10.5 | 397.9 | | | 88.6 | all but unassigned |
| avg/sum | 1.9 | 90184 | 9816 | 0 | 65.2 | 8.7 | 50.1 | 24.4 | 9.2 | 348.3 | | | 90.2 | all with unassigned |

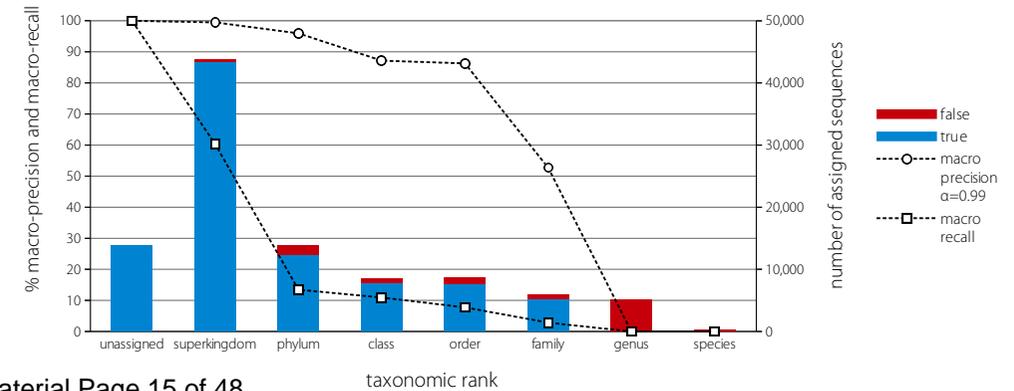

*Supplementary Figure 3: Simulated 500 bp sequence assignment with taxator-tk* **(e) new family scenario**

| rank | depth | true | false | unknown | macro precision α=0.99 | stdev | pred. bins | macro recall | stdev | real bins | sum true | sum false | overall prec. | description |
|---|---|---|---|---|---|---|---|---|---|---|---|---|---|---|
| unassigned | 0 | 18731 | 0 | 0 | 100.0 | 0.0 | 1 | 100.0 | 0.0 | 1 | 114197 | 702 | 99.4 | root+superkingdom |
| superkingdom | 1 | 47733 | 702 | 0 | 98.9 | 0.0 | 1 | 47.8 | 34.0 | 2 | | | | |
| phylum | 2 | 12581 | 2006 | 0 | 88.9 | 4.2 | 6 | 5.9 | 10.5 | 32 | 25162 | 4675 | 84.3 | phylum+class+order |
| class | 3 | 7177 | 1099 | 0 | 64.4 | 31.5 | 16 | 4.0 | 6.7 | 52 | | | | |
| order | 4 | 5404 | 1570 | 0 | 40.0 | 36.6 | 42 | 1.6 | 4.2 | 110 | | | | |
| family | 5 | 0 | 1251 | 0 | 0.0 | 0.0 | 103 | 0.0 | 0.0 | 240 | 0 | 2997 | 0.0 | family+genus+species |
| genus | 6 | 0 | 1492 | 0 | 0.0 | 0.0 | 122 | 0.0 | 0.0 | 656 | | | | |
| species | 7 | 0 | 254 | 0 | 0.0 | 0.0 | 82 | 0.0 | 0.0 | 1693 | | | | |
| avg/sum | 1.8 | 72895 | 8374 | 0 | 41.8 | 10.3 | 53.1 | 8.5 | 7.9 | 397.9 | | | 89.7 | all but unassigned |
| avg/sum | 1.5 | 91626 | 8374 | 0 | 49.0 | 9.0 | 46.6 | 19.9 | 6.9 | 348.3 | | | 91.6 | all with unassigned |

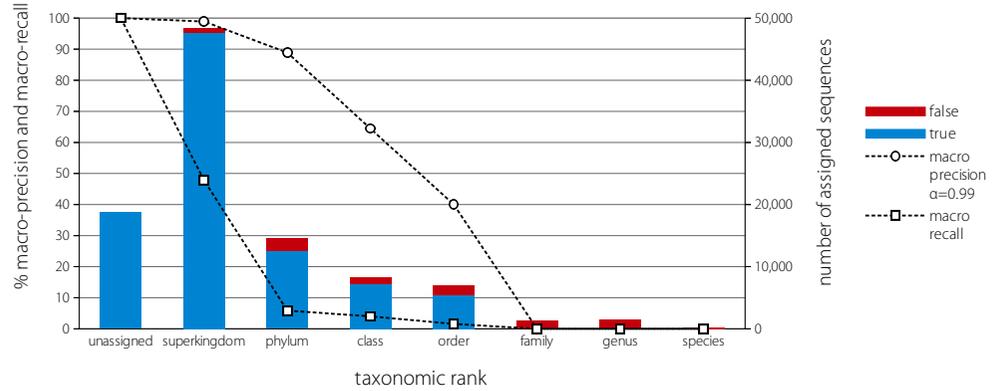

*Supplementary Figure 3: Simulated 500 bp sequence assignment with taxator-tk* **(f) new order scenario**

| rank | depth | true | false | unknown | macro precision α=0.99 | stdev | pred. bins | macro recall | stdev | real bins | sum true | sum false | overall prec. | description |
|---|---|---|---|---|---|---|---|---|---|---|---|---|---|---|
| unassigned | 0 | 21881 | 0 | 0 | 100.0 | 0.0 | 1 | 100.0 | 0.0 | 1 | 121329 | 771 | 99.4 | root+superkingdom |
| superkingdom | 1 | 49724 | 771 | 0 | 98.8 | 0.0 | 1 | 44.4 | 34.2 | 2 | | | | |
| phylum | 2 | 11475 | 4991 | 0 | 71.7 | 17.0 | 6 | 3.6 | 7.8 | 32 | 15178 | 10362 | 59.4 | phylum+class+order |
| class | 3 | 3703 | 1888 | 0 | 31.9 | 29.3 | 18 | 1.1 | 2.5 | 52 | | | | |
| order | 4 | 0 | 3483 | 0 | 0.0 | 0.0 | 56 | 0.0 | 0.0 | 110 | | | | |
| family | 5 | 0 | 1003 | 0 | 0.0 | 0.0 | 106 | 0.0 | 0.0 | 240 | 0 | 2084 | 0.0 | family+genus+species |
| genus | 6 | 0 | 868 | 0 | 0.0 | 0.0 | 110 | 0.0 | 0.0 | 656 | | | | |
| species | 7 | 0 | 213 | 0 | 0.0 | 0.0 | 63 | 0.0 | 0.0 | 1693 | | | | |
| avg/sum | 1.6 | 64902 | 13217 | 0 | 28.9 | 6.6 | 51.4 | 7.0 | 6.4 | 397.9 | | | 83.1 | all but unassigned |
| avg/sum | 1.2 | 86783 | 13217 | 0 | 37.8 | 5.8 | 45.1 | 18.6 | 5.6 | 348.3 | | | 86.8 | all with unassigned |

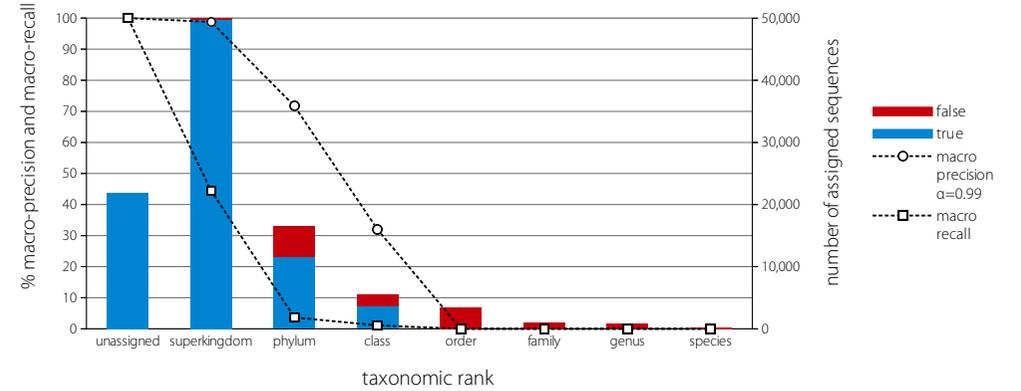

*Supplementary Figure 3: Simulated 500 bp sequence assignment with taxator-tk* **(g) new class scenario**

| rank | depth | true | false | unknown | macro precision α=0.99 | stdev | pred. bins | macro recall | stdev | real bins | sum true | sum false | overall prec. | description |
|---|---|---|---|---|---|---|---|---|---|---|---|---|---|---|
| unassigned | 0 | 28770 | 0 | 0 | 100.0 | 0.0 | 1 | 100.0 | 0.0 | 1 | 127418 | 838 | 99.3 | root+superkingdom |
| superkingdom | 1 | 49324 | 838 | 0 | 98.5 | 0.0 | 1 | 39.9 | 31.6 | 2 | | | | |
| phylum | 2 | 6263 | 6679 | 0 | 34.0 | 28.9 | 7 | 1.5 | 3.8 | 32 | 6263 | 12967 | 32.6 | phylum+class+order |
| class | 3 | 0 | 3676 | 0 | 0.0 | 0.0 | 21 | 0.0 | 0.0 | 52 | | | | |
| order | 4 | 0 | 2612 | 0 | 0.0 | 0.0 | 52 | 0.0 | 0.0 | 110 | | | | |
| family | 5 | 0 | 852 | 0 | 0.0 | 0.0 | 108 | 0.0 | 0.0 | 240 | 0 | 1838 | 0.0 | family+genus+species |
| genus | 6 | 0 | 834 | 0 | 0.0 | 0.0 | 103 | 0.0 | 0.0 | 656 | | | | |
| species | 7 | 0 | 152 | 0 | 0.0 | 0.0 | 56 | 0.0 | 0.0 | 1693 | | | | |
| avg/sum | 1.5 | 55587 | 15643 | 0 | 18.9 | 4.1 | 49.7 | 5.9 | 5.1 | 397.9 | | | 78.0 | all but unassigned |
| avg/sum | 1.1 | 84357 | 15643 | 0 | 29.1 | 3.6 | 43.6 | 17.7 | 4.4 | 348.3 | | | 84.4 | all with unassigned |

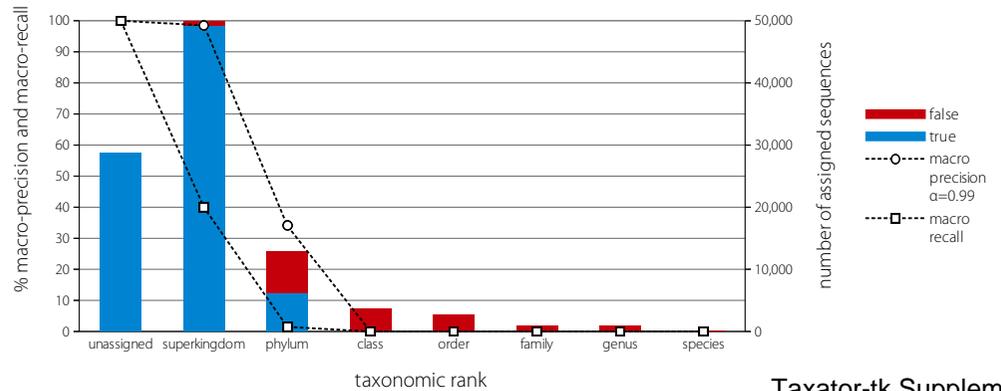

*Supplementary Figure 3: Simulated 500 bp sequence assignment with taxator-tk* **(h) new phylum scenario**

| rank | depth | true | false | unknown | macro precision α=0.99 | stdev | pred. bins | macro recall | stdev | real bins | sum true | sum false | overall prec. | description |
|---|---|---|---|---|---|---|---|---|---|---|---|---|---|---|
| unassigned | 0 | 38223 | 0 | 0 | 100.0 | 0.0 | 1 | 100.0 | 0.0 | 1 | 120513 | 1182 | 99.0 | root+superkingdom |
| superkingdom | 1 | 41145 | 1182 | 0 | 97.7 | 0.0 | 1 | 32.6 | 28.9 | 2 | | | | |
| phylum | 2 | 0 | 8732 | 0 | 0.0 | 0.0 | 11 | 0.0 | 0.0 | 32 | 0 | 17649 | 0.0 | phylum+class+order |
| class | 3 | 0 | 3127 | 0 | 0.0 | 0.0 | 22 | 0.0 | 0.0 | 52 | | | | |
| order | 4 | 0 | 5790 | 0 | 0.0 | 0.0 | 44 | 0.0 | 0.0 | 110 | | | | |
| family | 5 | 0 | 798 | 0 | 0.0 | 0.0 | 106 | 0.0 | 0.0 | 240 | 0 | 1801 | 0.0 | family+genus+species |
| genus | 6 | 0 | 831 | 0 | 0.0 | 0.0 | 93 | 0.0 | 0.0 | 656 | | | | |
| species | 7 | 0 | 172 | 0 | 0.0 | 0.0 | 45 | 0.0 | 0.0 | 1693 | | | | |
| avg/sum | 1.6 | 41145 | 20632 | 0 | 14.0 | 0.0 | 46.0 | 4.7 | 4.1 | 397.9 | | | 66.6 | all but unassigned |
| avg/sum | 1.0 | 79368 | 20632 | 0 | 24.7 | 0.0 | 40.4 | 16.6 | 3.6 | 348.3 | | | 79.4 | all with unassigned |

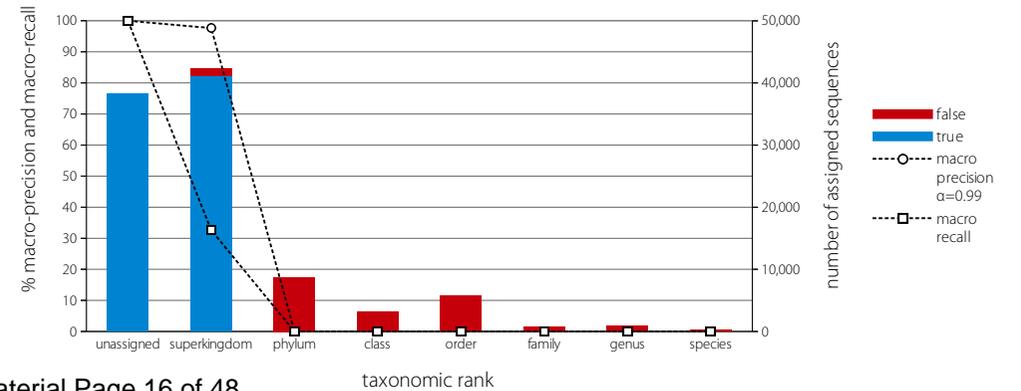

*Supplementary Figure 4: Simulated 1000 bp sequence assignment with taxator-tk* **(a) summary scenario**

| rank | depth | true | false | unknown | macro precision α=0.99 | stdev | pred. bins | macro recall | stdev | real bins | sum true | sum false | overall prec. | description |
|---|---|---|---|---|---|---|---|---|---|---|---|---|---|---|
| unassigned | 0 | 18217.3 | 0.0 | 0 | 100.0 | 0.0 | 1 | 100.0 | 0.0 | 1 | 93809.9 | 550.7 | 99.4 | root+superkingdom |
| superkingdom | 1 | 37796.3 | 550.7 | 0 | 99.2 | 0.0 | 1 | 38.1 | 33.8 | 3 | | | | |
| phylum | 2 | 9465.1 | 3300.7 | 0 | 87.0 | 12.2 | 12 | 15.2 | 12.7 | 32 | 18677.1 | 6634.6 | 73.8 | phylum+class+order |
| class | 3 | 4795.0 | 1367.4 | 0 | 83.2 | 14.7 | 25 | 13.4 | 10.3 | 52 | | | | |
| order | 4 | 4417.0 | 1966.7 | 0 | 84.5 | 15.1 | 57 | 10.8 | 9.2 | 110 | | | | |
| family | 5 | 3498.1 | 817.1 | 0 | 84.4 | 17.9 | 106 | 7.5 | 8.1 | 240 | 15169.9 | 2954.1 | 83.7 | family+genus+species |
| genus | 6 | 7834.3 | 1397.6 | 0 | 86.4 | 19.6 | 219 | 4.3 | 6.3 | 653 | | | | |
| species | 7 | 3837.4 | 739.4 | 0 | 77.2 | 34.2 | 472 | 1.5 | 3.5 | 1690 | | | | |
| avg/sum | 2.4 | 71643.3 | 10139.4 | 0 | 86.0 | 16.2 | 127.4 | 13.0 | 12.0 | 397.1 | | | 87.6 | all but unassigned |
| avg/sum | 1.9 | 89860.6 | 10139.4 | 0 | 87.7 | 14.2 | 111.6 | 23.8 | 10.5 | 347.6 | | | 89.9 | all with unassigned |

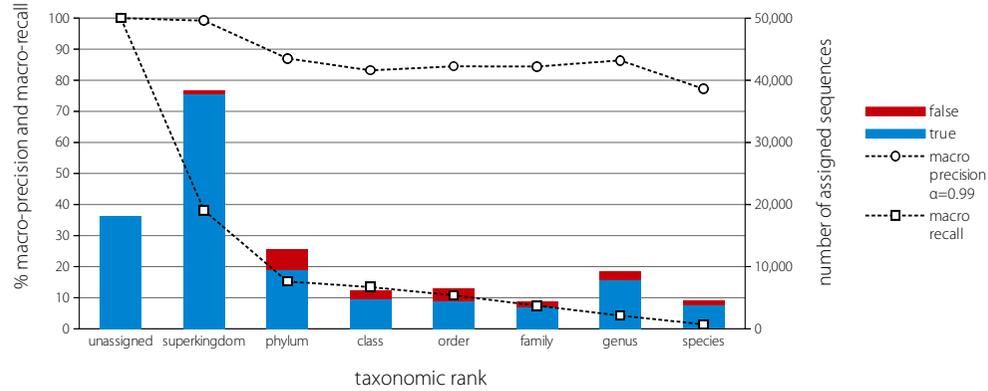

*Supplementary Figure 4: Simulated 1000 bp sequence assignment with taxator-tk* **(b) all reference scenario**

| rank | depth | true | false | unknown | macro precision α=0.99 | stdev | pred. bins | macro recall | stdev | real bins | sum true | sum false | overall prec. | description |
|---|---|---|---|---|---|---|---|---|---|---|---|---|---|---|
| unassigned | 0 | 7256 | 0 | 0 | 100.0 | 0.0 | 1 | 100.0 | 0.0 | 1 | 40990 | 0 | 100.0 | root+superkingdom |
| superkingdom | 1 | 16867 | 0 | 0 | 100.0 | 0.0 | 2 | 55.1 | 39.9 | 3 | | | | |
| phylum | 2 | 4739 | 0 | 0 | 100.0 | 0.0 | 14 | 45.2 | 30.5 | 32 | 12595 | 0 | 100.0 | phylum+class+order |
| class | 3 | 3024 | 0 | 0 | 100.0 | 0.0 | 27 | 44.7 | 29.7 | 52 | | | | |
| order | 4 | 4832 | 0 | 0 | 100.0 | 0.0 | 59 | 40.6 | 31.1 | 110 | | | | |
| family | 5 | 8132 | 0 | 0 | 100.0 | 0.0 | 112 | 33.1 | 32.0 | 240 | 63282 | 0 | 100.0 | family+genus+species |
| genus | 6 | 28288 | 0 | 0 | 100.0 | 0.0 | 221 | 23.3 | 32.2 | 653 | | | | |
| species | 7 | 26862 | 0 | 0 | 100.0 | 0.0 | 402 | 10.1 | 24.8 | 1690 | | | | |
| avg/sum | 4.0 | 92744 | 0 | 0 | 100.0 | 0.0 | 119.6 | 36.0 | 31.5 | 397.1 | | | 100.0 | all but unassigned |
| avg/sum | 3.6 | 100000 | 0 | 0 | 100.0 | 0.0 | 104.8 | 44.0 | 27.5 | 347.6 | | | 100.0 | all with unassigned |

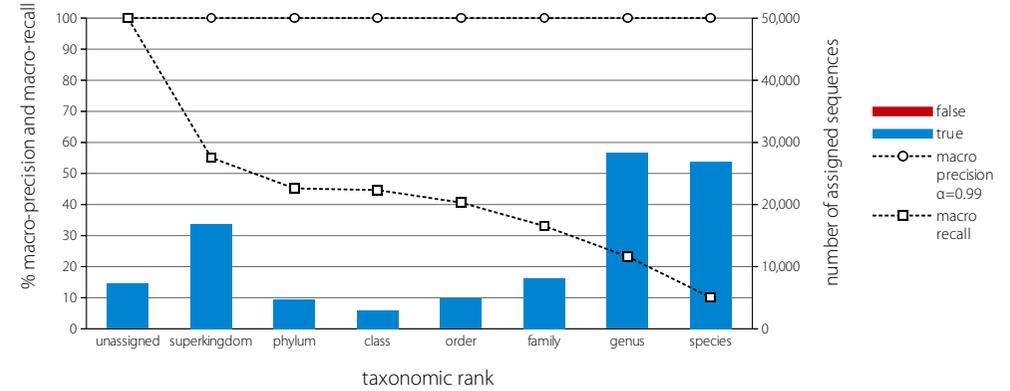

*Supplementary Figure 4: Simulated 1000 bp sequence assignment with taxator-tk* **(c) new species scenario**

| rank | depth | true | false | unknown | macro precision α=0.99 | stdev | pred. bins | macro recall | stdev | real bins | sum true | sum false | overall prec. | description |
|---|---|---|---|---|---|---|---|---|---|---|---|---|---|---|
| unassigned | 0 | 7557 | 0 | 0 | 100.0 | 0.0 | 1 | 100.0 | 0.0 | 1 | 60223 | 191 | 99.7 | root+superkingdom |
| superkingdom | 1 | 26333 | 191 | 0 | 99.5 | 0.2 | 2 | 49.7 | 38.2 | 3 | | | | |
| phylum | 2 | 9128 | 523 | 0 | 99.2 | 0.6 | 12 | 28.6 | 25.3 | 32 | 23034 | 1142 | 95.3 | phylum+class+order |
| class | 3 | 6028 | 250 | 0 | 98.4 | 2.4 | 24 | 28.3 | 25.2 | 52 | | | | |
| order | 4 | 7878 | 369 | 0 | 97.8 | 4.4 | 52 | 23.0 | 23.6 | 110 | | | | |
| family | 5 | 9803 | 325 | 0 | 96.3 | 9.0 | 83 | 15.4 | 22.6 | 240 | 36355 | 5388 | 87.1 | family+genus+species |
| genus | 6 | 26552 | 898 | 0 | 90.5 | 22.4 | 107 | 6.5 | 16.7 | 653 | | | | |
| species | 7 | 0 | 4165 | 0 | 0.0 | 0.0 | 230 | 0.0 | 0.0 | 1690 | | | | |
| avg/sum | 3.5 | 85722 | 6721 | 0 | 83.1 | 5.6 | 72.9 | 21.6 | 21.7 | 397.1 | | | 92.7 | all but unassigned |
| avg/sum | 3.3 | 93279 | 6721 | 0 | 85.2 | 4.9 | 63.9 | 31.4 | 19.0 | 347.6 | | | 93.3 | all with unassigned |

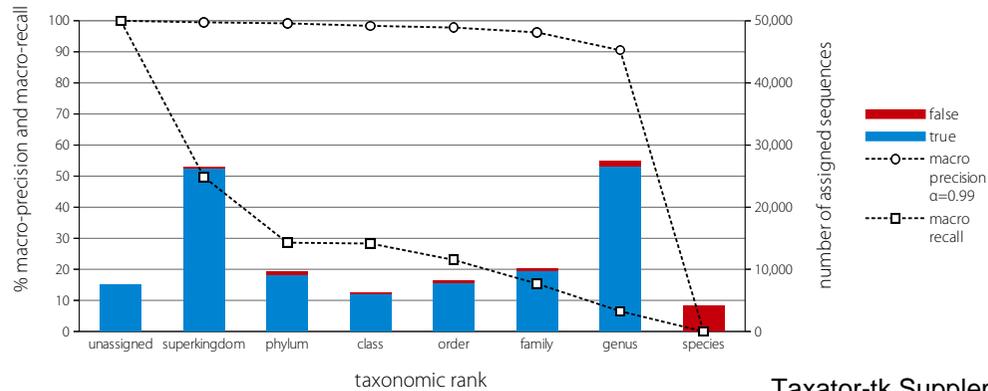

*Supplementary Figure 4: Simulated 1000 bp sequence assignment with taxator-tk* **(d) new genus scenario**

| rank | depth | true | false | unknown | macro precision α=0.99 | stdev | pred. bins | macro recall | stdev | real bins | sum true | sum false | overall prec. | description |
|---|---|---|---|---|---|---|---|---|---|---|---|---|---|---|
| unassigned | 0 | 9974 | 0 | 0 | 100.0 | 0.0 | 1 | 100.0 | 0.0 | 1 | 89246 | 293 | 99.7 | root+superkingdom |
| superkingdom | 1 | 39636 | 293 | 0 | 98.8 | 0.8 | 2 | 44.0 | 37.1 | 3 | | | | |
| phylum | 2 | 14673 | 1006 | 0 | 94.6 | 9.8 | 10 | 17.7 | 18.6 | 32 | 34900 | 2418 | 93.5 | phylum+class+order |
| class | 3 | 9782 | 538 | 0 | 90.4 | 11.6 | 22 | 14.2 | 15.4 | 52 | | | | |
| order | 4 | 10445 | 874 | 0 | 90.2 | 10.8 | 47 | 9.7 | 13.5 | 110 | | | | |
| family | 5 | 6552 | 955 | 0 | 59.0 | 39.9 | 82 | 3.7 | 9.4 | 240 | 6552 | 6227 | 51.3 | family+genus+species |
| genus | 6 | 0 | 4978 | 0 | 0.0 | 0.0 | 143 | 0.0 | 0.0 | 653 | | | | |
| species | 7 | 0 | 294 | 0 | 0.0 | 0.0 | 94 | 0.0 | 0.0 | 1690 | | | | |
| avg/sum | 2.4 | 81088 | 8938 | 0 | 61.9 | 10.4 | 57.1 | 12.8 | 13.4 | 397.1 | | | 90.1 | all but unassigned |
| avg/sum | 2.2 | 91062 | 8938 | 0 | 66.6 | 9.1 | 50.1 | 23.7 | 11.7 | 347.6 | | | 91.1 | all with unassigned |

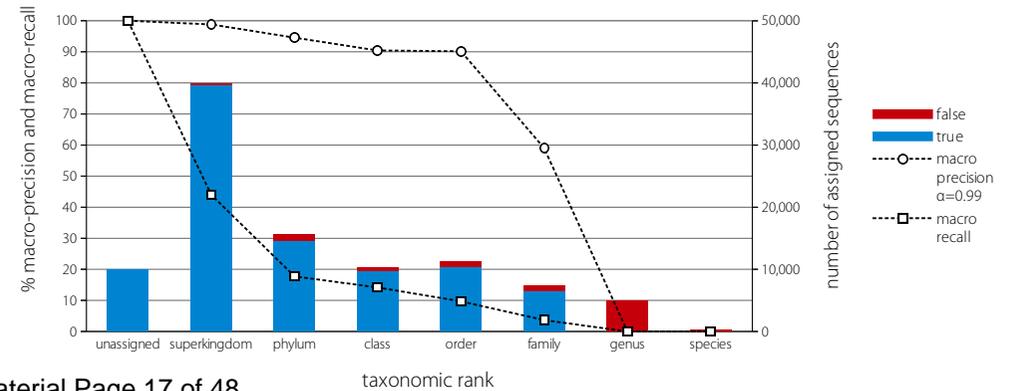

*Supplementary Figure 4: Simulated 1000 bp sequence assignment with taxator-tk* **(e) new family scenario**

| rank | depth | true | false | unknown | macro precision α=0.99 | stdev | pred. bins | macro recall | stdev | real bins | sum true | sum false | overall prec. | description |
|---|---|---|---|---|---|---|---|---|---|---|---|---|---|---|
| unassigned | 0 | 15429 | 0 | 0 | 100.0 | 0.0 | 1 | 100.0 | 0.0 | 1 | 103635 | 644 | 99.4 | root+superkingdom |
| superkingdom | 1 | 44103 | 644 | 0 | 99.1 | 0.0 | 1 | 35.3 | 36.3 | 3 | | | | |
| phylum | 2 | 15330 | 1458 | 0 | 94.1 | 2.4 | 6 | 8.0 | 13.7 | 32 | 32792 | 3711 | 89.8 | phylum+class+order |
| class | 3 | 9698 | 849 | 0 | 79.4 | 24.5 | 15 | 5.4 | 9.1 | 52 | | | | |
| order | 4 | 7764 | 1404 | 0 | 45.8 | 38.6 | 40 | 2.1 | 5.6 | 110 | | | | |
| family | 5 | 0 | 1548 | 0 | 0.0 | 0.0 | 104 | 0.0 | 0.0 | 240 | 0 | 3321 | 0.0 | family+genus+species |
| genus | 6 | 0 | 1507 | 0 | 0.0 | 0.0 | 127 | 0.0 | 0.0 | 653 | | | | |
| species | 7 | 0 | 266 | 0 | 0.0 | 0.0 | 65 | 0.0 | 0.0 | 1690 | | | | |
| avg/sum | 1.9 | 76895 | 7676 | 0 | 45.5 | 9.4 | 51.1 | 7.3 | 9.3 | 397.1 | | | 90.9 | all but unassigned |
| avg/sum | 1.6 | 92324 | 7676 | 0 | 52.3 | 8.2 | 44.9 | 18.9 | 8.1 | 347.6 | | | 92.3 | all with unassigned |

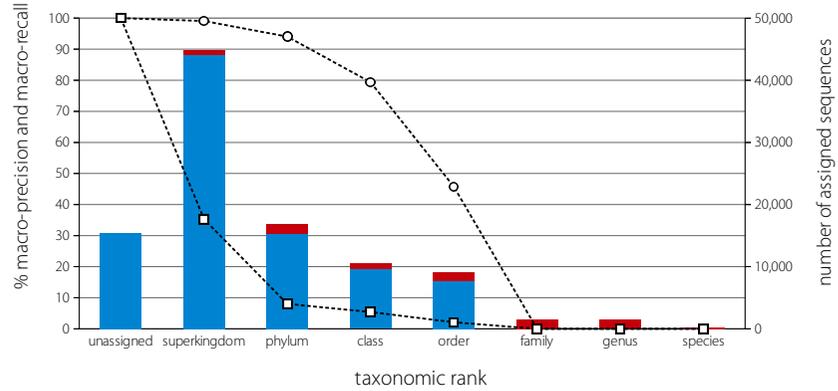

*Supplementary Figure 4: Simulated 1000 bp sequence assignment with taxator-tk* **(f) new order scenario**

| rank | depth | true | false | unknown | macro precision α=0.99 | stdev | pred. bins | macro recall | stdev | real bins | sum true | sum false | overall prec. | description |
|---|---|---|---|---|---|---|---|---|---|---|---|---|---|---|
| unassigned | 0 | 19932 | 0 | 0 | 100.0 | 0.0 | 1 | 100.0 | 0.0 | 1 | 115360 | 725 | 99.4 | root+superkingdom |
| superkingdom | 1 | 47714 | 725 | 0 | 98.9 | 0.0 | 1 | 32.4 | 34.7 | 3 | | | | |
| phylum | 2 | 14366 | 4764 | 0 | 81.4 | 11.9 | 6 | 5.0 | 10.2 | 32 | 19399 | 9995 | 66.0 | phylum+class+order |
| class | 3 | 5033 | 1596 | 0 | 40.2 | 33.9 | 17 | 1.5 | 3.5 | 52 | | | | |
| order | 4 | 0 | 3635 | 0 | 0.0 | 0.0 | 49 | 0.0 | 0.0 | 110 | | | | |
| family | 5 | 0 | 1133 | 0 | 0.0 | 0.0 | 80 | 0.0 | 0.0 | 240 | 0 | 2235 | 0.0 | family+genus+species |
| genus | 6 | 0 | 883 | 0 | 0.0 | 0.0 | 98 | 0.0 | 0.0 | 653 | | | | |
| species | 7 | 0 | 219 | 0 | 0.0 | 0.0 | 41 | 0.0 | 0.0 | 1690 | | | | |
| avg/sum | 1.7 | 67113 | 12955 | 0 | 31.5 | 6.5 | 43.0 | 5.6 | 6.9 | 397.1 | | | 83.8 | all but unassigned |
| avg/sum | 1.3 | 87045 | 12955 | 0 | 40.1 | 5.7 | 37.8 | 17.4 | 6.0 | 347.6 | | | 87.0 | all with unassigned |

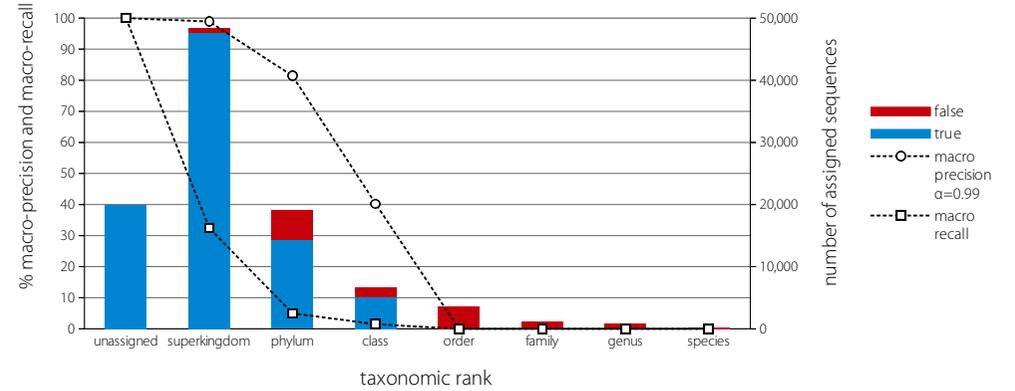

*Supplementary Figure 4: Simulated 1000 bp sequence assignment with taxator-tk* **(g) new class scenario**

| rank | depth | true | false | unknown | macro precision α=0.99 | stdev | pred. bins | macro recall | stdev | real bins | sum true | sum false | overall prec. | description |
|---|---|---|---|---|---|---|---|---|---|---|---|---|---|---|
| unassigned | 0 | 28057 | 0 | 0 | 100.0 | 0.0 | 1 | 100.0 | 0.0 | 1 | 125231 | 824 | 99.3 | root+superkingdom |
| superkingdom | 1 | 48587 | 824 | 0 | 98.6 | 0.0 | 1 | 28.5 | 31.4 | 3 | | | | |
| phylum | 2 | 8020 | 6696 | 0 | 39.8 | 34.0 | 7 | 2.2 | 5.1 | 32 | 8020 | 12645 | 38.8 | phylum+class+order |
| class | 3 | 0 | 3649 | 0 | 0.0 | 0.0 | 22 | 0.0 | 0.0 | 52 | | | | |
| order | 4 | 0 | 2300 | 0 | 0.0 | 0.0 | 48 | 0.0 | 0.0 | 110 | | | | |
| family | 5 | 0 | 939 | 0 | 0.0 | 0.0 | 94 | 0.0 | 0.0 | 240 | 0 | 1867 | 0.0 | family+genus+species |
| genus | 6 | 0 | 835 | 0 | 0.0 | 0.0 | 94 | 0.0 | 0.0 | 653 | | | | |
| species | 7 | 0 | 93 | 0 | 0.0 | 0.0 | 38 | 0.0 | 0.0 | 1690 | | | | |
| avg/sum | 1.5 | 56607 | 15336 | 0 | 19.8 | 4.9 | 43.4 | 4.4 | 5.2 | 397.1 | | | 78.7 | all but unassigned |
| avg/sum | 1.1 | 84664 | 15336 | 0 | 29.8 | 4.2 | 38.1 | 16.3 | 4.6 | 347.6 | | | 84.7 | all with unassigned |

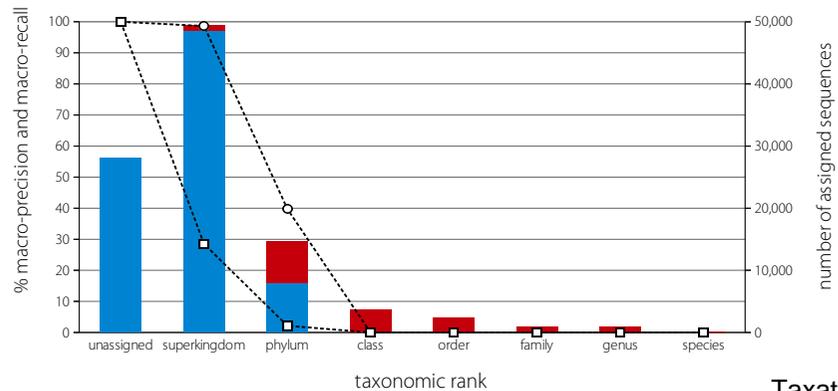

*Supplementary Figure 4: Simulated 1000 bp sequence assignment with taxator-tk* **(h) new phylum scenario**

| rank | depth | true | false | unknown | macro precision α=0.99 | stdev | pred. bins | macro recall | stdev | real bins | sum true | sum false | overall prec. | description |
|---|---|---|---|---|---|---|---|---|---|---|---|---|---|---|
| unassigned | 0 | 39316 | 0 | 0 | 100.0 | 0.0 | 1 | 100.0 | 0.0 | 1 | 121984 | 1178 | 99.0 | root+superkingdom |
| superkingdom | 1 | 41334 | 1178 | 0 | 97.7 | 0.0 | 1 | 21.9 | 27.3 | 3 | | | | |
| phylum | 2 | 0 | 8658 | 0 | 0.0 | 0.0 | 12 | 0.0 | 0.0 | 32 | 0 | 16531 | 0.0 | phylum+class+order |
| class | 3 | 0 | 2688 | 0 | 0.0 | 0.0 | 21 | 0.0 | 0.0 | 52 | | | | |
| order | 4 | 0 | 5185 | 0 | 0.0 | 0.0 | 44 | 0.0 | 0.0 | 110 | | | | |
| family | 5 | 0 | 820 | 0 | 0.0 | 0.0 | 98 | 0.0 | 0.0 | 240 | 0 | 1641 | 0.0 | family+genus+species |
| genus | 6 | 0 | 682 | 0 | 0.0 | 0.0 | 99 | 0.0 | 0.0 | 653 | | | | |
| species | 7 | 0 | 139 | 0 | 0.0 | 0.0 | 50 | 0.0 | 0.0 | 1690 | | | | |
| avg/sum | 1.6 | 41334 | 19350 | 0 | 14.0 | 0.0 | 45.1 | 3.1 | 3.9 | 397.1 | | | 68.1 | all but unassigned |
| avg/sum | 1.0 | 80650 | 19350 | 0 | 24.7 | 0.0 | 39.6 | 15.2 | 3.4 | 347.6 | | | 80.7 | all with unassigned |

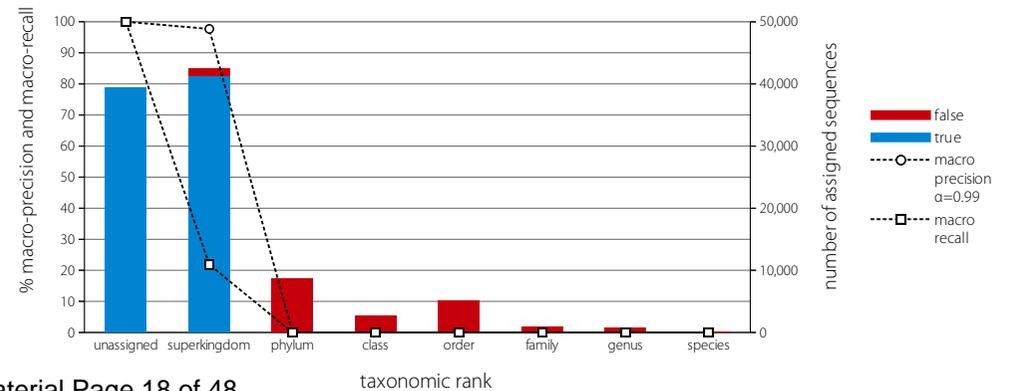

*Supplementary Figure 5: MEGAN binning for FAMeS SimMC*   **(a) summary scenario**

| rank | depth | true | false | unknown | macro precision α=0.99 | stdev | pred. bins | macro recall | stdev | real bins | sum true | sum false | overall prec. | description |
|---|---|---|---|---|---|---|---|---|---|---|---|---|---|---|
| unassigned | 0 | 877.9 | 0.0 | 0 | 100.0 | 0.0 | 1 | 100.0 | 0.0 | 1 | 5735.3 | 7.5 | 99.9 | root+superkingdom |
| superkingdom | 1 | 2428.7 | 7.5 | 0 | 100.0 | 0.0 | 1 | 45.0 | 45.0 | 2 | | | | |
| phylum | 2 | 2508.3 | 60.0 | 0 | 18.7 | 32.6 | 8 | 35.4 | 23.9 | 8 | 4604.0 | 1095.1 | 80.8 | phylum+class+order |
| class | 3 | 1611.6 | 389.1 | 0 | 14.8 | 29.4 | 17 | 24.3 | 19.7 | 12 | | | | |
| order | 4 | 484.1 | 646.1 | 0 | 9.8 | 23.8 | 39 | 15.6 | 16.5 | 23 | | | | |
| family | 5 | 1590.7 | 617.3 | 0 | 6.1 | 21.5 | 69 | 7.5 | 13.0 | 30 | 4734.4 | 3292.7 | 59.0 | family+genus+species |
| genus | 6 | 811.4 | 1102.6 | 0 | 3.9 | 18.0 | 131 | 3.5 | 7.2 | 37 | | | | |
| species | 7 | 2332.3 | 1572.8 | 0 | 3.0 | 16.5 | 188 | 1.8 | 4.7 | 47 | | | | |
| avg/sum | 3.3 | 11767.1 | 4395.3 | 0 | 22.3 | 20.2 | 64.7 | 19.0 | 18.6 | 22.7 | | | 72.8 | all but unassigned |
| avg/sum | 3.1 | 12645.0 | 4395.3 | 0 | 32.0 | 17.7 | 56.8 | 29.1 | 16.2 | 20.0 | | | 74.2 | all with unassigned |

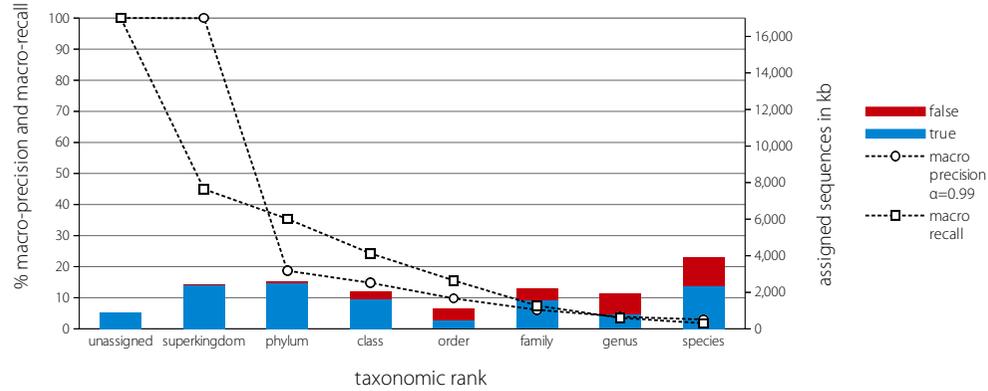

*Supplementary Figure 5: MEGAN binning for FAMeS SimMC*   **(b) all reference scenario**

| rank | depth | true | false | unknown | macro precision α=0.99 | stdev | pred. bins | macro recall | stdev | real bins | sum true | sum false | overall prec. | description |
|---|---|---|---|---|---|---|---|---|---|---|---|---|---|---|
| unassigned | 0 | 2.03 | 0 | 0 | 100.0 | 0.0 | 1 | 100.0 | 0.0 | 1 | 20.91 | 0 | 100.0 | root+superkingdom |
| superkingdom | 1 | 9.44 | 0 | 0 | 100.0 | 0.0 | 1 | 50.0 | 50.0 | 2 | | | | |
| phylum | 2 | 21.2 | 0 | 0 | 100.0 | 0.0 | 2 | 49.3 | 49.3 | 8 | 72.15 | 0 | 100.0 | phylum+class+order |
| class | 3 | 34.76 | 0 | 0 | 100.0 | 0.0 | 2 | 49.2 | 49.2 | 12 | | | | |
| order | 4 | 16.19 | 0 | 0 | 100.0 | 0.0 | 3 | 33.9 | 46.5 | 23 | | | | |
| family | 5 | 28.28 | 0 | 0 | 100.0 | 0.0 | 3 | 19.7 | 39.5 | 30 | 16956.66 | 0 | 100.0 | family+genus+species |
| genus | 6 | 602.48 | 0 | 0 | 100.0 | 0.0 | 4 | 18.7 | 38.7 | 37 | | | | |
| species | 7 | 16325.9 | 0 | 0 | 100.0 | 0.0 | 4 | 12.5 | 32.6 | 47 | | | | |
| avg/sum | 5.6 | 17038.25 | 0 | 0 | 100.0 | 0.0 | 2.6 | 33.3 | 43.7 | 22.7 | | | 100.0 | all but unassigned |
| avg/sum | 5.6 | 17040.28 | 0 | 0 | 100.0 | 0.0 | 2.4 | 41.7 | 38.2 | 20.0 | | | 100.0 | all with unassigned |

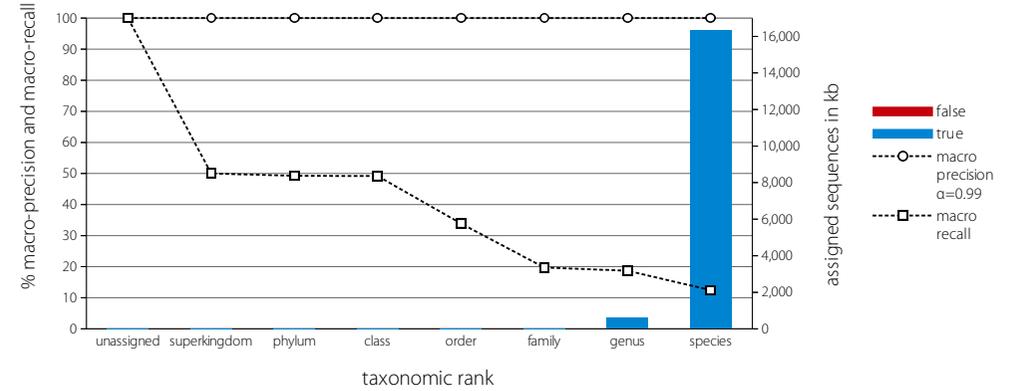

*Supplementary Figure 5: MEGAN binning for FAMeS SimMC*   **(c) new species scenario**

| rank | depth | true | false | unknown | macro precision α=0.99 | stdev | pred. bins | macro recall | stdev | real bins | sum true | sum false | overall prec. | description |
|---|---|---|---|---|---|---|---|---|---|---|---|---|---|---|
| unassigned | 0 | 234.56 | 0 | 0 | 100.0 | 0.0 | 1 | 100.0 | 0.0 | 1 | 781.74 | 0 | 100.0 | root+superkingdom |
| superkingdom | 1 | 273.59 | 0 | 0 | 100.0 | 0.0 | 1 | 48.4 | 48.4 | 2 | | | | |
| phylum | 2 | 1162.26 | 2.62 | 0 | 100.0 | 0.0 | 1 | 55.4 | 44.1 | 8 | 2423.08 | 125.55 | 95.1 | phylum+class+order |
| class | 3 | 683.93 | 59.82 | 0 | 66.8 | 44.7 | 3 | 45.6 | 37.9 | 12 | | | | |
| order | 4 | 576.89 | 63.11 | 0 | 36.1 | 40.3 | 10 | 36.2 | 39.5 | 23 | | | | |
| family | 5 | 4640.58 | 256.49 | 0 | 21.0 | 38.4 | 18 | 21.3 | 36.6 | 30 | 9717.71 | 4265.77 | 69.5 | family+genus+species |
| genus | 6 | 5077.13 | 1966.57 | 0 | 11.0 | 28.8 | 24 | 6.1 | 18.7 | 37 | | | | |
| species | 7 | 0 | 2042.71 | 0 | 0.0 | 0.0 | 32 | 0.0 | 0.0 | 47 | | | | |
| avg/sum | 5.0 | 12414.38 | 4391.32 | 0 | 47.8 | 21.7 | 12.7 | 30.4 | 32.2 | 22.7 | | | 73.9 | all but unassigned |
| avg/sum | 4.9 | 12648.94 | 4391.32 | 0 | 54.4 | 19.0 | 11.3 | 39.1 | 28.2 | 20.0 | | | 74.2 | all with unassigned |

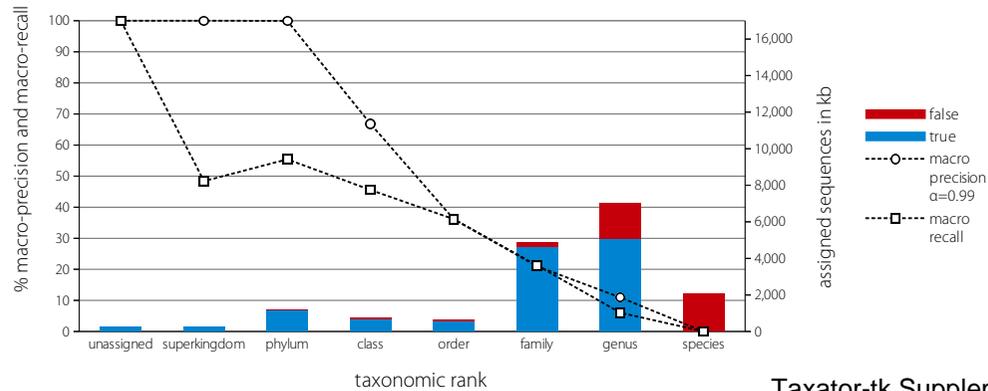

*Supplementary Figure 5: MEGAN binning for FAMeS SimMC*   **(d) new genus scenario**

| rank | depth | true | false | unknown | macro precision α=0.99 | stdev | pred. bins | macro recall | stdev | real bins | sum true | sum false | overall prec. | description |
|---|---|---|---|---|---|---|---|---|---|---|---|---|---|---|
| unassigned | 0 | 358.62 | 0 | 0 | 100.0 | 0.0 | 1 | 100.0 | 0.0 | 1 | 1411.58 | 0 | 100.0 | root+superkingdom |
| superkingdom | 1 | 526.48 | 0 | 0 | 100.0 | 0.0 | 1 | 47.7 | 47.7 | 2 | | | | |
| phylum | 2 | 1889.35 | 2.62 | 0 | 100.0 | 0.0 | 1 | 54.3 | 43.3 | 8 | 4564.44 | 220.44 | 95.4 | phylum+class+order |
| class | 3 | 1360.44 | 89.37 | 0 | 65.6 | 45.6 | 3 | 40.3 | 35.8 | 12 | | | | |
| order | 4 | 1314.65 | 128.45 | 0 | 31.8 | 40.5 | 11 | 26.2 | 31.9 | 23 | | | | |
| family | 5 | 6466.24 | 303.88 | 0 | 13.9 | 32.2 | 17 | 11.2 | 28.1 | 30 | 6466.24 | 4904.05 | 56.9 | family+genus+species |
| genus | 6 | 0 | 2126.87 | 0 | 0.0 | 0.0 | 39 | 0.0 | 0.0 | 37 | | | | |
| species | 7 | 0 | 2473.3 | 0 | 0.0 | 0.0 | 45 | 0.0 | 0.0 | 47 | | | | |
| avg/sum | 4.3 | 11557.16 | 5124.49 | 0 | 44.5 | 16.9 | 16.7 | 25.7 | 26.7 | 22.7 | | | 69.3 | all but unassigned |
| avg/sum | 4.2 | 11915.78 | 5124.49 | 0 | 51.4 | 14.8 | 14.8 | 34.9 | 23.3 | 20.0 | | | 69.9 | all with unassigned |

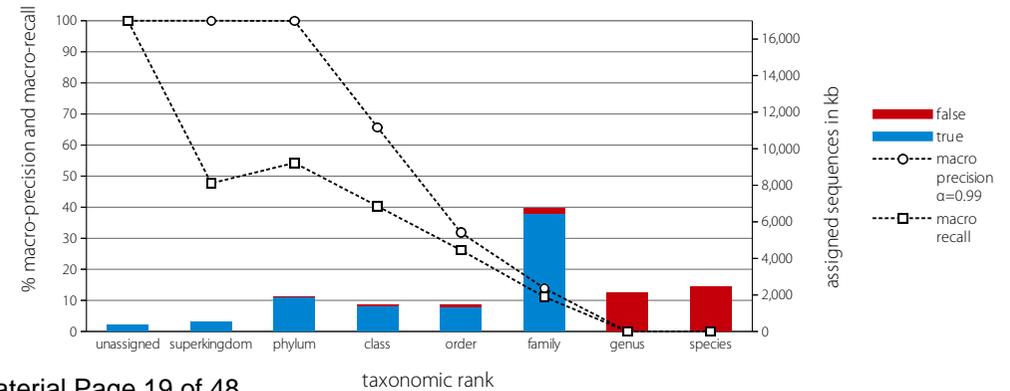

*Supplementary Figure 5: MEGAN binning for FAMeS SimMC*  **(e) new family scenario**

| rank | depth | true | false | unknown | macro precision α=0.99 | stdev | pred. bins | macro recall | stdev | real bins | sum true | sum false | overall prec. | description |
|---|---|---|---|---|---|---|---|---|---|---|---|---|---|---|
| unassigned | 0 | 663.34 | 0 | 0 | 100.0 | 0.0 | 1 | 100.0 | 0.0 | 1 | 4214.02 | 0 | 100.0 | root+superkingdom |
| superkingdom | 1 | 1775.34 | 0 | 0 | 100.0 | 0.0 | 1 | 45.8 | 45.8 | 2 | | | | |
| phylum | 2 | 4398.98 | 13.81 | 0 | 52.5 | 47.5 | 2 | 25.8 | 33.8 | 8 | 10748.33 | 470.17 | 95.8 | phylum+class+order |
| class | 3 | 4868.55 | 130.75 | 0 | 38.0 | 42.5 | 5 | 23.4 | 24.1 | 12 | | | | |
| order | 4 | 1480.8 | 325.61 | 0 | 9.8 | 23.2 | 18 | 12.7 | 21.0 | 23 | | | | |
| family | 5 | 0 | 1031.59 | 0 | 0.0 | 0.0 | 28 | 0.0 | 0.0 | 30 | 0 | 3383.11 | 0.0 | family+genus+species |
| genus | 6 | 0 | 839.84 | 0 | 0.0 | 0.0 | 47 | 0.0 | 0.0 | 37 | | | | |
| species | 7 | 0 | 1511.68 | 0 | 0.0 | 0.0 | 47 | 0.0 | 0.0 | 47 | | | | |
| avg/sum | 2.9 | 12523.67 | 3853.28 | 0 | 28.6 | 16.2 | 21.1 | 15.4 | 17.8 | 22.7 | | | 76.5 | all but unassigned |
| avg/sum | 2.8 | 13187.01 | 3853.28 | 0 | 37.5 | 14.1 | 18.6 | 26.0 | 15.6 | 20.0 | | | 77.4 | all with unassigned |

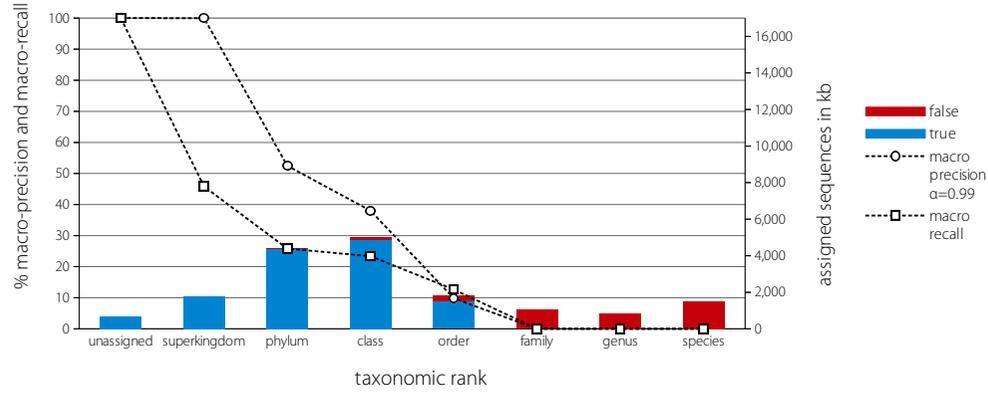

*Supplementary Figure 5: MEGAN binning for FAMeS SimMC*  **(f) new order scenario**

| rank | depth | true | false | unknown | macro precision α=0.99 | stdev | pred. bins | macro recall | stdev | real bins | sum true | sum false | overall prec. | description |
|---|---|---|---|---|---|---|---|---|---|---|---|---|---|---|
| unassigned | 0 | 767.74 | 0 | 0 | 100.0 | 0.0 | 1 | 100.0 | 0.0 | 1 | 5311.66 | 0 | 100.0 | root+superkingdom |
| superkingdom | 1 | 2271.96 | 0 | 0 | 100.0 | 0.0 | 1 | 45.1 | 45.1 | 2 | | | | |
| phylum | 2 | 5432.33 | 39.29 | 0 | 38.6 | 43.7 | 3 | 41.6 | 40.0 | 8 | 9765.97 | 793.72 | 92.5 | phylum+class+order |
| class | 3 | 4333.64 | 193 | 0 | 25.7 | 36.9 | 7 | 11.5 | 13.9 | 12 | | | | |
| order | 4 | 0 | 561.43 | 0 | 0.0 | 0.0 | 21 | 0.0 | 0.0 | 23 | | | | |
| family | 5 | 0 | 1006.4 | 0 | 0.0 | 0.0 | 30 | 0.0 | 0.0 | 30 | 0 | 3440.89 | 0.0 | family+genus+species |
| genus | 6 | 0 | 817.67 | 0 | 0.0 | 0.0 | 43 | 0.0 | 0.0 | 37 | | | | |
| species | 7 | 0 | 1616.82 | 0 | 0.0 | 0.0 | 39 | 0.0 | 0.0 | 47 | | | | |
| avg/sum | 2.7 | 12037.93 | 4234.61 | 0 | 23.5 | 11.5 | 20.6 | 14.0 | 14.1 | 22.7 | | | 74.0 | all but unassigned |
| avg/sum | 2.5 | 12805.67 | 4234.61 | 0 | 33.0 | 10.1 | 18.1 | 24.8 | 12.4 | 20.0 | | | 75.1 | all with unassigned |

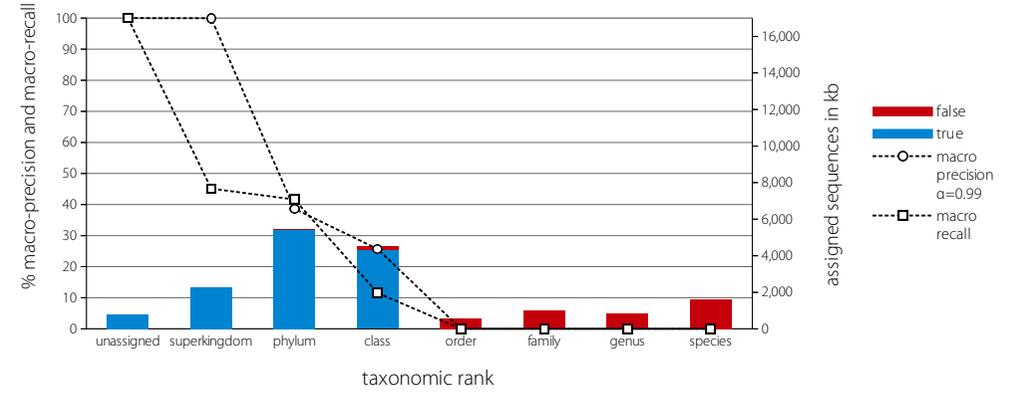

*Supplementary Figure 5: MEGAN binning for FAMeS SimMC*  **(g) new class scenario**

| rank | depth | true | false | unknown | macro precision α=0.99 | stdev | pred. bins | macro recall | stdev | real bins | sum true | sum false | overall prec. | description |
|---|---|---|---|---|---|---|---|---|---|---|---|---|---|---|
| unassigned | 0 | 1274.66 | 0 | 0 | 100.0 | 0.0 | 1 | 100.0 | 0.0 | 1 | 11193.42 | 0 | 100.0 | root+superkingdom |
| superkingdom | 1 | 4959.38 | 0 | 0 | 100.0 | 0.0 | 1 | 42.4 | 42.4 | 2 | | | | |
| phylum | 2 | 4654.01 | 87.88 | 0 | 15.0 | 34.7 | 7 | 21.4 | 34.5 | 8 | 4654.01 | 3194.98 | 59.3 | phylum+class+order |
| class | 3 | 0 | 1661.84 | 0 | 0.0 | 0.0 | 12 | 0.0 | 0.0 | 12 | | | | |
| order | 4 | 0 | 1445.26 | 0 | 0.0 | 0.0 | 31 | 0.0 | 0.0 | 23 | | | | |
| family | 5 | 0 | 960.19 | 0 | 0.0 | 0.0 | 40 | 0.0 | 0.0 | 30 | 0 | 2957.25 | 0.0 | family+genus+species |
| genus | 6 | 0 | 1106.84 | 0 | 0.0 | 0.0 | 57 | 0.0 | 0.0 | 37 | | | | |
| species | 7 | 0 | 890.22 | 0 | 0.0 | 0.0 | 46 | 0.0 | 0.0 | 47 | | | | |
| avg/sum | 2.5 | 9613.39 | 6152.23 | 0 | 16.4 | 5.0 | 27.7 | 9.1 | 11.0 | 22.7 | | | 61.0 | all but unassigned |
| avg/sum | 2.3 | 10888.05 | 6152.23 | 0 | 26.9 | 4.3 | 24.4 | 20.5 | 9.6 | 20.0 | | | 63.9 | all with unassigned |

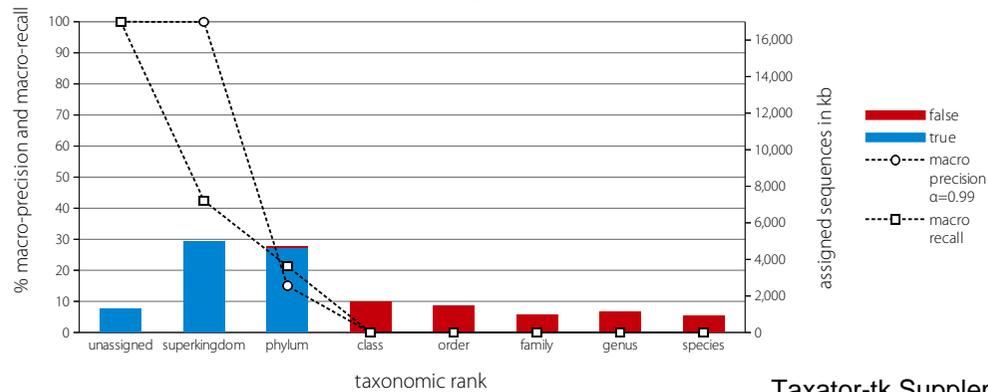

*Supplementary Figure 5: MEGAN binning for FAMeS SimMC*  **(h) new phylum scenario**

| rank | depth | true | false | unknown | macro precision α=0.99 | stdev | pred. bins | macro recall | stdev | real bins | sum true | sum false | overall prec. | description |
|---|---|---|---|---|---|---|---|---|---|---|---|---|---|---|
| unassigned | 0 | 2844.15 | 0 | 0 | 100.0 | 0.0 | 1 | 100.0 | 0.0 | 1 | 17213.77 | 52.43 | 99.7 | root+superkingdom |
| superkingdom | 1 | 7184.81 | 52.43 | 0 | 100.0 | 0.0 | 1 | 35.3 | 35.3 | 2 | | | | |
| phylum | 2 | 0 | 273.68 | 0 | 0.0 | 0.0 | 14 | 0.0 | 0.0 | 8 | 0 | 2861.02 | 0.0 | phylum+class+order |
| class | 3 | 0 | 588.78 | 0 | 0.0 | 0.0 | 26 | 0.0 | 0.0 | 12 | | | | |
| order | 4 | 0 | 1998.56 | 0 | 0.0 | 0.0 | 40 | 0.0 | 0.0 | 23 | | | | |
| family | 5 | 0 | 762.4 | 0 | 0.0 | 0.0 | 61 | 0.0 | 0.0 | 30 | 0 | 4097.87 | 0.0 | family+genus+species |
| genus | 6 | 0 | 860.34 | 0 | 0.0 | 0.0 | 72 | 0.0 | 0.0 | 37 | | | | |
| species | 7 | 0 | 2475.13 | 0 | 0.0 | 0.0 | 69 | 0.0 | 0.0 | 47 | | | | |
| avg/sum | 2.3 | 7184.81 | 7011.32 | 0 | 14.3 | 0.0 | 40.4 | 5.0 | 5.0 | 22.7 | | | 50.6 | all but unassigned |
| avg/sum | 1.8 | 10028.96 | 7011.32 | 0 | 25.0 | 0.0 | 35.5 | 16.9 | 4.4 | 20.0 | | | 58.9 | all with unassigned |

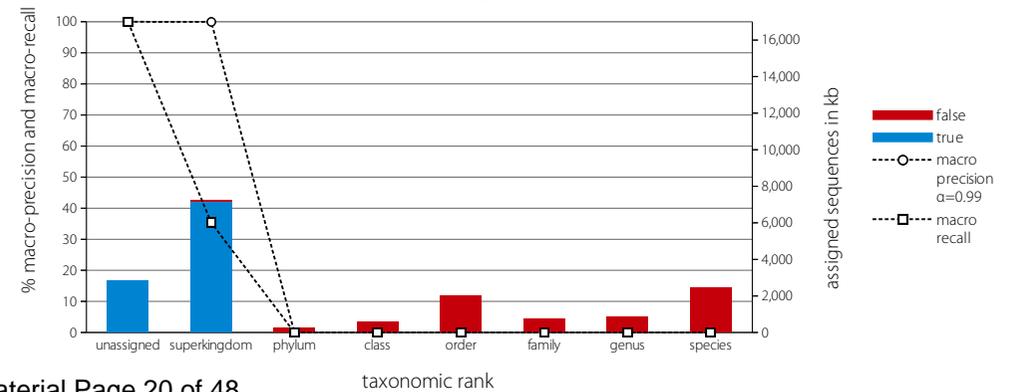

*Supplementary Figure 6: Taxator-tk binning for FAMeS SimMC* **(a) summary scenario**

| rank | depth | true | false | unknown | macro precision α=0.99 | stdev | pred. bins | macro recall | stdev | real bins | sum true | sum false | overall prec. | description |
|---|---|---|---|---|---|---|---|---|---|---|---|---|---|---|
| unassigned | 0 | 2083.8 | 0.0 | 0 | 100.0 | 0.0 | 1 | 100.0 | 0.0 | 1 | 11492.5 | 0.6 | 100.0 | root+superkingdom |
| superkingdom | 1 | 4704.3 | 0.6 | 0 | 100.0 | 0.0 | 1 | 62.5 | 19.6 | 2 | | | | |
| phylum | 2 | 3460.1 | 26.9 | 0 | 52.0 | 36.2 | 4 | 38.1 | 13.3 | 8 | 5881.5 | 466.4 | 92.7 | phylum+class+order |
| class | 3 | 1860.6 | 182.0 | 0 | 49.0 | 47.6 | 4 | 24.0 | 14.2 | 12 | | | | |
| order | 4 | 560.8 | 257.2 | 0 | 40.7 | 41.6 | 15 | 18.6 | 12.0 | 23 | | | | |
| family | 5 | 1573.3 | 89.4 | 0 | 22.8 | 38.3 | 19 | 12.8 | 10.8 | 30 | 3564.0 | 339.7 | 91.3 | family+genus+species |
| genus | 6 | 1012.7 | 196.9 | 0 | 37.5 | 45.7 | 19 | 8.0 | 7.9 | 37 | | | | |
| species | 7 | 978.0 | 53.3 | 0 | 39.2 | 48.4 | 54 | 5.0 | 6.3 | 47 | | | | |
| avg/sum | 2.6 | 14149.8 | 806.6 | 0 | 48.7 | 36.8 | 16.6 | 24.1 | 12.0 | 22.7 | | | 94.6 | all but unassigned |
| avg/sum | 2.3 | 16233.6 | 806.6 | 0 | 55.2 | 32.2 | 14.6 | 33.6 | 10.5 | 20.0 | | | 95.3 | all with unassigned |

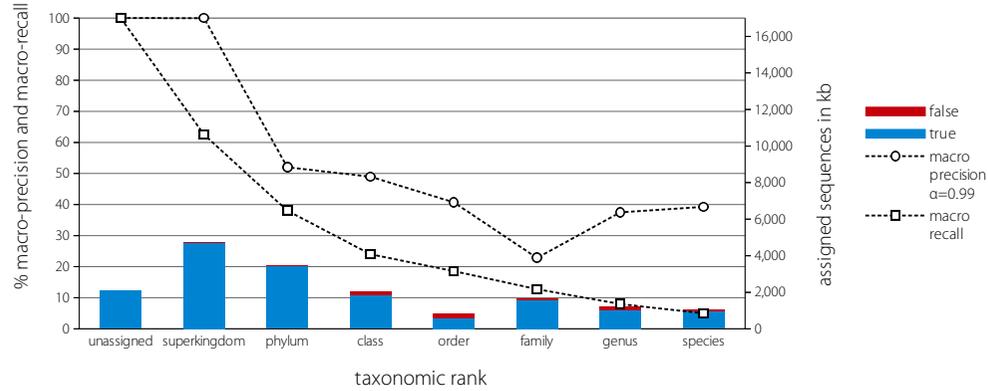

*Supplementary Figure 6: Taxator-tk binning for FAMeS SimMC* **(b) all reference scenario**

| rank | depth | true | false | unknown | macro precision α=0.99 | stdev | pred. bins | macro recall | stdev | real bins | sum true | sum false | overall prec. | description |
|---|---|---|---|---|---|---|---|---|---|---|---|---|---|---|
| unassigned | 0 | 251.07 | 0 | 0 | 100.0 | 0.0 | 1 | 100.0 | 0.0 | 1 | 2857.85 | 0 | 100.0 | root+superkingdom |
| superkingdom | 1 | 1303.39 | 0 | 0 | 100.0 | 0.0 | 1 | 98.8 | 1.2 | 2 | | | | |
| phylum | 2 | 1673.66 | 0 | 0 | 100.0 | 0.0 | 1 | 85.8 | 15.8 | 8 | 3450.59 | 0 | 100.0 | phylum+class+order |
| class | 3 | 1129.62 | 0 | 0 | 100.0 | 0.0 | 2 | 73.6 | 22.1 | 12 | | | | |
| order | 4 | 647.31 | 0 | 0 | 100.0 | 0.0 | 1 | 68.7 | 29.4 | 23 | | | | |
| family | 5 | 1728.38 | 0 | 0 | 100.0 | 0.0 | 3 | 59.0 | 42.0 | 30 | 12035.22 | 0 | 100.0 | family+genus+species |
| genus | 6 | 3460.93 | 0 | 0 | 100.0 | 0.0 | 5 | 47.1 | 43.6 | 37 | | | | |
| species | 7 | 6845.91 | 0 | 0 | 100.0 | 0.0 | 5 | 34.9 | 44.2 | 47 | | | | |
| avg/sum | 4.0 | 16789.2 | 0 | 0 | 100.0 | 0.0 | 3.0 | 66.8 | 28.3 | 22.7 | | | 100.0 | all but unassigned |
| avg/sum | 3.9 | 17040.27 | 0 | 0 | 100.0 | 0.0 | 2.8 | 71.0 | 24.8 | 20.0 | | | 100.0 | all with unassigned |

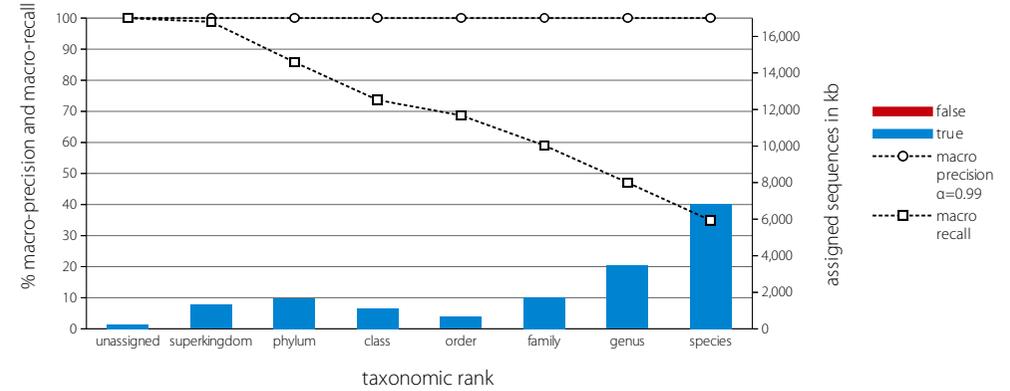

*Supplementary Figure 6: Taxator-tk binning for FAMeS SimMC* **(c) new species scenario**

| rank | depth | true | false | unknown | macro precision α=0.99 | stdev | pred. bins | macro recall | stdev | real bins | sum true | sum false | overall prec. | description |
|---|---|---|---|---|---|---|---|---|---|---|---|---|---|---|
| unassigned | 0 | 1558.82 | 0 | 0 | 100.0 | 0.0 | 1 | 100.0 | 0.0 | 1 | 4785.94 | 0 | 100.0 | root+superkingdom |
| superkingdom | 1 | 1613.56 | 0 | 0 | 100.0 | 0.0 | 1 | 95.8 | 4.2 | 2 | | | | |
| phylum | 2 | 2761.05 | 2.62 | 0 | 100.0 | 0.0 | 1 | 69.6 | 30.2 | 8 | 5592.14 | 50.24 | 99.1 | phylum+class+order |
| class | 3 | 1806.58 | 27.87 | 0 | 69.1 | 42.1 | 3 | 39.9 | 35.3 | 12 | | | | |
| order | 4 | 1024.51 | 19.75 | 0 | 75.1 | 35.6 | 7 | 32.4 | 34.2 | 23 | | | | |
| family | 5 | 3915.66 | 33.96 | 0 | 81.0 | 36.6 | 6 | 22.0 | 34.7 | 30 | 7543.8 | 681.73 | 91.7 | family+genus+species |
| genus | 6 | 3628.14 | 630.72 | 0 | 63.1 | 44.8 | 3 | 8.8 | 24.1 | 37 | | | | |
| species | 7 | 0 | 17.05 | 0 | 0.0 | 0.0 | 11 | 0.0 | 0.0 | 47 | | | | |
| avg/sum | 4.0 | 14749.5 | 731.97 | 0 | 69.7 | 22.7 | 4.6 | 38.4 | 23.2 | 22.7 | | | 95.3 | all but unassigned |
| avg/sum | 3.6 | 16308.32 | 731.97 | 0 | 73.5 | 19.9 | 4.1 | 46.1 | 20.3 | 20.0 | | | 95.7 | all with unassigned |

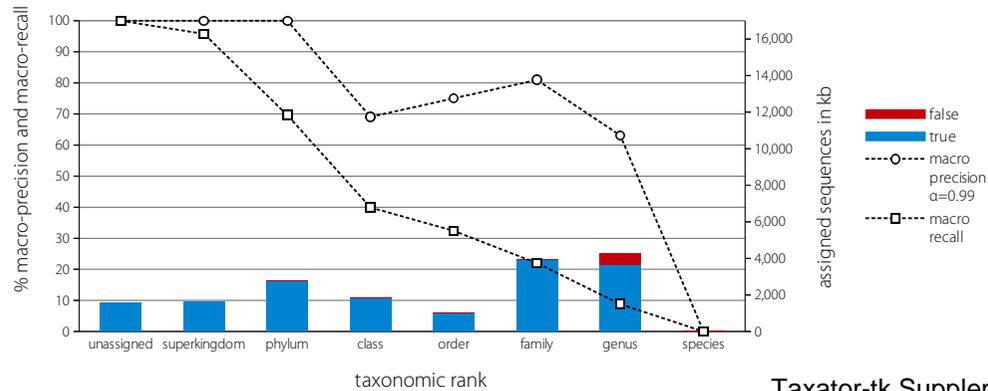

*Supplementary Figure 6: Taxator-tk binning for FAMeS SimMC* **(d) new genus scenario**

| rank | depth | true | false | unknown | macro precision α=0.99 | stdev | pred. bins | macro recall | stdev | real bins | sum true | sum false | overall prec. | description |
|---|---|---|---|---|---|---|---|---|---|---|---|---|---|---|
| unassigned | 0 | 1745.53 | 0 | 0 | 100.0 | 0.0 | 1 | 100.0 | 0.0 | 1 | 5698.63 | 0 | 100.0 | root+superkingdom |
| superkingdom | 1 | 1976.55 | 0 | 0 | 100.0 | 0.0 | 1 | 94.6 | 5.4 | 2 | | | | |
| phylum | 2 | 3398.22 | 2.62 | 0 | 100.0 | 0.0 | 1 | 63.2 | 32.7 | 8 | 6900.35 | 75.58 | 98.9 | phylum+class+order |
| class | 3 | 2312.46 | 32.37 | 0 | 66.9 | 44.9 | 3 | 30.2 | 32.2 | 12 | | | | |
| order | 4 | 1189.02 | 40.59 | 0 | 55.5 | 42.8 | 7 | 22.2 | 32.0 | 23 | | | | |
| family | 5 | 5368.74 | 54.74 | 0 | 50.0 | 46.3 | 7 | 8.3 | 22.3 | 30 | 5368.74 | 973.52 | 84.7 | family+genus+species |
| genus | 6 | 0 | 636.48 | 0 | 0.0 | 0.0 | 15 | 0.0 | 0.0 | 37 | | | | |
| species | 7 | 0 | 282.3 | 0 | 0.0 | 0.0 | 9 | 0.0 | 0.0 | 47 | | | | |
| avg/sum | 3.4 | 14245.64 | 1049.1 | 0 | 53.2 | 19.2 | 6.1 | 31.2 | 17.8 | 22.7 | | | 93.1 | all but unassigned |
| avg/sum | 3.1 | 15991.17 | 1049.1 | 0 | 59.0 | 16.8 | 5.5 | 39.8 | 15.6 | 20.0 | | | 93.8 | all with unassigned |

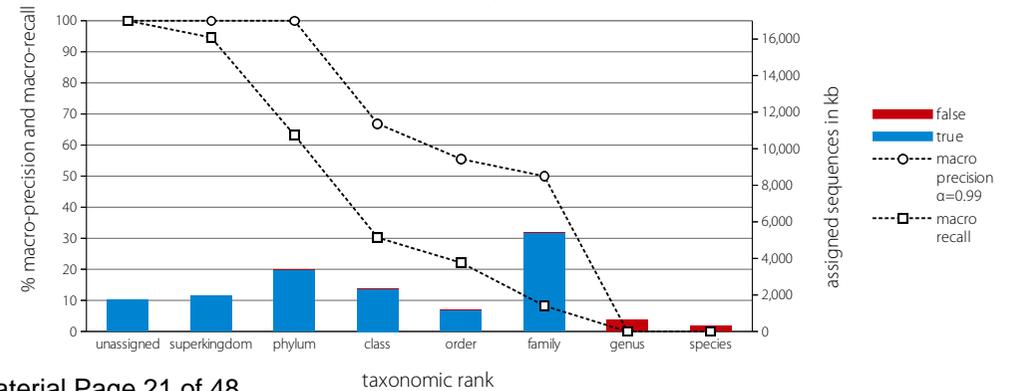

*Supplementary Figure 6: Taxator-tk binning for FAMeS SimMC* **(e) new family scenario**

| rank | depth | true | false | unknown | macro precision α=0.99 | stdev | pred. bins | macro recall | stdev | real bins | sum true | sum false | overall prec. | description |
|---|---|---|---|---|---|---|---|---|---|---|---|---|---|---|
| unassigned | 0 | 1444.21 | 0 | 0 | 100.0 | 0.0 | 1 | 100.0 | 0.0 | 1 | | | 100.0 | root+superkingdom |
| superkingdom | 1 | 4049.76 | 0 | 0 | 100.0 | 0.0 | 1 | 42.8 | 42.8 | 2 | 9543.73 | 0 | | |
| phylum | 2 | 5463.82 | 11.04 | 0 | 100.0 | 0.0 | 1 | 31.1 | 36.5 | 8 | | | | |
| class | 3 | 4607.11 | 80.28 | 0 | 61.6 | 43.9 | 3 | 21.0 | 31.0 | 12 | 11135.95 | 187.88 | 98.3 | phylum+class+order |
| order | 4 | 1065.02 | 96.56 | 0 | 20.6 | 37.4 | 18 | 6.8 | 17.0 | 23 | | | | |
| family | 5 | 0 | 179.84 | 0 | 0.0 | 0.0 | 21 | 0.0 | 0.0 | 30 | | | | |
| genus | 6 | 0 | 32.8 | 0 | 0.0 | 0.0 | 14 | 0.0 | 0.0 | 37 | 0 | 222.49 | 0.0 | family+genus+species |
| species | 7 | 0 | 9.85 | 0 | 0.0 | 0.0 | 9 | 0.0 | 0.0 | 47 | | | | |
| avg/sum | 2.2 | 15185.71 | 410.37 | 0 | 40.3 | 11.6 | 9.3 | 14.5 | 18.2 | 22.7 | | | 97.4 | all but unassigned |
| avg/sum | 2.0 | 16629.92 | 410.37 | 0 | 47.8 | 10.2 | 8.3 | 25.2 | 15.9 | 20.0 | | | 97.6 | all with unassigned |

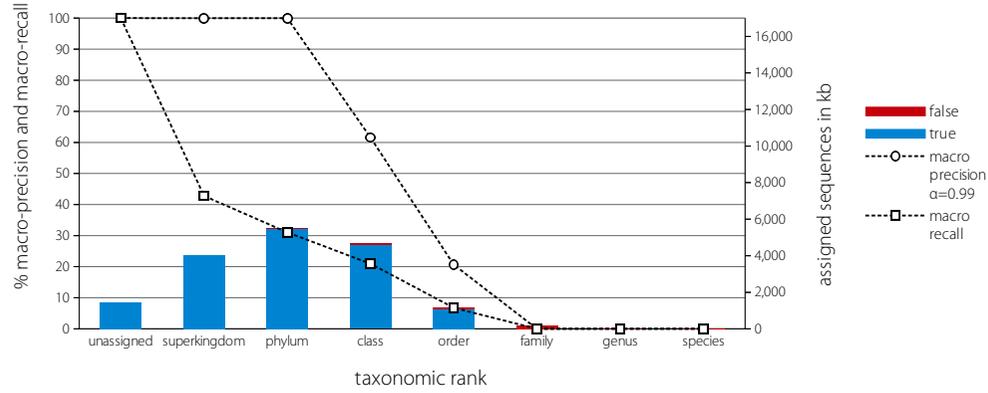

*Supplementary Figure 6: Taxator-tk binning for FAMeS SimMC* **(f) new order scenario**

| rank | depth | true | false | unknown | macro precision α=0.99 | stdev | pred. bins | macro recall | stdev | real bins | sum true | sum false | overall prec. | description |
|---|---|---|---|---|---|---|---|---|---|---|---|---|---|---|
| unassigned | 0 | 1665.24 | 0 | 0 | 100.0 | 0.0 | 1 | 100.0 | 0.0 | 1 | | | 100.0 | root+superkingdom |
| superkingdom | 1 | 5067.52 | 0 | 0 | 100.0 | 0.0 | 1 | 41.4 | 41.4 | 2 | 11800.28 | 0 | | |
| phylum | 2 | 6525.5 | 14.12 | 0 | 100.0 | 0.0 | 1 | 13.5 | 20.3 | 8 | | | | |
| class | 3 | 3168.64 | 86.66 | 0 | 45.2 | 45.6 | 4 | 3.3 | 6.6 | 12 | 9694.14 | 388.89 | 96.1 | phylum+class+order |
| order | 4 | 0 | 288.11 | 0 | 0.0 | 0.0 | 19 | 0.0 | 0.0 | 23 | | | | |
| family | 5 | 0 | 169.88 | 0 | 0.0 | 0.0 | 17 | 0.0 | 0.0 | 30 | | | | |
| genus | 6 | 0 | 40.01 | 0 | 0.0 | 0.0 | 14 | 0.0 | 0.0 | 37 | 0 | 224.49 | 0.0 | family+genus+species |
| species | 7 | 0 | 14.6 | 0 | 0.0 | 0.0 | 9 | 0.0 | 0.0 | 47 | | | | |
| avg/sum | 2.0 | 14761.66 | 613.38 | 0 | 35.0 | 6.5 | 9.3 | 8.3 | 9.8 | 22.7 | | | 96.0 | all but unassigned |
| avg/sum | 1.8 | 16426.9 | 613.38 | 0 | 43.1 | 5.7 | 8.3 | 19.8 | 8.5 | 20.0 | | | 96.4 | all with unassigned |

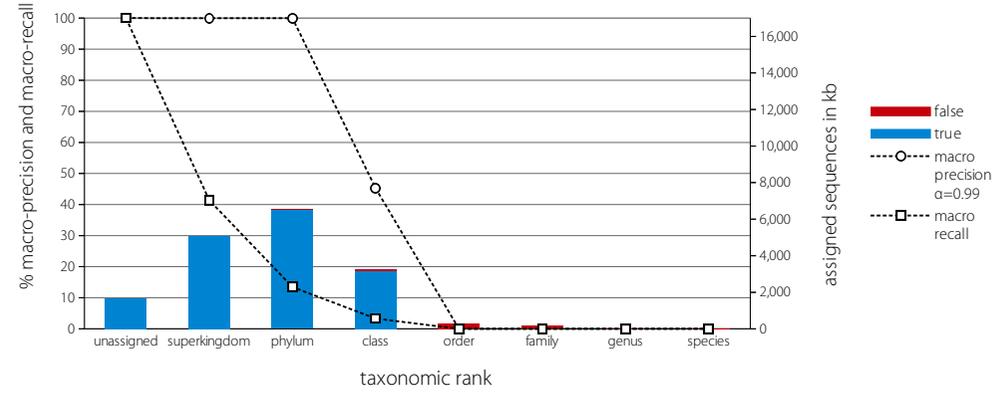

*Supplementary Figure 6: Taxator-tk binning for FAMeS SimMC* **(g) new class scenario**

| rank | depth | true | false | unknown | macro precision α=0.99 | stdev | pred. bins | macro recall | stdev | real bins | sum true | sum false | overall prec. | description |
|---|---|---|---|---|---|---|---|---|---|---|---|---|---|---|
| unassigned | 0 | 2853.36 | 0 | 0 | 100.0 | 0.0 | 1 | 100.0 | 0.0 | 1 | | | 100.0 | root+superkingdom |
| superkingdom | 1 | 8356.71 | 0 | 0 | 100.0 | 0.0 | 1 | 36.9 | 36.9 | 2 | 19566.78 | 0 | | |
| phylum | 2 | 4397.58 | 25.84 | 0 | 33.3 | 47.1 | 3 | 3.2 | 8.5 | 8 | | | | |
| class | 3 | 0 | 659.76 | 0 | 0.0 | 0.0 | 11 | 0.0 | 0.0 | 12 | 4397.58 | 1282.66 | 77.4 | phylum+class+order |
| order | 4 | 0 | 597.06 | 0 | 0.0 | 0.0 | 18 | 0.0 | 0.0 | 23 | | | | |
| family | 5 | 0 | 108.29 | 0 | 0.0 | 0.0 | 21 | 0.0 | 0.0 | 30 | | | | |
| genus | 6 | 0 | 23.82 | 0 | 0.0 | 0.0 | 14 | 0.0 | 0.0 | 37 | 0 | 149.96 | 0.0 | family+genus+species |
| species | 7 | 0 | 17.85 | 0 | 0.0 | 0.0 | 9 | 0.0 | 0.0 | 47 | | | | |
| avg/sum | 1.6 | 12754.29 | 1432.62 | 0 | 19.0 | 6.7 | 11.0 | 5.7 | 6.5 | 22.7 | | | 89.9 | all but unassigned |
| avg/sum | 1.3 | 15607.65 | 1432.62 | 0 | 29.2 | 5.9 | 9.8 | 17.5 | 5.7 | 20.0 | | | 91.6 | all with unassigned |

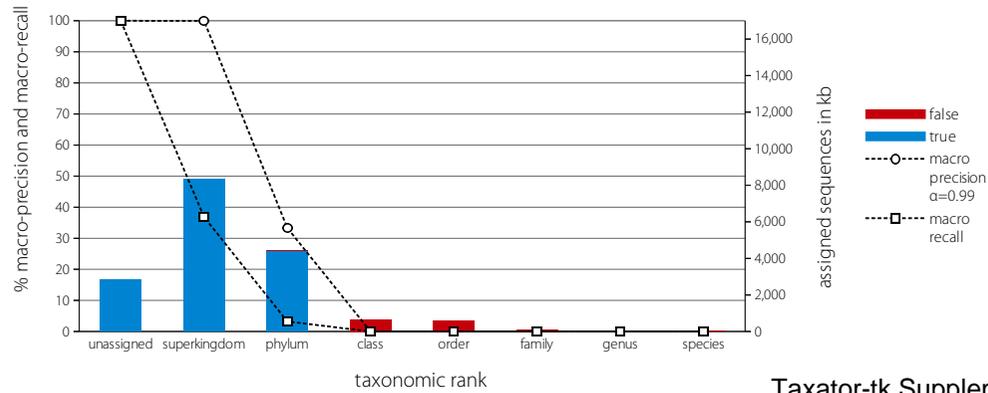

*Supplementary Figure 6: Taxator-tk binning for FAMeS SimMC* **(h) new phylum scenario**

| rank | depth | true | false | unknown | macro precision α=0.99 | stdev | pred. bins | macro recall | stdev | real bins | sum true | sum false | overall prec. | description |
|---|---|---|---|---|---|---|---|---|---|---|---|---|---|---|
| unassigned | 0 | 5068.27 | 0 | 0 | 100.0 | 0.0 | 1 | 100.0 | 0.0 | 1 | | | 100.0 | root+superkingdom |
| superkingdom | 1 | 10562.94 | 4.05 | 0 | 100.0 | 0.0 | 1 | 27.3 | 27.3 | 2 | 26194.15 | 4.05 | | |
| phylum | 2 | 0 | 132.26 | 0 | 0.0 | 0.0 | 10 | 0.0 | 0.0 | 8 | | | | |
| class | 3 | 0 | 388.44 | 0 | 0.0 | 0.0 | 18 | 0.0 | 0.0 | 12 | 0 | 1279.35 | 0.0 | phylum+class+order |
| order | 4 | 0 | 758.65 | 0 | 0.0 | 0.0 | 19 | 0.0 | 0.0 | 23 | | | | |
| family | 5 | 0 | 79.35 | 0 | 0.0 | 0.0 | 24 | 0.0 | 0.0 | 30 | | | | |
| genus | 6 | 0 | 14.55 | 0 | 0.0 | 0.0 | 17 | 0.0 | 0.0 | 37 | 0 | 125.67 | 0.0 | family+genus+species |
| species | 7 | 0 | 31.77 | 0 | 0.0 | 0.0 | 14 | 0.0 | 0.0 | 47 | | | | |
| avg/sum | 1.3 | 10562.94 | 1409.07 | 0 | 14.3 | 0.0 | 14.7 | 3.9 | 3.9 | 22.7 | | | 88.2 | all but unassigned |
| avg/sum | 0.9 | 15631.21 | 1409.07 | 0 | 25.0 | 0.0 | 13.0 | 15.9 | 3.4 | 20.0 | | | 91.7 | all with unassigned |

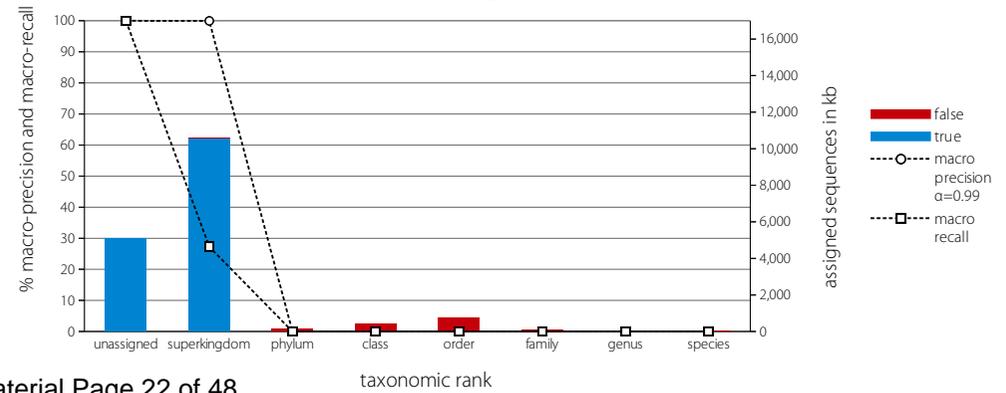

*Supplementary Figure 7: MEGAN binning for FAMeS SimHC* **(a) summary scenario**

| rank | depth | true | false | unknown | macro precision α=0.99 | stdev | pred. bins | macro recall | stdev | real bins | sum true | sum false | overall prec | description |
|---|---|---|---|---|---|---|---|---|---|---|---|---|---|---|
| unassigned | 0 | 135097.7 | 0.0 | 0.0 | 100.0 | 0.0 | 1 | 100.0 | 0.0 | 1 | 504932.0 | 1240.4 | 99.8 | root+superkingdom |
| superkingdom | 1 | 184917.1 | 1240.4 | 0.0 | 99.5 | 0.0 | 1 | 65.7 | 21.3 | 2 | | | | |
| phylum | 2 | 126186.3 | 17996.0 | 0.0 | 54.3 | 45.2 | 10 | 39.8 | 24.2 | 8 | | | | |
| class | 3 | 105637.4 | 52554.0 | 2704.9 | 61.0 | 36.9 | 12 | 32.6 | 19.7 | 12 | 297237.9 | 124491.1 | 70.5 | phylum+class+order |
| order | 4 | 65414.1 | 53941.1 | 0.0 | 60.5 | 41.5 | 27 | 17.4 | 17.2 | 36 | | | | |
| family | 5 | 47775.7 | 34408.6 | 0.0 | 68.1 | 38.6 | 36 | 13.1 | 13.3 | 52 | | | | |
| genus | 6 | 70368.1 | 42132.0 | 382.3 | 65.3 | 43.0 | 47 | 8.4 | 9.4 | 72 | 228155.2 | 135050.2 | 62.8 | family+genus+species |
| species | 7 | 110011.3 | 58509.7 | 356.5 | 69.9 | 45.6 | 47 | 4.7 | 6.5 | 96 | | | | |
| avg/sum | 3.1 | 710310.2 | 260781.8 | 3443.6 | 68.4 | 35.8 | 25.7 | 25.9 | 15.9 | 39.7 | | | 73.1 | all but unassigned |
| avg/sum | 2.7 | 845407.9 | 260781.8 | 3443.6 | 72.3 | 31.3 | 22.6 | 35.2 | 14.0 | 34.9 | | | 76.4 | all with unassigned |

*Supplementary Figure 7: MEGAN binning for FAMeS SimHC* **(b) all reference scenario**

| rank | depth | true | false | unknown | macro precision α=0.99 | stdev | pred. bins | macro recall | stdev | real bins | sum true | sum false | overall prec | description |
|---|---|---|---|---|---|---|---|---|---|---|---|---|---|---|
| unassigned | 0 | 0 | 0 | 0 | 100.0 | 0.0 | 1 | 100.0 | 0.0 | 1 | 29660 | 0 | 100.0 | root+superkingdom |
| superkingdom | 1 | 14830 | 0 | 0 | 100.0 | 0.0 | 2 | 100.0 | 0.0 | 2 | | | | |
| phylum | 2 | 27071 | 0 | 0 | 99.6 | 1.0 | 6 | 74.9 | 43.2 | 8 | | | | |
| class | 3 | 58344 | 0 | 0 | 99.6 | 0.8 | 9 | 74.0 | 42.8 | 12 | 123849 | 0 | 100.0 | phylum+class+order |
| order | 4 | 38434 | 0 | 0 | 99.8 | 0.7 | 20 | 55.0 | 49.3 | 36 | | | | |
| family | 5 | 58139 | 0 | 0 | 99.4 | 3.1 | 29 | 54.8 | 49.1 | 52 | | | | |
| genus | 6 | 223807 | 0 | 2676 | 99.5 | 2.8 | 34 | 46.7 | 49.5 | 72 | 942014 | 0 | 100.0 | family+genus+species |
| species | 7 | 660068 | 0 | 2139 | 99.5 | 2.9 | 33 | 32.8 | 45.8 | 96 | | | | |
| avg/sum | 4.8 | 1080693 | 0 | 4815 | 99.6 | 1.6 | 19.0 | 62.6 | 39.9 | 39.7 | | | 100.0 | all but unassigned |
| avg/sum | 4.8 | 1080693 | 0 | 4815 | 99.7 | 1.4 | 16.8 | 67.3 | 35.0 | 34.9 | | | 100.0 | all with unassigned |

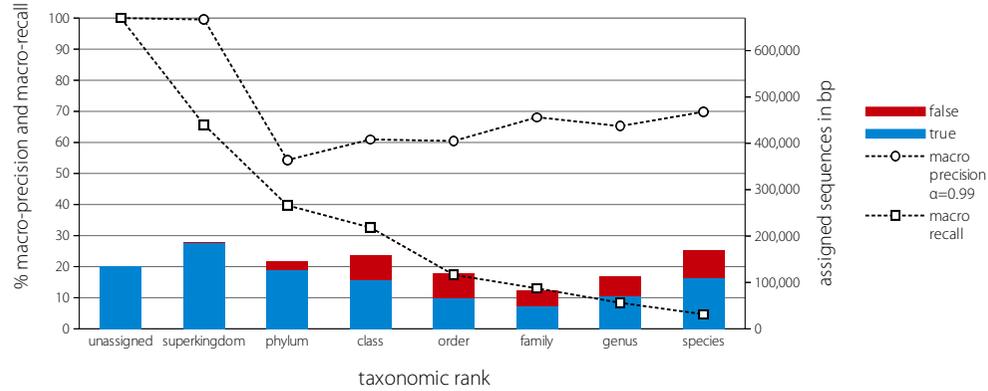

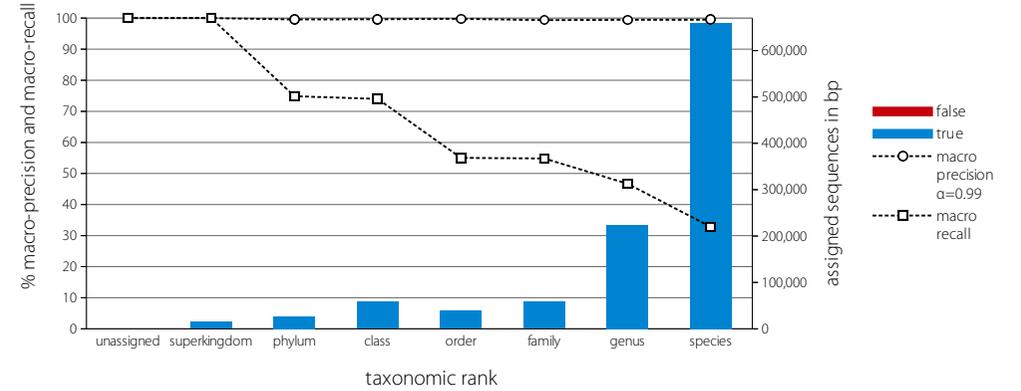

*Supplementary Figure 7: MEGAN binning for FAMeS SimHC* **(c) new species scenario**

| rank | depth | true | false | unknown | macro precision α=0.99 | stdev | pred. bins | macro recall | stdev | real bins | sum true | sum false | overall prec | description |
|---|---|---|---|---|---|---|---|---|---|---|---|---|---|---|
| unassigned | 0 | 53833 | 0 | 0 | 100.0 | 0.0 | 1 | 100.0 | 0.0 | 1 | 196059 | 0 | 100.0 | root+superkingdom |
| superkingdom | 1 | 71113 | 0 | 0 | 99.9 | 0.1 | 2 | 86.2 | 8.4 | 2 | | | | |
| phylum | 2 | 85694 | 0 | 0 | 99.7 | 0.5 | 6 | 65.7 | 38.2 | 8 | | | | |
| class | 3 | 111601 | 14697 | 0 | 96.4 | 4.6 | 9 | 57.5 | 34.8 | 12 | 311019 | 25259 | 92.5 | phylum+class+order |
| order | 4 | 113724 | 10562 | 0 | 86.9 | 29.7 | 18 | 33.0 | 38.4 | 36 | | | | |
| family | 5 | 140308 | 25633 | 0 | 83.4 | 27.0 | 20 | 24.5 | 35.6 | 52 | | | | |
| genus | 6 | 268770 | 68288 | 0 | 56.8 | 44.8 | 17 | 11.9 | 29.1 | 72 | 409078 | 215206 | 65.5 | family+genus+species |
| species | 7 | 0 | 121285 | 0 | 0.0 | 0.0 | 8 | 0.0 | 0.0 | 96 | | | | |
| avg/sum | 4.4 | 791210 | 240465 | 0 | 74.7 | 15.2 | 11.4 | 39.8 | 26.4 | 39.7 | | | 76.7 | all but unassigned |
| avg/sum | 4.1 | 845043 | 240465 | 0 | 77.9 | 13.3 | 10.1 | 47.3 | 23.1 | 34.9 | | | 77.8 | all with unassigned |

*Supplementary Figure 7: MEGAN binning for FAMeS SimHC* **(d) new genus scenario**

| rank | depth | true | false | unknown | macro precision α=0.99 | stdev | pred. bins | macro recall | stdev | real bins | sum true | sum false | overall prec | description |
|---|---|---|---|---|---|---|---|---|---|---|---|---|---|---|
| unassigned | 0 | 60008 | 0 | 0 | 100.0 | 0.0 | 1 | 100.0 | 0.0 | 1 | 266134 | 0 | 100.0 | root+superkingdom |
| superkingdom | 1 | 103063 | 0 | 0 | 99.9 | 0.1 | 2 | 85.7 | 8.0 | 2 | | | | |
| phylum | 2 | 142075 | 0 | 0 | 98.3 | 2.3 | 6 | 62.6 | 36.7 | 8 | | | | |
| class | 3 | 171151 | 27602 | 0 | 92.0 | 6.7 | 9 | 49.7 | 31.2 | 12 | 481779 | 57207 | 89.4 | phylum+class+order |
| order | 4 | 168553 | 29605 | 0 | 73.6 | 34.2 | 18 | 25.2 | 32.4 | 36 | | | | |
| family | 5 | 135983 | 53126 | 0 | 55.4 | 39.5 | 16 | 12.0 | 25.2 | 52 | | | | |
| genus | 6 | 0 | 123176 | 0 | 0.0 | 0.0 | 12 | 0.0 | 0.0 | 72 | 135983 | 247468 | 35.5 | family+genus+species |
| species | 7 | 0 | 71166 | 0 | 0.0 | 0.0 | 5 | 0.0 | 0.0 | 96 | | | | |
| avg/sum | 3.6 | 720825 | 304675 | 0 | 59.9 | 11.8 | 9.7 | 33.6 | 19.1 | 39.7 | | | 70.3 | all but unassigned |
| avg/sum | 3.4 | 780833 | 304675 | 0 | 64.9 | 10.4 | 8.6 | 41.9 | 16.7 | 34.9 | | | 71.9 | all with unassigned |

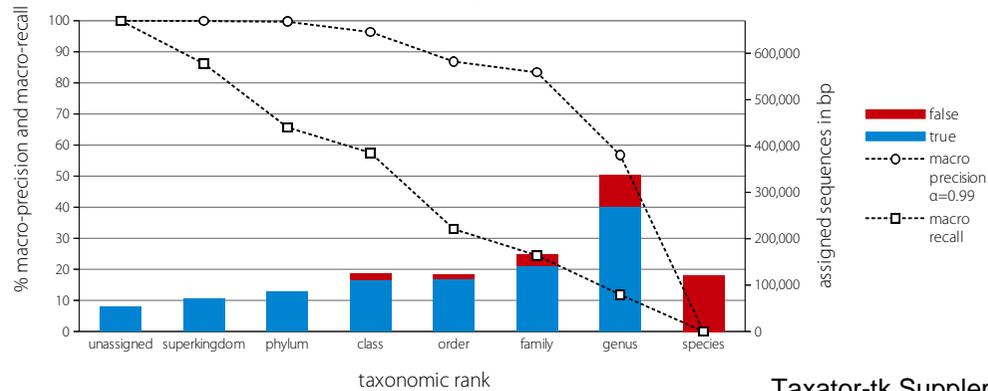

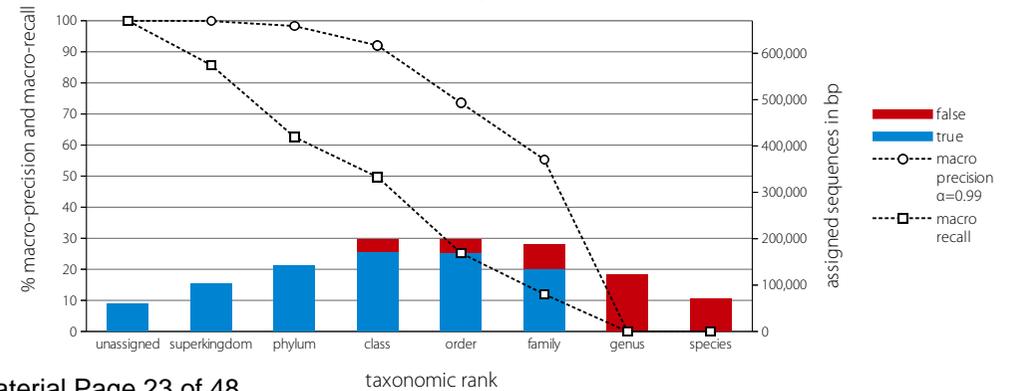

*Supplementary Figure 7: MEGAN binning for FAMeS SimHC* **(e) new family scenario**

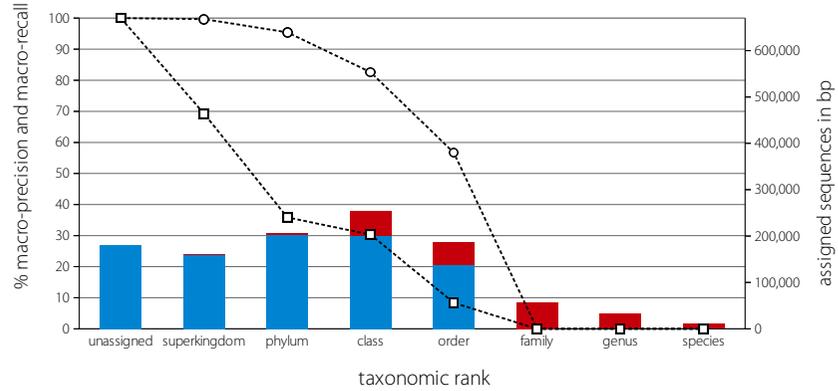

*Supplementary Figure 7: MEGAN binning for FAMeS SimHC* **(f) new order scenario**

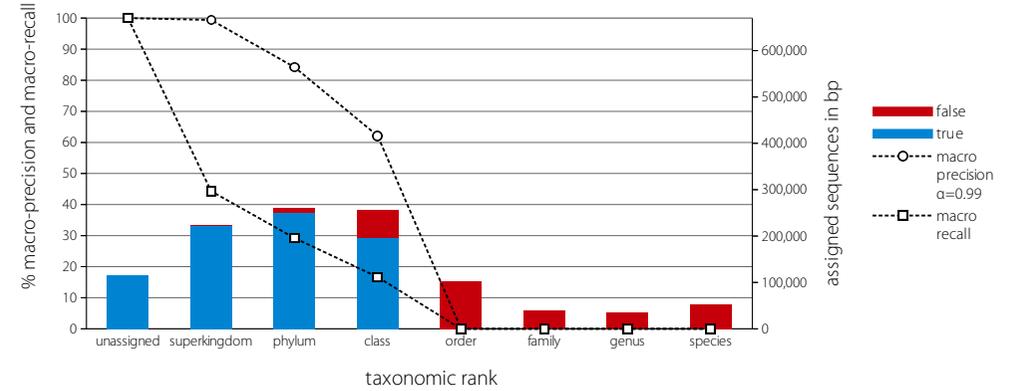

*Supplementary Figure 7: MEGAN binning for FAMeS SimHC* **(g) new class scenario**

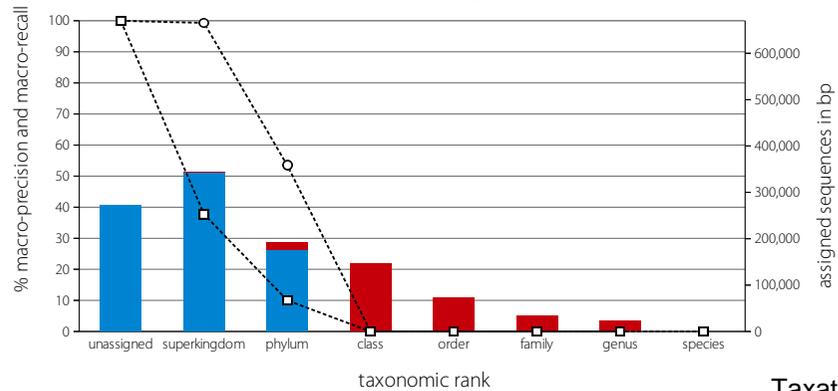

*Supplementary Figure 7: MEGAN binning for FAMeS SimHC* **(h) new phylum scenario**

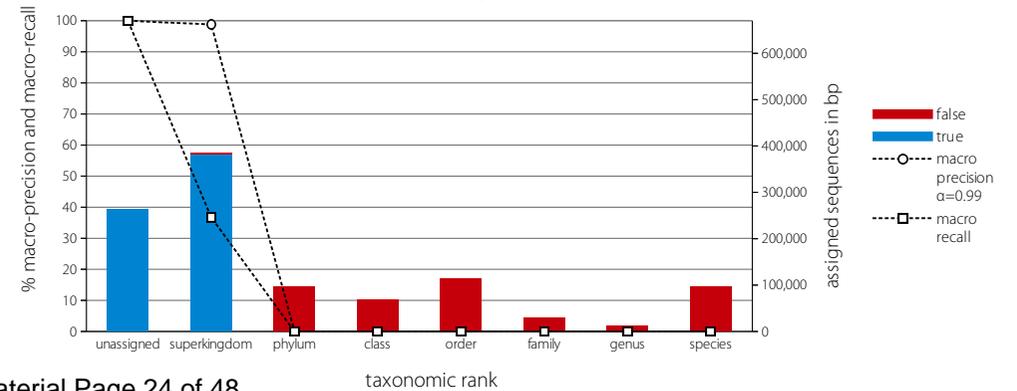

*Supplementary Figure 8: Taxator-tk binning for FAMeS SimHC* **(a) summary scenario**

| rank | depth | true | false | unknown | macro precision α=0.99 | stdev | pred. bins | macro recall | stdev | real bins | sum true | sum false | overall prec. | description |
|---|---|---|---|---|---|---|---|---|---|---|---|---|---|---|
| unassigned | 0 | 213863.4 | 0.0 | 0.0 | 100.0 | 0.0 | 1 | 100.0 | 0.0 | 1 | 993072.9 | 1240.4 | 99.9 | root+superkingdom |
| superkingdom | 1 | 389604.7 | 1240.4 | 0.0 | 99.9 | 0.1 | 2 | 64.1 | 14.9 | 2 | | | | |
| phylum | 2 | 144608.1 | 5934.9 | 0.0 | 96.5 | 5.1 | 7 | 34.9 | 7.9 | 8 | | | | |
| class | 3 | 74893.3 | 12043.4 | 757.0 | 92.7 | 8.4 | 11 | 24.4 | 10.3 | 12 | 259408.7 | 32094.3 | 89.0 | phylum+class+order |
| order | 4 | 39907.3 | 14116.0 | 0.0 | 65.6 | 44.8 | 47 | 15.3 | 10.0 | 36 | | | | |
| family | 5 | 31822.9 | 11110.7 | 0.0 | 68.9 | 43.0 | 58 | 11.7 | 9.7 | 52 | | | | |
| genus | 6 | 59831.9 | 13687.3 | 382.3 | 75.1 | 40.3 | 68 | 8.2 | 7.5 | 72 | 153059.7 | 34715.1 | 81.5 | family+genus+species |
| species | 7 | 61405.0 | 9917.1 | 382.3 | 76.6 | 41.5 | 66 | 4.1 | 4.8 | 96 | | | | |
| avg/sum | 2.3 | 802073.1 | 68049.9 | 1521.6 | 82.2 | 26.2 | 37.0 | 23.2 | 9.3 | 39.7 | | | 92.2 | all but unassigned |
| avg/sum | 1.8 | 1015936.6 | 68049.9 | 1521.6 | 84.4 | 22.9 | 32.5 | 32.8 | 8.1 | 34.9 | | | 93.7 | all with unassigned |

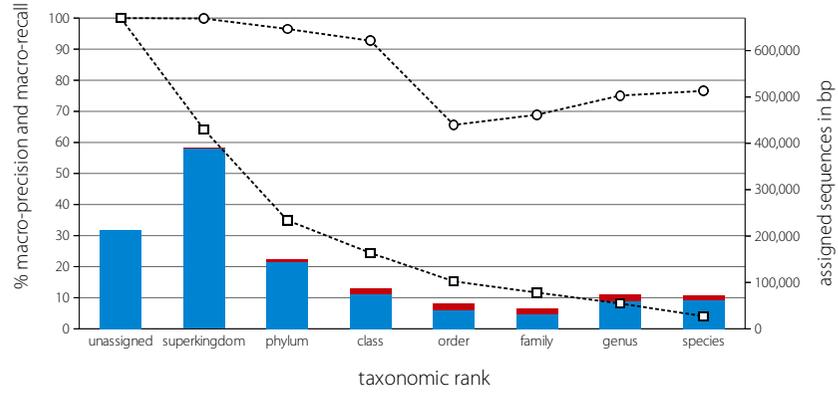

*Supplementary Figure 8: Taxator-tk binning for FAMeS SimHC* **(b) all reference scenario**

| rank | depth | true | false | unknown | macro precision α=0.99 | stdev | pred. bins | macro recall | stdev | real bins | sum true | sum false | overall prec. | description |
|---|---|---|---|---|---|---|---|---|---|---|---|---|---|---|
| unassigned | 0 | 47885 | 0 | 0 | 100.0 | 0.0 | 1 | 100.0 | 0.0 | 1 | 375435 | 0 | 100.0 | root+superkingdom |
| superkingdom | 1 | 163775 | 0 | 0 | 100.0 | 0.0 | 2 | 97.6 | 2.4 | 2 | | | | |
| phylum | 2 | 70870 | 0 | 0 | 99.4 | 1.2 | 6 | 70.3 | 28.5 | 8 | | | | |
| class | 3 | 65908 | 0 | 2139 | 99.5 | 1.1 | 9 | 59.3 | 28.9 | 12 | 185117 | 0 | 100.0 | phylum+class+order |
| order | 4 | 48339 | 0 | 0 | 99.9 | 0.4 | 32 | 55.5 | 30.6 | 36 | | | | |
| family | 5 | 72448 | 0 | 0 | 99.9 | 0.4 | 43 | 49.9 | 34.0 | 52 | | | | |
| genus | 6 | 179113 | 0 | 2676 | 99.9 | 0.3 | 55 | 43.1 | 35.7 | 72 | 681396 | 2520 | 99.6 | family+genus+species |
| species | 7 | 429835 | 2520 | 0 | 100.0 | 0.3 | 52 | 28.5 | 33.5 | 96 | | | | |
| avg/sum | 3.6 | 1030288 | 2520 | 4815 | 99.8 | 0.5 | 28.4 | 57.8 | 27.7 | 39.7 | | | 99.8 | all but unassigned |
| avg/sum | 3.3 | 1078173 | 2520 | 4815 | 99.8 | 0.5 | 25.0 | 63.0 | 24.2 | 34.9 | | | 99.8 | all with unassigned |

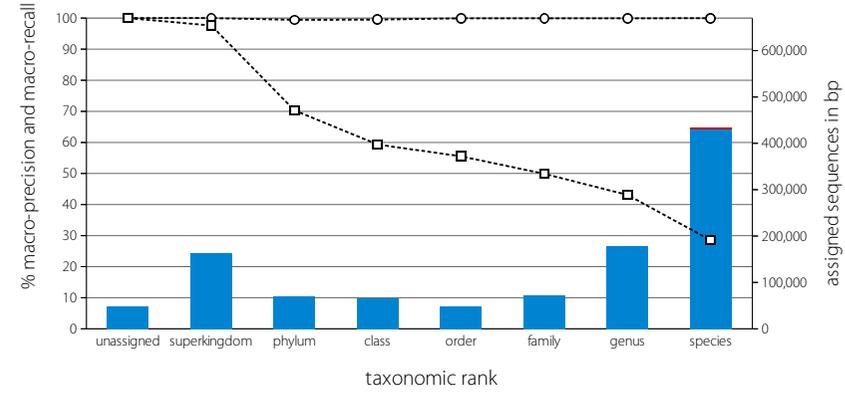

*Supplementary Figure 8: Taxator-tk binning for FAMeS SimHC* **(c) new species scenario**

| rank | depth | true | false | unknown | macro precision α=0.99 | stdev | pred. bins | macro recall | stdev | real bins | sum true | sum false | overall prec. | description |
|---|---|---|---|---|---|---|---|---|---|---|---|---|---|---|
| unassigned | 0 | 92565 | 0 | 0 | 100.0 | 0.0 | 1 | 100.0 | 0.0 | 1 | 608487 | 0 | 100.0 | root+superkingdom |
| superkingdom | 1 | 257961 | 0 | 0 | 100.0 | 0.0 | 2 | 84.1 | 6.4 | 2 | | | | |
| phylum | 2 | 152494 | 0 | 0 | 100.0 | 0.0 | 8 | 62.6 | 26.6 | 8 | | | | |
| class | 3 | 94075 | 0 | 0 | 100.0 | 0.0 | 10 | 48.6 | 28.5 | 12 | 343148 | 6930 | 98.0 | phylum+class+order |
| order | 4 | 96579 | 6930 | 0 | 88.6 | 28.8 | 26 | 30.0 | 30.1 | 36 | | | | |
| family | 5 | 80651 | 10331 | 0 | 86.9 | 30.4 | 30 | 23.7 | 31.8 | 52 | | | | |
| genus | 6 | 239710 | 34526 | 0 | 65.5 | 46.1 | 31 | 13.9 | 26.1 | 72 | 320361 | 64543 | 83.2 | family+genus+species |
| species | 7 | 0 | 19686 | 0 | 0.0 | 0.0 | 9 | 0.0 | 0.0 | 96 | | | | |
| avg/sum | 3.5 | 921470 | 71473 | 0 | 77.3 | 15.0 | 16.4 | 37.6 | 21.3 | 39.7 | | | 92.8 | all but unassigned |
| avg/sum | 3.2 | 1014035 | 71473 | 0 | 80.1 | 13.2 | 14.5 | 45.4 | 18.7 | 34.9 | | | 93.4 | all with unassigned |

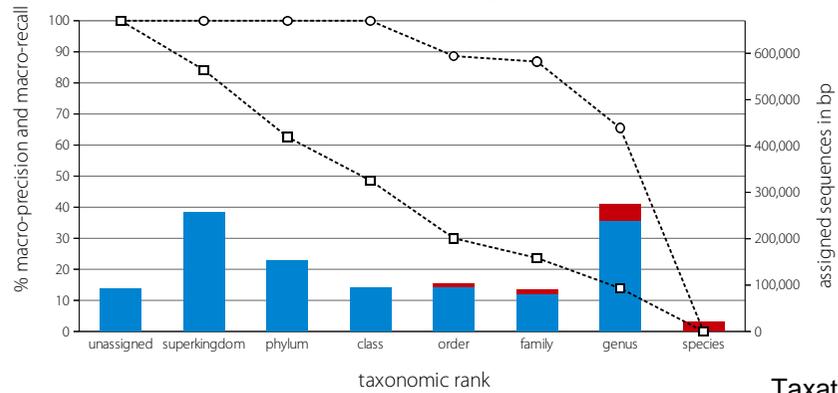

*Supplementary Figure 8: Taxator-tk binning for FAMeS SimHC* **(d) new genus scenario**

| rank | depth | true | false | unknown | macro precision α=0.99 | stdev | pred. bins | macro recall | stdev | real bins | sum true | sum false | overall prec. | description |
|---|---|---|---|---|---|---|---|---|---|---|---|---|---|---|
| unassigned | 0 | 121975 | 0 | 0 | 100.0 | 0.0 | 1 | 100.0 | 0.0 | 1 | 804007 | 0 | 100.0 | root+superkingdom |
| superkingdom | 1 | 341016 | 0 | 0 | 100.0 | 0.0 | 2 | 82.7 | 4.9 | 2 | | | | |
| phylum | 2 | 232318 | 0 | 0 | 100.0 | 0.0 | 7 | 57.5 | 14.0 | 8 | | | | |
| class | 3 | 148392 | 5857 | 0 | 98.6 | 2.6 | 11 | 37.0 | 20.9 | 12 | 468720 | 17552 | 96.4 | phylum+class+order |
| order | 4 | 88010 | 11695 | 0 | 77.2 | 38.5 | 23 | 17.8 | 24.3 | 36 | | | | |
| family | 5 | 69661 | 14936 | 0 | 58.7 | 46.8 | 20 | 8.1 | 19.0 | 52 | | | | |
| genus | 6 | 0 | 39229 | 0 | 0.0 | 0.0 | 14 | 0.0 | 0.0 | 72 | 69661 | 66584 | 51.1 | family+genus+species |
| species | 7 | 0 | 12419 | 0 | 0.0 | 0.0 | 7 | 0.0 | 0.0 | 96 | | | | |
| avg/sum | 2.4 | 879397 | 84136 | 0 | 62.1 | 12.6 | 12.0 | 29.0 | 11.9 | 39.7 | | | 91.3 | all but unassigned |
| avg/sum | 2.2 | 1001372 | 84136 | 0 | 66.8 | 11.0 | 10.6 | 37.9 | 10.4 | 34.9 | | | 92.2 | all with unassigned |

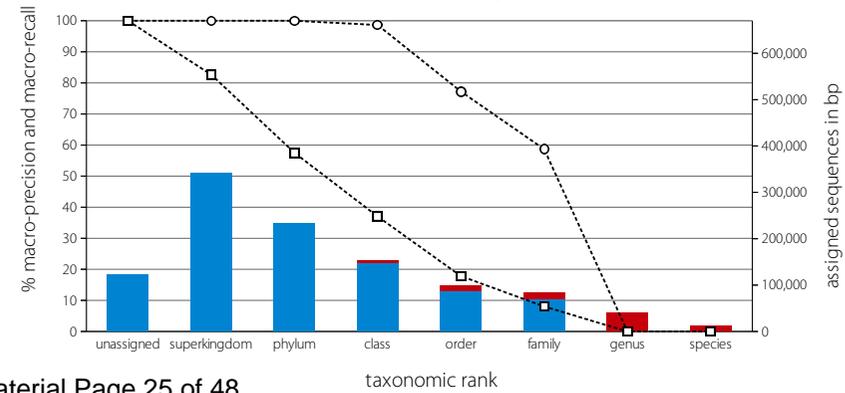

*Supplementary Figure 8: Taxator-tk binning for FAMeS SimHC* **(e) new family scenario**

| rank | depth | true | false | unknown | macro precision α=0.99 | stdev | pred. bins | macro recall | stdev | real bins | sum true | sum false | overall prec. | description |
|---|---|---|---|---|---|---|---|---|---|---|---|---|---|---|
| unassigned | 0 | 243406 | 0 | 0 | 100.0 | 0.0 | 1 | 100.0 | 0.0 | 1 | 1010134 | 1776 | 99.8 | root+superkingdom |
| superkingdom | 1 | 383364 | 1776 | 0 | 99.6 | 0.0 | 1 | 55.2 | 21.8 | 2 | | | | |
| phylum | 2 | 215052 | 1898 | 0 | 98.5 | 1.9 | 7 | 28.4 | 17.2 | 8 | 400621 | 21498 | 94.9 | phylum+class+order |
| class | 3 | 139146 | 11207 | 0 | 92.4 | 11.7 | 10 | 20.6 | 14.9 | 12 | | | | |
| order | 4 | 46423 | 8393 | 0 | 57.8 | 44.5 | 12 | 3.6 | 9.5 | 36 | | | | |
| family | 5 | 0 | 19955 | 0 | 0.0 | 0.0 | 11 | 0.0 | 0.0 | 52 | 0 | 34843 | 0.0 | family+genus+species |
| genus | 6 | 0 | 7389 | 0 | 0.0 | 0.0 | 6 | 0.0 | 0.0 | 72 | | | | |
| species | 7 | 0 | 7499 | 0 | 0.0 | 0.0 | 3 | 0.0 | 0.0 | 96 | | | | |
| avg/sum | 2.0 | 783985 | 58117 | 0 | 49.8 | 8.3 | 7.1 | 15.4 | 9.1 | 39.7 | | | 93.1 | all but unassigned |
| avg/sum | 1.5 | 1027391 | 58117 | 0 | 56.0 | 7.3 | 6.4 | 26.0 | 7.9 | 34.9 | | | 94.6 | all with unassigned |

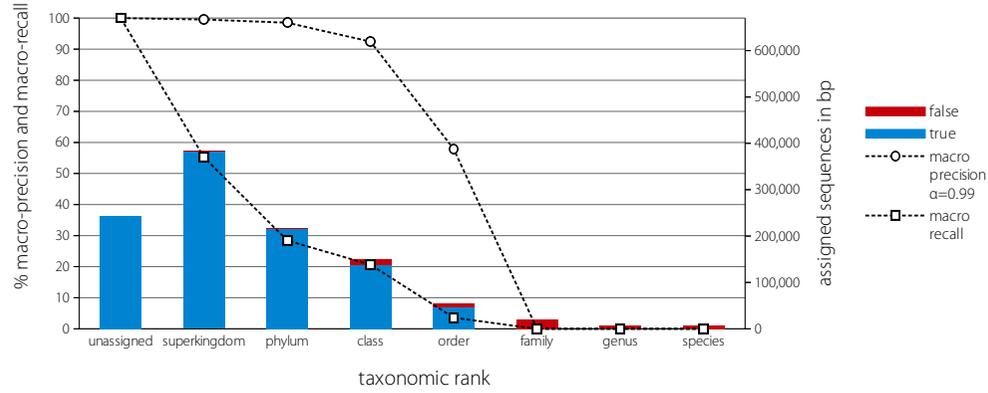

*Supplementary Figure 8: Taxator-tk binning for FAMeS SimHC* **(f) new order scenario**

| rank | depth | true | false | unknown | macro precision α=0.99 | stdev | pred. bins | macro recall | stdev | real bins | sum true | sum false | overall prec. | description |
|---|---|---|---|---|---|---|---|---|---|---|---|---|---|---|
| unassigned | 0 | 197736 | 0 | 0 | 100.0 | 0.0 | 1 | 100.0 | 0.0 | 1 | 1157742 | 1776 | 99.8 | root+superkingdom |
| superkingdom | 1 | 480003 | 1776 | 0 | 99.6 | 0.0 | 1 | 56.7 | 23.4 | 2 | | | | |
| phylum | 2 | 238652 | 3399 | 0 | 91.7 | 12.7 | 5 | 21.9 | 17.7 | 8 | 315384 | 53711 | 85.4 | phylum+class+order |
| class | 3 | 76732 | 11575 | 0 | 70.7 | 25.1 | 7 | 5.3 | 5.9 | 12 | | | | |
| order | 4 | 0 | 38737 | 0 | 0.0 | 0.0 | 13 | 0.0 | 0.0 | 36 | | | | |
| family | 5 | 0 | 16635 | 0 | 0.0 | 0.0 | 9 | 0.0 | 0.0 | 52 | 0 | 36898 | 0.0 | family+genus+species |
| genus | 6 | 0 | 11103 | 0 | 0.0 | 0.0 | 6 | 0.0 | 0.0 | 72 | | | | |
| species | 7 | 0 | 9160 | 0 | 0.0 | 0.0 | 3 | 0.0 | 0.0 | 96 | | | | |
| avg/sum | 1.7 | 795387 | 92385 | 0 | 37.4 | 5.4 | 6.3 | 12.0 | 6.7 | 39.7 | | | 89.6 | all but unassigned |
| avg/sum | 1.4 | 993123 | 92385 | 0 | 45.2 | 4.7 | 5.6 | 23.0 | 5.9 | 34.9 | | | 91.5 | all with unassigned |

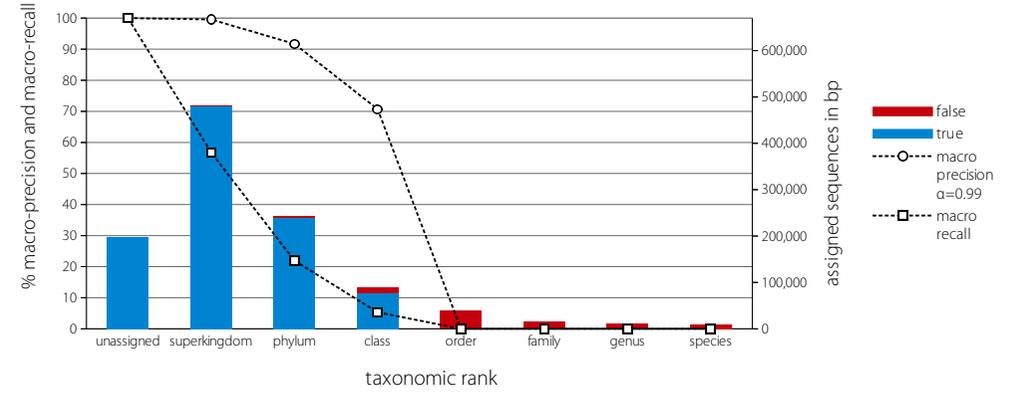

*Supplementary Figure 8: Taxator-tk binning for FAMeS SimHC* **(g) new class scenario**

| rank | depth | true | false | unknown | macro precision α=0.99 | stdev | pred. bins | macro recall | stdev | real bins | sum true | sum false | overall prec. | description |
|---|---|---|---|---|---|---|---|---|---|---|---|---|---|---|
| unassigned | 0 | 366106 | 0 | 0 | 100.0 | 0.0 | 1 | 100.0 | 0.0 | 1 | 1442954 | 1776 | 99.9 | root+superkingdom |
| superkingdom | 1 | 538424 | 1776 | 0 | 99.5 | 0.0 | 1 | 38.5 | 27.4 | 2 | | | | |
| phylum | 2 | 102871 | 9200 | 0 | 42.6 | 42.8 | 4 | 3.8 | 7.3 | 8 | 102871 | 60229 | 63.1 | phylum+class+order |
| class | 3 | 0 | 39497 | 0 | 0.0 | 0.0 | 9 | 0.0 | 0.0 | 12 | | | | |
| order | 4 | 0 | 11532 | 0 | 0.0 | 0.0 | 8 | 0.0 | 0.0 | 36 | | | | |
| family | 5 | 0 | 7485 | 0 | 0.0 | 0.0 | 5 | 0.0 | 0.0 | 52 | 0 | 16102 | 0.0 | family+genus+species |
| genus | 6 | 0 | 3564 | 0 | 0.0 | 0.0 | 2 | 0.0 | 0.0 | 72 | | | | |
| species | 7 | 0 | 5053 | 0 | 0.0 | 0.0 | 2 | 0.0 | 0.0 | 96 | | | | |
| avg/sum | 1.4 | 641295 | 78107 | 0 | 20.3 | 6.1 | 4.6 | 6.0 | 5.0 | 39.7 | | | 89.1 | all but unassigned |
| avg/sum | 0.9 | 1007401 | 78107 | 0 | 30.3 | 5.3 | 4.1 | 17.8 | 4.3 | 34.9 | | | 92.8 | all with unassigned |

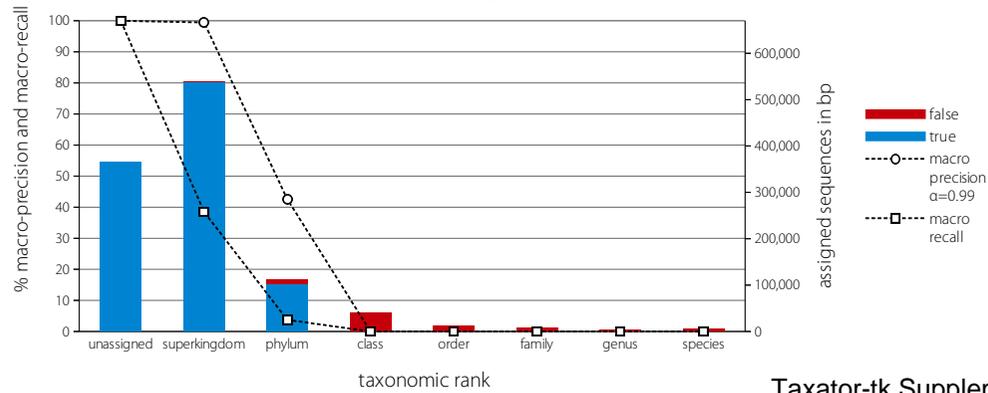

*Supplementary Figure 8: Taxator-tk binning for FAMeS SimHC* **(h) new phylum scenario**

| rank | depth | true | false | unknown | macro precision α=0.99 | stdev | pred. bins | macro recall | stdev | real bins | sum true | sum false | overall prec. | description |
|---|---|---|---|---|---|---|---|---|---|---|---|---|---|---|
| unassigned | 0 | 427371 | 0 | 0 | 100.0 | 0.0 | 1 | 100.0 | 0.0 | 1 | 1552751 | 3355 | 99.8 | root+superkingdom |
| superkingdom | 1 | 562690 | 3355 | 0 | 99.1 | 0.0 | 1 | 34.1 | 23.0 | 2 | | | | |
| phylum | 2 | 0 | 27047 | 0 | 0.0 | 0.0 | 11 | 0.0 | 0.0 | 8 | 0 | 64740 | 0.0 | phylum+class+order |
| class | 3 | 0 | 16168 | 3160 | 0.0 | 0.0 | 11 | 0.0 | 0.0 | 12 | | | | |
| order | 4 | 0 | 21525 | 0 | 0.0 | 0.0 | 10 | 0.0 | 0.0 | 36 | | | | |
| family | 5 | 0 | 8433 | 0 | 0.0 | 0.0 | 7 | 0.0 | 0.0 | 52 | 0 | 21516 | 0.0 | family+genus+species |
| genus | 6 | 0 | 0 | 0 | 0.0 | 0.0 | 3 | 0.0 | 0.0 | 72 | | | | |
| species | 7 | 0 | 13083 | 2676 | 0.0 | 0.0 | 5 | 0.0 | 0.0 | 96 | | | | |
| avg/sum | 1.2 | 562690 | 89611 | 5836 | 14.2 | 0.0 | 6.9 | 4.9 | 3.3 | 39.7 | | | 86.3 | all but unassigned |
| avg/sum | 0.7 | 990061 | 89611 | 5836 | 24.9 | 0.0 | 6.1 | 16.8 | 2.9 | 34.9 | | | 91.7 | all with unassigned |

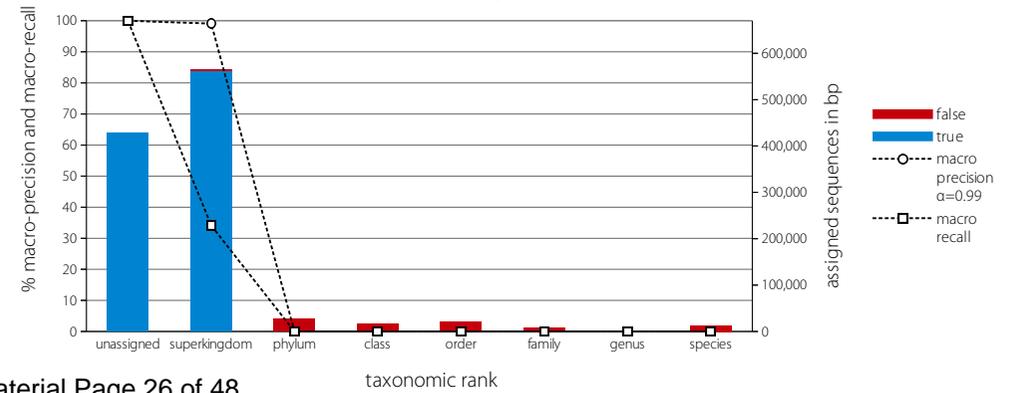

*Supplementary Figure 9: CARMA binning of simulated metagenome with 49 species (simArt49e)* **(a) summary scenario**

| rank | depth | true | false | unknown | macro precision α=0.99 | stdev | pred. bins | macro recall | stdev | real bins | sum true | sum false | overall prec. | description |
|---|---|---|---|---|---|---|---|---|---|---|---|---|---|---|
| unassigned | 0 | 97460.6 | 0.0 | 0 | 100.0 | 0.0 | 1 | 100.0 | 0.0 | 1 | 115012.6 | 500.1 | 99.6 | root+superkingdom |
| superkingdom | 1 | 8776.0 | 500.1 | 0 | 93.6 | 3.0 | 2 | 64.1 | 17.1 | 2 | | | | |
| phylum | 2 | 5011.0 | 7085.1 | 0 | 69.7 | 22.7 | 20 | 36.9 | 15.2 | 20 | 17937.9 | 28325.0 | 38.8 | phylum+class+order |
| class | 3 | 4568.4 | 9153.1 | 0 | 47.0 | 38.5 | 36 | 33.4 | 11.8 | 23 | | | | |
| order | 4 | 8358.4 | 12086.7 | 0 | 31.8 | 39.5 | 78 | 29.2 | 9.5 | 32 | | | | |
| family | 5 | 10303.4 | 12858.1 | 0 | 16.6 | 33.6 | 176 | 24.7 | 7.1 | 36 | 59286.4 | 54892.0 | 51.9 | family+genus+species |
| genus | 6 | 17193.4 | 13745.6 | 0 | 6.7 | 24.1 | 553 | 19.3 | 5.1 | 41 | | | | |
| species | 7 | 31789.6 | 28288.3 | 0 | 2.9 | 16.5 | 1672 | 11.8 | 2.6 | 49 | | | | |
| avg/sum | 4.2 | 86000.3 | 83717.1 | 0 | 38.3 | 25.4 | 362.4 | 31.3 | 9.8 | 29.0 | | | 50.7 | all but unassigned |
| avg/sum | 2.2 | 183460.9 | 83717.1 | 0 | 46.0 | 22.2 | 317.3 | 39.9 | 8.5 | 25.5 | | | 68.7 | all with unassigned |

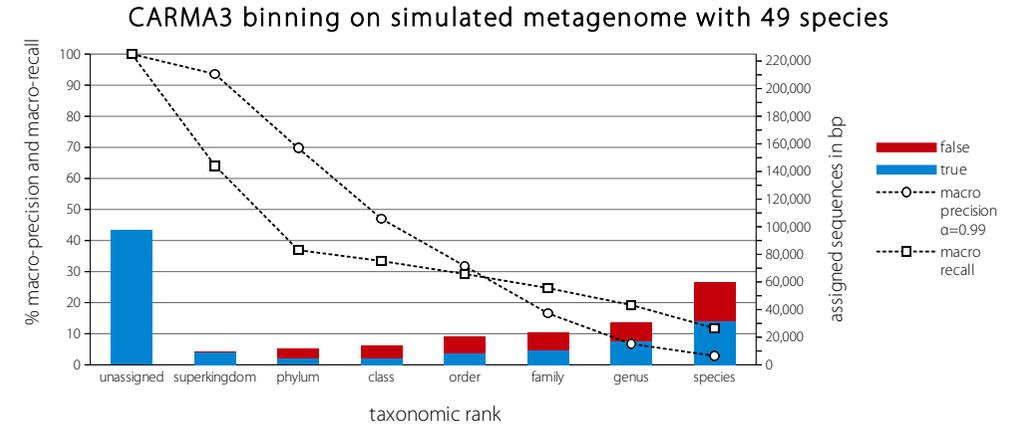

*Supplementary Figure 9: CARMA binning of simulated metagenome with 49 species (simArt49e)* **(b) all reference scenario**

| rank | depth | true | false | unknown | macro precision α=0.99 | stdev | pred. bins | macro recall | stdev | real bins | sum true | sum false | overall prec. | description |
|---|---|---|---|---|---|---|---|---|---|---|---|---|---|---|
| unassigned | 0 | 1071 | 0 | 0 | 100.0 | 0.0 | 1 | 100.0 | 0.0 | 1 | 1403 | 0 | 100.0 | root+superkingdom |
| superkingdom | 1 | 166 | 0 | 0 | 100.0 | 0.0 | 2 | 99.9 | 0.1 | 2 | | | | |
| phylum | 2 | 130 | 1 | 0 | 100.0 | 0.0 | 19 | 99.9 | 0.2 | 20 | 852 | 1 | 99.9 | phylum+class+order |
| class | 3 | 108 | 0 | 0 | 100.0 | 0.0 | 23 | 99.9 | 0.2 | 23 | | | | |
| order | 4 | 614 | 0 | 0 | 100.0 | 0.0 | 31 | 99.9 | 0.2 | 32 | | | | |
| family | 5 | 1000 | 0 | 0 | 100.0 | 0.0 | 35 | 99.9 | 0.2 | 36 | 263301 | 1787 | 99.3 | family+genus+species |
| genus | 6 | 39774 | 30 | 0 | 100.0 | 0.0 | 40 | 99.8 | 0.2 | 41 | | | | |
| species | 7 | 222527 | 1757 | 0 | 99.5 | 2.7 | 46 | 82.6 | 18.3 | 49 | | | | |
| avg/sum | 5.9 | 264319 | 1788 | 0 | 99.9 | 0.4 | 28.0 | 97.4 | 2.8 | 29.0 | | | 99.3 | all but unassigned |
| avg/sum | 5.8 | 265390 | 1788 | 0 | 99.9 | 0.3 | 24.6 | 97.7 | 2.4 | 25.5 | | | 99.3 | all with unassigned |

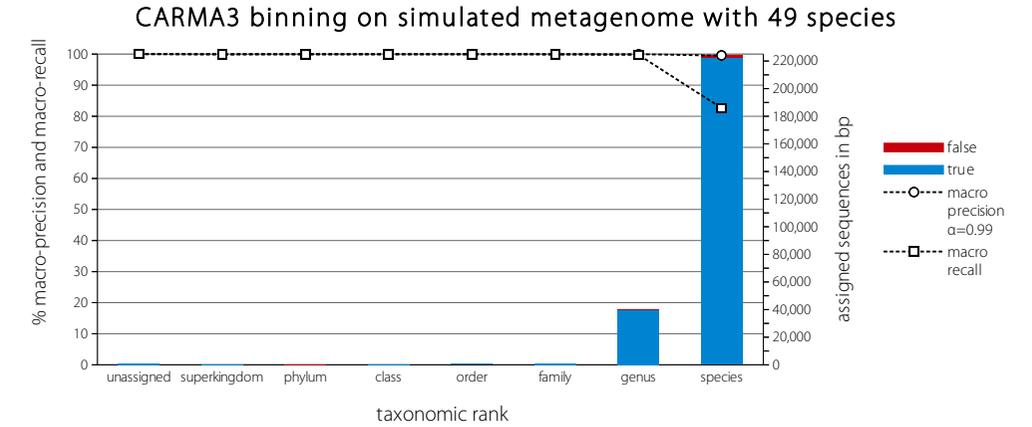

*Supplementary Figure 9: CARMA binning of simulated metagenome with 49 species (simArt49e)* **(c) new species scenario**

| rank | depth | true | false | unknown | macro precision α=0.99 | stdev | pred. bins | macro recall | stdev | real bins | sum true | sum false | overall prec. | description |
|---|---|---|---|---|---|---|---|---|---|---|---|---|---|---|
| unassigned | 0 | 48039 | 0 | 0 | 100.0 | 0.0 | 1 | 100.0 | 0.0 | 1 | 55507 | 174 | 99.7 | root+superkingdom |
| superkingdom | 1 | 3734 | 174 | 0 | 98.9 | 0.2 | 2 | 82.3 | 8.3 | 2 | | | | |
| phylum | 2 | 5195 | 1718 | 0 | 93.8 | 5.8 | 17 | 65.6 | 31.8 | 20 | 29216 | 7007 | 80.7 | phylum+class+order |
| class | 3 | 6657 | 2267 | 0 | 81.4 | 31.7 | 24 | 63.8 | 28.3 | 23 | | | | |
| order | 4 | 17364 | 3022 | 0 | 56.4 | 46.1 | 48 | 59.4 | 32.1 | 32 | | | | |
| family | 5 | 43855 | 4055 | 0 | 32.0 | 44.7 | 96 | 53.0 | 33.4 | 36 | 124435 | 54573 | 69.5 | family+genus+species |
| genus | 6 | 80580 | 10693 | 0 | 12.5 | 32.0 | 216 | 35.0 | 35.5 | 41 | | | | |
| species | 7 | 0 | 39825 | 0 | 0.0 | 0.0 | 1153 | 0.0 | 0.0 | 49 | | | | |
| avg/sum | 5.1 | 157385 | 61754 | 0 | 53.6 | 22.9 | 222.3 | 51.3 | 24.2 | 29.0 | | | 71.8 | all but unassigned |
| avg/sum | 4.0 | 205424 | 61754 | 0 | 59.4 | 20.0 | 194.6 | 57.4 | 21.2 | 25.5 | | | 76.9 | all with unassigned |

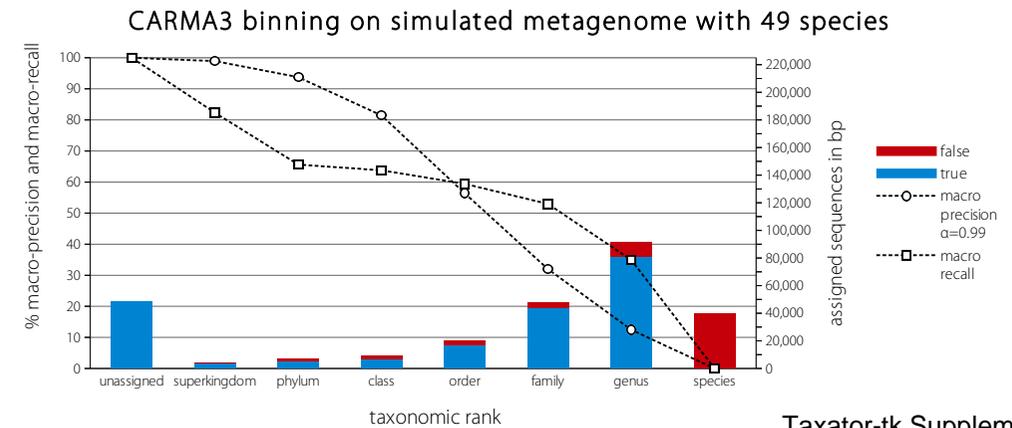

*Supplementary Figure 9: CARMA binning of simulated metagenome with 49 species (simArt49e)* **(d) new genus scenario**

| rank | depth | true | false | unknown | macro precision α=0.99 | stdev | pred. bins | macro recall | stdev | real bins | sum true | sum false | overall prec. | description |
|---|---|---|---|---|---|---|---|---|---|---|---|---|---|---|
| unassigned | 0 | 101939 | 0 | 0 | 100.0 | 0.0 | 1 | 100.0 | 0.0 | 1 | 116715 | 386 | 99.7 | root+superkingdom |
| superkingdom | 1 | 7388 | 386 | 0 | 96.1 | 1.9 | 2 | 67.2 | 14.9 | 2 | | | | |
| phylum | 2 | 7629 | 4042 | 0 | 78.4 | 17.6 | 17 | 40.7 | 29.4 | 20 | 36454 | 18595 | 66.2 | phylum+class+order |
| class | 3 | 8751 | 5633 | 0 | 53.1 | 39.5 | 31 | 38.3 | 26.6 | 23 | | | | |
| order | 4 | 20074 | 8920 | 0 | 33.8 | 39.9 | 65 | 32.7 | 27.5 | 32 | | | | |
| family | 5 | 27269 | 13450 | 0 | 12.5 | 29.1 | 156 | 20.1 | 23.1 | 36 | 27269 | 75147 | 26.6 | family+genus+species |
| genus | 6 | 0 | 33904 | 0 | 0.0 | 0.0 | 535 | 0.0 | 0.0 | 41 | | | | |
| species | 7 | 0 | 27793 | 0 | 0.0 | 0.0 | 1788 | 0.0 | 0.0 | 49 | | | | |
| avg/sum | 4.3 | 71111 | 94128 | 0 | 39.1 | 18.3 | 370.6 | 28.4 | 17.4 | 29.0 | | | 43.0 | all but unassigned |
| avg/sum | 2.5 | 173050 | 94128 | 0 | 46.7 | 16.0 | 324.4 | 37.4 | 15.2 | 25.5 | | | 64.8 | all with unassigned |

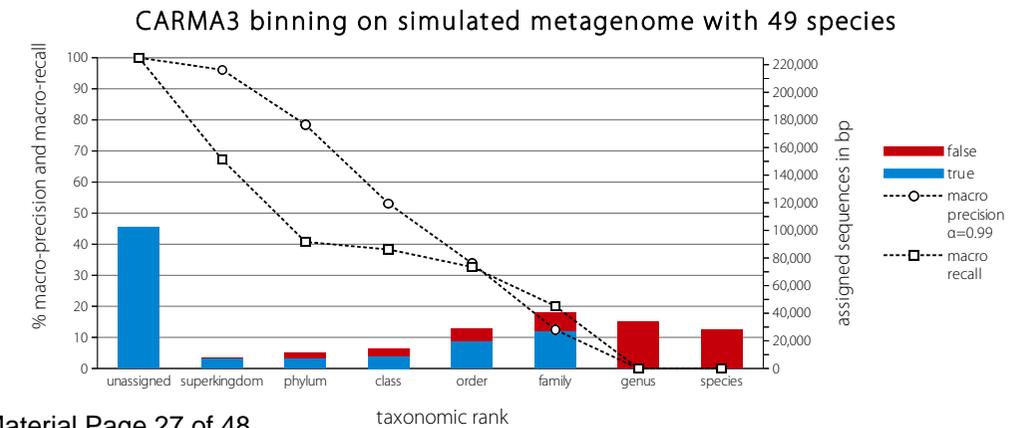

*Supplementary Figure 9: CARMA binning of simulated metagenome with 49 species (simArt49e)* **(e) new family scenario**

| rank | depth | true | false | unknown | macro precision α=0.99 | stdev | pred. bins | macro recall | stdev | real bins | sum true | sum false | overall prec. | description |
|---|---|---|---|---|---|---|---|---|---|---|---|---|---|---|
| unassigned | 0 | 114225 | 0 | 0 | 100.0 | 0.0 | 1 | 100.0 | 0.0 | 1 | 136949 | 536 | 99.6 | root+superkingdom |
| superkingdom | 1 | 11362 | 536 | 0 | 92.1 | 3.8 | 2 | 57.6 | 21.4 | 2 | | | | |
| phylum | 2 | 9527 | 6860 | 0 | 48.3 | 32.1 | 18 | 26.3 | 27.3 | 20 | | | | |
| class | 3 | 9232 | 8904 | 0 | 23.8 | 32.6 | 36 | 21.8 | 25.3 | 23 | 39216 | 29057 | 57.4 | phylum+class+order |
| order | 4 | 20457 | 13293 | 0 | 9.7 | 22.8 | 81 | 12.7 | 19.9 | 32 | | | | |
| family | 5 | 0 | 24317 | 0 | 0.0 | 0.0 | 196 | 0.0 | 0.0 | 36 | | | | |
| genus | 6 | 0 | 18709 | 0 | 0.0 | 0.0 | 625 | 0.0 | 0.0 | 41 | 0 | 72782 | 0.0 | family+genus+species |
| species | 7 | 0 | 29756 | 0 | 0.0 | 0.0 | 1816 | 0.0 | 0.0 | 49 | | | | |
| avg/sum | 3.8 | 50578 | 102375 | 0 | 24.8 | 13.0 | 396.3 | 16.9 | 13.4 | 29.0 | | | 33.1 | all but unassigned |
| avg/sum | 2.0 | 164803 | 102375 | 0 | 34.2 | 11.4 | 346.9 | 27.3 | 11.7 | 25.5 | | | 61.7 | all with unassigned |

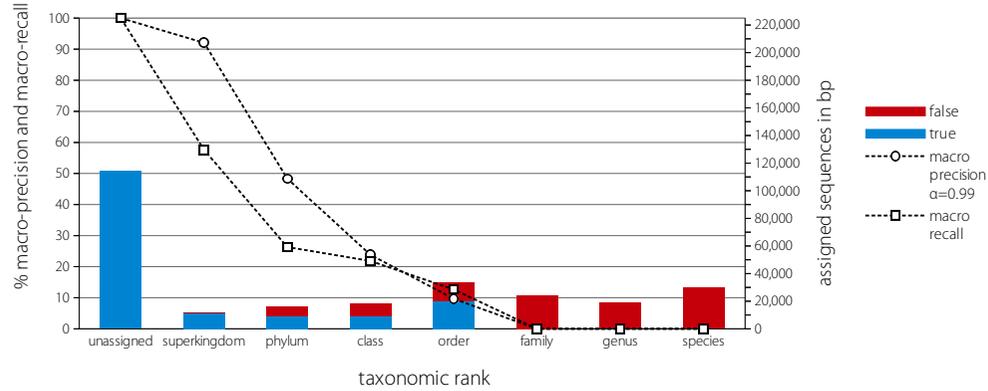

*Supplementary Figure 9: CARMA binning of simulated metagenome with 49 species (simArt49e)* **(f) new order scenario**

| rank | depth | true | false | unknown | macro precision α=0.99 | stdev | pred. bins | macro recall | stdev | real bins | sum true | sum false | overall prec. | description |
|---|---|---|---|---|---|---|---|---|---|---|---|---|---|---|
| unassigned | 0 | 130123 | 0 | 0 | 100.0 | 0.0 | 1 | 100.0 | 0.0 | 1 | 154807 | 706 | 99.5 | root+superkingdom |
| superkingdom | 1 | 12342 | 706 | 0 | 87.7 | 7.2 | 2 | 52.3 | 23.4 | 2 | | | | |
| phylum | 2 | 8019 | 10411 | 0 | 27.9 | 30.5 | 21 | 17.0 | 23.2 | 20 | | | | |
| class | 3 | 7231 | 14435 | 0 | 10.0 | 21.5 | 39 | 10.1 | 16.8 | 23 | 15250 | 45492 | 25.1 | phylum+class+order |
| order | 4 | 0 | 20646 | 0 | 0.0 | 0.0 | 90 | 0.0 | 0.0 | 32 | | | | |
| family | 5 | 0 | 18233 | 0 | 0.0 | 0.0 | 203 | 0.0 | 0.0 | 36 | | | | |
| genus | 6 | 0 | 12779 | 0 | 0.0 | 0.0 | 652 | 0.0 | 0.0 | 41 | 0 | 63265 | 0.0 | family+genus+species |
| species | 7 | 0 | 32253 | 0 | 0.0 | 0.0 | 1810 | 0.0 | 0.0 | 49 | | | | |
| avg/sum | 3.5 | 27592 | 109463 | 0 | 17.9 | 8.5 | 402.4 | 11.4 | 9.0 | 29.0 | | | 20.1 | all but unassigned |
| avg/sum | 1.6 | 157715 | 109463 | 0 | 28.2 | 7.4 | 352.3 | 22.4 | 7.9 | 25.5 | | | 59.0 | all with unassigned |

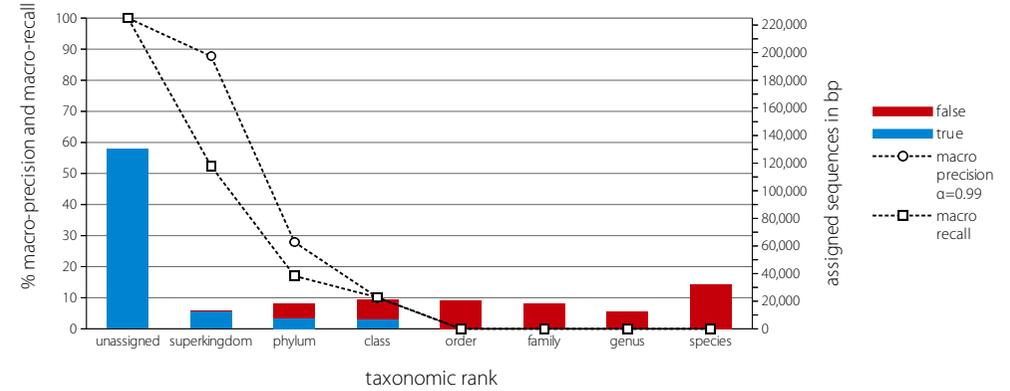

*Supplementary Figure 9: CARMA binning of simulated metagenome with 49 species (simArt49e)* **(g) new class scenario**

| rank | depth | true | false | unknown | macro precision α=0.99 | stdev | pred. bins | macro recall | stdev | real bins | sum true | sum false | overall prec. | description |
|---|---|---|---|---|---|---|---|---|---|---|---|---|---|---|
| unassigned | 0 | 139175 | 0 | 0 | 100.0 | 0.0 | 1 | 100.0 | 0.0 | 1 | 165277 | 778 | 99.5 | root+superkingdom |
| superkingdom | 1 | 13051 | 778 | 0 | 84.3 | 9.8 | 2 | 48.2 | 24.2 | 2 | | | | |
| phylum | 2 | 4577 | 12085 | 0 | 12.9 | 20.1 | 22 | 8.9 | 15.4 | 20 | | | | |
| class | 3 | 0 | 18605 | 0 | 0.0 | 0.0 | 41 | 0.0 | 0.0 | 23 | 4577 | 50298 | 8.3 | phylum+class+order |
| order | 4 | 0 | 19608 | 0 | 0.0 | 0.0 | 91 | 0.0 | 0.0 | 32 | | | | |
| family | 5 | 0 | 15771 | 0 | 0.0 | 0.0 | 206 | 0.0 | 0.0 | 36 | | | | |
| genus | 6 | 0 | 10539 | 0 | 0.0 | 0.0 | 657 | 0.0 | 0.0 | 41 | 0 | 59299 | 0.0 | family+genus+species |
| species | 7 | 0 | 32989 | 0 | 0.0 | 0.0 | 1814 | 0.0 | 0.0 | 49 | | | | |
| avg/sum | 3.4 | 17628 | 110375 | 0 | 13.9 | 4.3 | 404.7 | 8.2 | 5.7 | 29.0 | | | 13.8 | all but unassigned |
| avg/sum | 1.4 | 156803 | 110375 | 0 | 24.6 | 3.7 | 354.3 | 19.6 | 5.0 | 25.5 | | | 58.7 | all with unassigned |

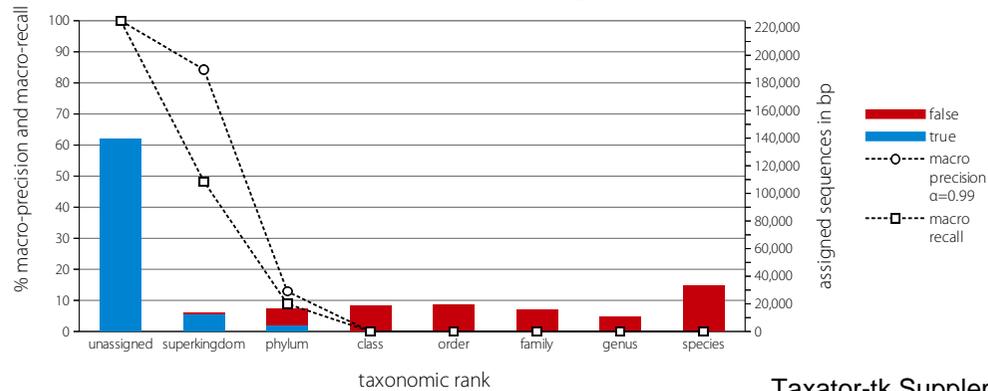

*Supplementary Figure 9: CARMA binning of simulated metagenome with 49 species (simArt49e)* **(h) new phylum scenario**

| rank | depth | true | false | unknown | macro precision α=0.99 | stdev | pred. bins | macro recall | stdev | real bins | sum true | sum false | overall prec. | description |
|---|---|---|---|---|---|---|---|---|---|---|---|---|---|---|
| unassigned | 0 | 147652 | 0 | 0 | 100.0 | 0.0 | 1 | 100.0 | 0.0 | 1 | 174430 | 921 | 99.5 | root+superkingdom |
| superkingdom | 1 | 13389 | 921 | 0 | 75.1 | 17.3 | 2 | 41.1 | 27.2 | 2 | | | | |
| phylum | 2 | 0 | 14479 | 0 | 0.0 | 0.0 | 24 | 0.0 | 0.0 | 20 | | | | |
| class | 3 | 0 | 14228 | 0 | 0.0 | 0.0 | 42 | 0.0 | 0.0 | 23 | 0 | 47825 | 0.0 | phylum+class+order |
| order | 4 | 0 | 19118 | 0 | 0.0 | 0.0 | 93 | 0.0 | 0.0 | 32 | | | | |
| family | 5 | 0 | 14181 | 0 | 0.0 | 0.0 | 214 | 0.0 | 0.0 | 36 | | | | |
| genus | 6 | 0 | 9565 | 0 | 0.0 | 0.0 | 664 | 0.0 | 0.0 | 41 | 0 | 57391 | 0.0 | family+genus+species |
| species | 7 | 0 | 33645 | 0 | 0.0 | 0.0 | 1820 | 0.0 | 0.0 | 49 | | | | |
| avg/sum | 3.4 | 13389 | 106137 | 0 | 10.7 | 2.5 | 408.4 | 5.9 | 3.9 | 29.0 | | | 11.2 | all but unassigned |
| avg/sum | 1.2 | 161041 | 106137 | 0 | 21.9 | 2.2 | 357.5 | 17.6 | 3.4 | 25.5 | | | 60.3 | all with unassigned |

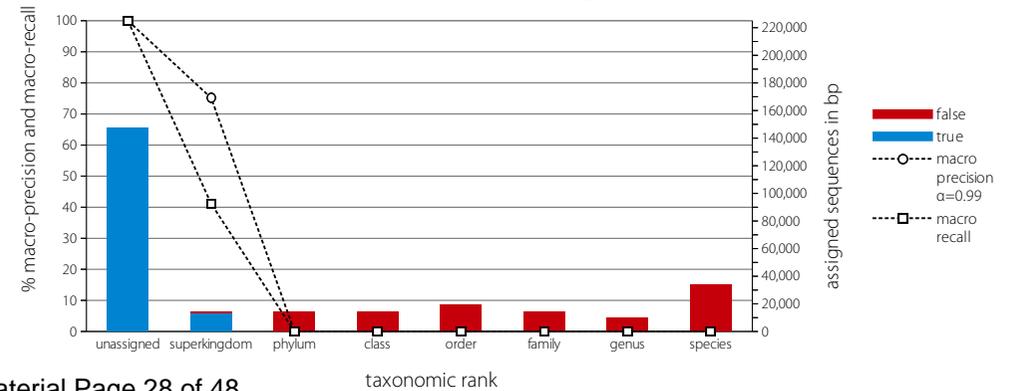

*Supplementary Figure 10: MEGAN binning of simulated metagenome with 49 species (simArt49e)* **(a) summary scenario**

| rank | depth | true | false | unknown | macro precision α=0.99 | stdev | pred. bins | macro recall | stdev | real bins | sum true | sum false | overall prec. | description |
|---|---|---|---|---|---|---|---|---|---|---|---|---|---|---|
| unassigned | 0 | 62255.4 | 0.0 | 0 | 100.0 | 0.0 | 1 | 100.0 | 0.0 | 1 | 232794.3 | 8388.1 | 96.5 | root+superkingdom |
| superkingdom | 1 | 85269.4 | 8388.1 | 0 | 97.8 | 0.1 | 2 | 64.8 | 21.3 | 2 | | | | |
| phylum | 2 | 5415.0 | 3937.6 | 0 | 89.4 | 8.3 | 19 | 35.9 | 14.3 | 20 | 14756.4 | 7006.7 | 67.8 | phylum+class+order |
| class | 3 | 3302.9 | 1523.4 | 0 | 61.7 | 41.3 | 33 | 34.3 | 9.9 | 23 | | | | |
| order | 4 | 6038.6 | 1545.7 | 0 | 43.3 | 45.1 | 66 | 32.3 | 9.3 | 32 | | | | |
| family | 5 | 6638.4 | 2415.9 | 0 | 22.4 | 38.7 | 139 | 28.0 | 8.0 | 36 | 53023.9 | 36478.0 | 59.2 | family+genus+species |
| genus | 6 | 18552.9 | 6525.4 | 0 | 9.3 | 27.9 | 400 | 21.2 | 5.9 | 41 | | | | |
| species | 7 | 27832.6 | 27536.7 | 0 | 5.4 | 21.9 | 824 | 11.8 | 4.5 | 49 | | | | |
| avg/sum | 2.4 | 153049.7 | 51872.9 | 0 | 47.0 | 26.2 | 211.9 | 32.6 | 10.4 | 29.0 | | | 74.7 | all but unassigned |
| avg/sum | 1.7 | 215305.1 | 51872.9 | 0 | 53.7 | 22.9 | 185.5 | 41.0 | 9.1 | 25.5 | | | 80.6 | all with unassigned |

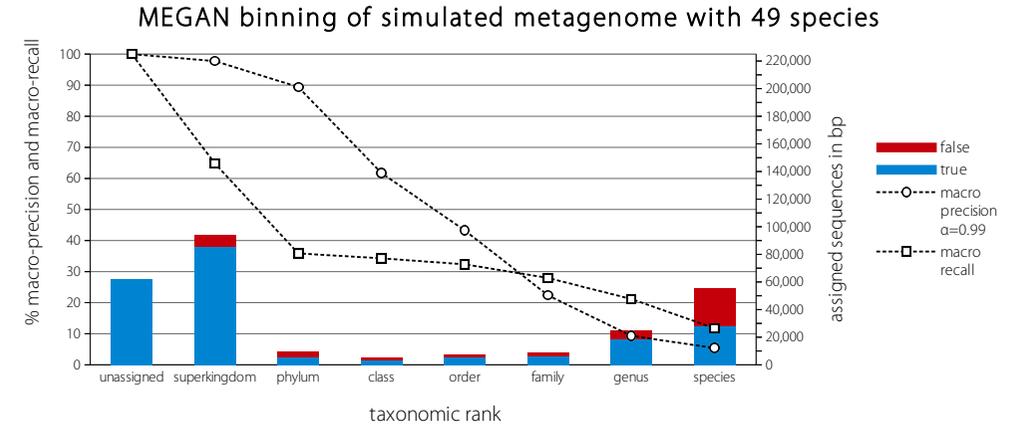

MEGAN binning of simulated metagenome with 49 species

*Supplementary Figure 10: MEGAN binning of simulated metagenome with 49 species (simArt49e)* **(b) all reference scenario**

| rank | depth | true | false | unknown | macro precision α=0.99 | stdev | pred. bins | macro recall | stdev | real bins | sum true | sum false | overall prec. | description |
|---|---|---|---|---|---|---|---|---|---|---|---|---|---|---|
| unassigned | 0 | 595 | 0 | 0 | 100.0 | 0.0 | 1 | 100.0 | 0.0 | 1 | 2301 | 0 | 100.0 | root+superkingdom |
| superkingdom | 1 | 853 | 0 | 0 | 100.0 | 0.0 | 2 | 99.9 | 0.1 | 2 | | | | |
| phylum | 2 | 515 | 1 | 0 | 100.0 | 0.0 | 19 | 99.9 | 0.1 | 20 | 2948 | 1 | 100.0 | phylum+class+order |
| class | 3 | 399 | 0 | 0 | 100.0 | 0.0 | 23 | 99.9 | 0.1 | 23 | | | | |
| order | 4 | 2034 | 0 | 0 | 100.0 | 0.0 | 31 | 99.9 | 0.1 | 32 | | | | |
| family | 5 | 5388 | 0 | 0 | 100.0 | 0.0 | 35 | 99.8 | 0.3 | 36 | 262771 | 10 | 100.0 | family+genus+species |
| genus | 6 | 62555 | 0 | 0 | 100.0 | 0.0 | 40 | 99.3 | 1.6 | 41 | | | | |
| species | 7 | 194828 | 10 | 0 | 100.0 | 0.0 | 44 | 82.5 | 31.3 | 49 | | | | |
| avg/sum | 5.8 | 266572 | 11 | 0 | 100.0 | 0.0 | 27.7 | 97.3 | 4.8 | 29.0 | | | 100.0 | all but unassigned |
| avg/sum | 5.7 | 267167 | 11 | 0 | 100.0 | 0.0 | 24.4 | 97.7 | 4.2 | 25.5 | | | 100.0 | all with unassigned |

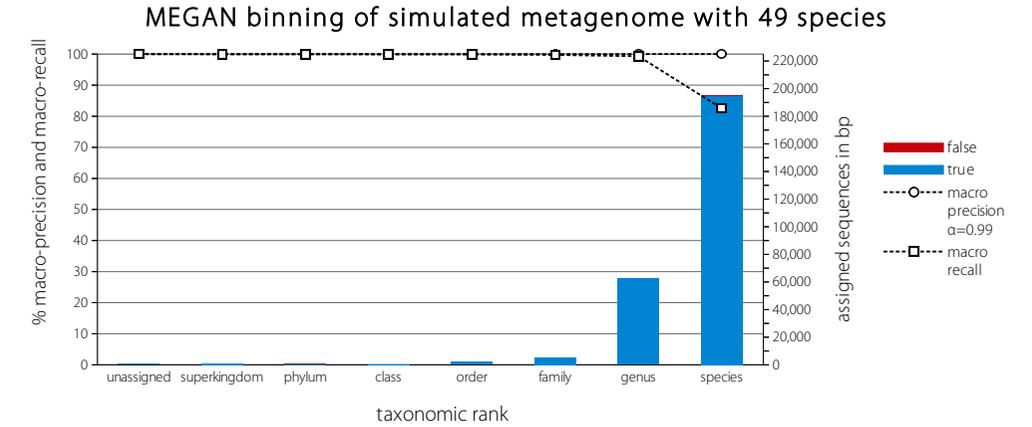

MEGAN binning of simulated metagenome with 49 species

*Supplementary Figure 10: MEGAN binning of simulated metagenome with 49 species (simArt49e)* **(c) new species scenario**

| rank | depth | true | false | unknown | macro precision α=0.99 | stdev | pred. bins | macro recall | stdev | real bins | sum true | sum false | overall prec. | description |
|---|---|---|---|---|---|---|---|---|---|---|---|---|---|---|
| unassigned | 0 | 22828 | 0 | 0 | 100.0 | 0.0 | 1 | 100.0 | 0.0 | 1 | 91304 | 3128 | 96.7 | root+superkingdom |
| superkingdom | 1 | 34238 | 3128 | 0 | 99.6 | 0.1 | 2 | 87.1 | 9.1 | 2 | | | | |
| phylum | 2 | 3671 | 867 | 0 | 98.4 | 1.9 | 17 | 67.3 | 33.8 | 20 | 15968 | 1537 | 91.2 | phylum+class+order |
| class | 3 | 3130 | 282 | 0 | 98.4 | 1.8 | 21 | 70.8 | 29.0 | 23 | | | | |
| order | 4 | 9167 | 388 | 0 | 83.5 | 34.4 | 35 | 70.5 | 31.9 | 32 | | | | |
| family | 5 | 20053 | 979 | 0 | 58.9 | 47.3 | 55 | 65.3 | 34.3 | 36 | 87368 | 102111 | 46.1 | family+genus+species |
| genus | 6 | 67315 | 3069 | 0 | 22.9 | 41.1 | 121 | 49.0 | 40.8 | 41 | | | | |
| species | 7 | 0 | 98063 | 0 | 0.0 | 0.0 | 218 | 0.0 | 0.0 | 49 | | | | |
| avg/sum | 4.3 | 137574 | 106776 | 0 | 66.0 | 18.1 | 67.0 | 58.6 | 25.6 | 29.0 | | | 56.3 | all but unassigned |
| avg/sum | 3.7 | 160402 | 106776 | 0 | 70.2 | 15.8 | 58.8 | 63.8 | 22.4 | 25.5 | | | 60.0 | all with unassigned |

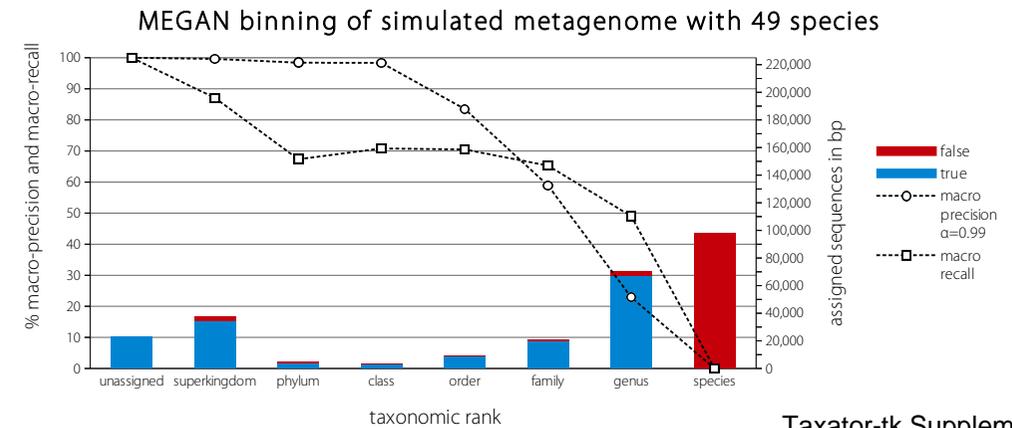

MEGAN binning of simulated metagenome with 49 species

*Supplementary Figure 10: MEGAN binning of simulated metagenome with 49 species (simArt49e)* **(d) new genus scenario**

| rank | depth | true | false | unknown | macro precision α=0.99 | stdev | pred. bins | macro recall | stdev | real bins | sum true | sum false | overall prec. | description |
|---|---|---|---|---|---|---|---|---|---|---|---|---|---|---|
| unassigned | 0 | 58109 | 0 | 0 | 100.0 | 0.0 | 1 | 100.0 | 0.0 | 1 | 238003 | 6636 | 97.3 | root+superkingdom |
| superkingdom | 1 | 89947 | 6636 | 0 | 98.8 | 0.0 | 2 | 74.5 | 15.4 | 2 | | | | |
| phylum | 2 | 8288 | 1861 | 0 | 93.1 | 9.2 | 15 | 40.0 | 29.9 | 20 | 29489 | 3515 | 89.3 | phylum+class+order |
| class | 3 | 6666 | 657 | 0 | 76.6 | 34.7 | 25 | 41.8 | 26.2 | 23 | | | | |
| order | 4 | 14535 | 997 | 0 | 52.0 | 45.1 | 50 | 41.1 | 29.8 | 32 | | | | |
| family | 5 | 21028 | 3013 | 0 | 22.3 | 38.5 | 105 | 30.6 | 30.6 | 36 | 21028 | 58454 | 26.5 | family+genus+species |
| genus | 6 | 0 | 20343 | 0 | 0.0 | 0.0 | 274 | 0.0 | 0.0 | 41 | | | | |
| species | 7 | 0 | 35098 | 0 | 0.0 | 0.0 | 430 | 0.0 | 0.0 | 49 | | | | |
| avg/sum | 2.5 | 140464 | 68605 | 0 | 49.0 | 18.2 | 128.7 | 32.6 | 18.8 | 29.0 | | | 67.2 | all but unassigned |
| avg/sum | 1.9 | 198573 | 68605 | 0 | 55.3 | 15.9 | 112.8 | 41.0 | 16.5 | 25.5 | | | 74.3 | all with unassigned |

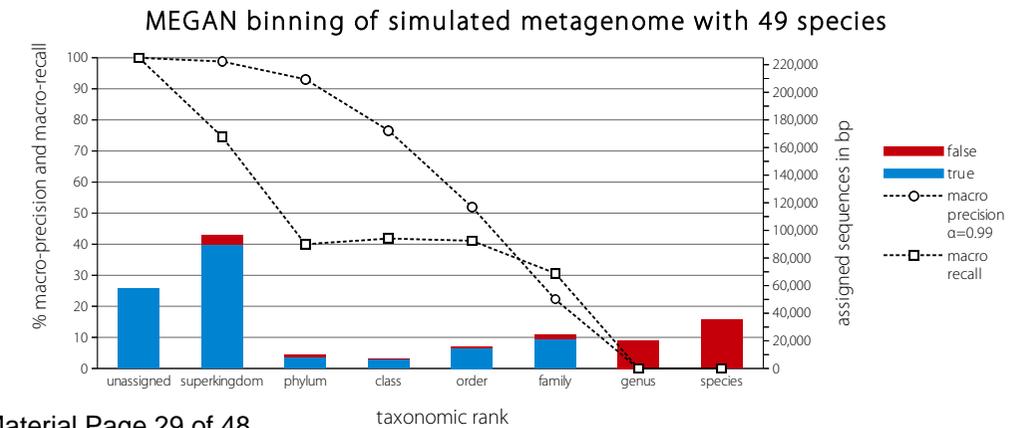

MEGAN binning of simulated metagenome with 49 species

*Supplementary Figure 10: MEGAN binning of simulated metagenome with 49 species (simArt49e)* **(e) new family scenario**

| rank | depth | true | false | unknown | macro precision α=0.99 | stdev | pred. bins | macro recall | stdev | real bins | sum true | sum false | overall prec. | description |
|---|---|---|---|---|---|---|---|---|---|---|---|---|---|---|
| unassigned | 0 | 73809 | 0 | 0 | 100.0 | 0.0 | 1 | 100.0 | 0.0 | 1 | 296855 | 8005 | 97.4 | root+superkingdom |
| superkingdom | 1 | 111523 | 8005 | 0 | 97.2 | 0.4 | 2 | 56.5 | 28.9 | 2 | | | | |
| phylum | 2 | 10028 | 2966 | 0 | 78.2 | 27.3 | 14 | 23.5 | 25.8 | 20 | | | | |
| class | 3 | 7514 | 1043 | 0 | 37.5 | 42.3 | 31 | 20.4 | 21.9 | 23 | 34076 | 5490 | 86.1 | phylum+class+order |
| order | 4 | 16534 | 1481 | 0 | 17.5 | 32.9 | 69 | 14.9 | 21.1 | 32 | | | | |
| family | 5 | 0 | 5906 | 0 | 0.0 | 0.0 | 161 | 0.0 | 0.0 | 36 | | | | |
| genus | 6 | 0 | 8487 | 0 | 0.0 | 0.0 | 366 | 0.0 | 0.0 | 41 | 0 | 34275 | 0.0 | family+genus+species |
| species | 7 | 0 | 19882 | 0 | 0.0 | 0.0 | 580 | 0.0 | 0.0 | 49 | | | | |
| avg/sum | 1.9 | 145599 | 47770 | 0 | 32.9 | 14.7 | 174.7 | 16.5 | 14.0 | 29.0 | | | 75.3 | all but unassigned |
| avg/sum | 1.3 | 219408 | 47770 | 0 | 41.3 | 12.9 | 153.0 | 26.9 | 12.2 | 25.5 | | | 82.1 | all with unassigned |

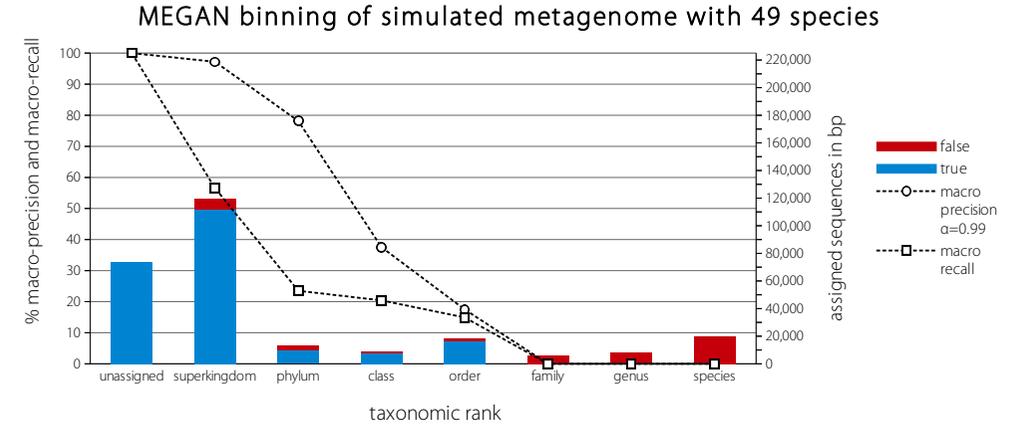

*Supplementary Figure 10: MEGAN binning of simulated metagenome with 49 species (simArt49e)* **(f) new order scenario**

| rank | depth | true | false | unknown | macro precision α=0.99 | stdev | pred. bins | macro recall | stdev | real bins | sum true | sum false | overall prec. | description |
|---|---|---|---|---|---|---|---|---|---|---|---|---|---|---|
| unassigned | 0 | 89682 | 0 | 0 | 100.0 | 0.0 | 1 | 100.0 | 0.0 | 1 | 326876 | 11005 | 96.7 | root+superkingdom |
| superkingdom | 1 | 118597 | 11005 | 0 | 95.5 | 1.3 | 2 | 49.8 | 30.9 | 2 | | | | |
| phylum | 2 | 10011 | 5581 | 0 | 47.4 | 37.8 | 17 | 13.8 | 21.7 | 20 | | | | |
| class | 3 | 5411 | 1881 | 0 | 14.4 | 28.3 | 39 | 7.0 | 11.7 | 23 | 15422 | 9908 | 60.9 | phylum+class+order |
| order | 4 | 0 | 2446 | 0 | 0.0 | 0.0 | 84 | 0.0 | 0.0 | 32 | | | | |
| family | 5 | 0 | 3416 | 0 | 0.0 | 0.0 | 167 | 0.0 | 0.0 | 36 | | | | |
| genus | 6 | 0 | 5382 | 0 | 0.0 | 0.0 | 401 | 0.0 | 0.0 | 41 | 0 | 22564 | 0.0 | family+genus+species |
| species | 7 | 0 | 13766 | 0 | 0.0 | 0.0 | 565 | 0.0 | 0.0 | 49 | | | | |
| avg/sum | 1.5 | 134019 | 43477 | 0 | 22.5 | 9.6 | 182.1 | 10.1 | 9.2 | 29.0 | | | 75.5 | all but unassigned |
| avg/sum | 1.0 | 223701 | 43477 | 0 | 32.2 | 8.4 | 159.5 | 21.3 | 8.0 | 25.5 | | | 83.7 | all with unassigned |

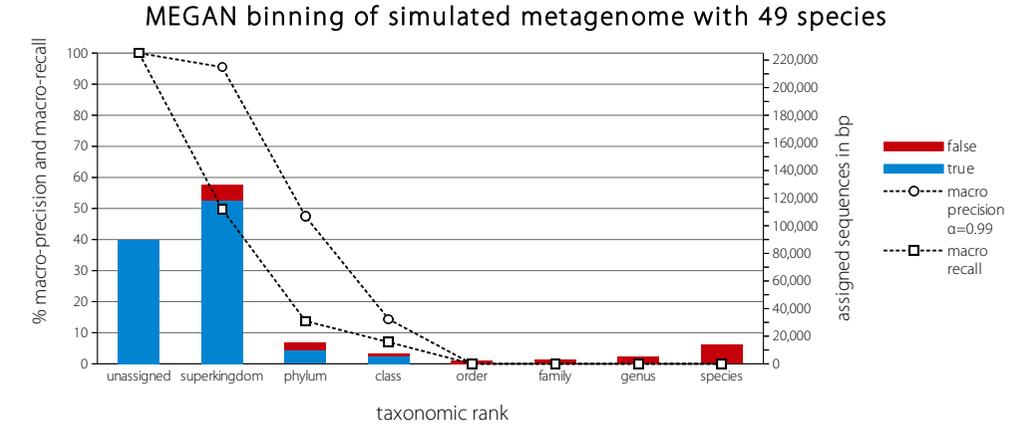

*Supplementary Figure 10: MEGAN binning of simulated metagenome with 49 species (simArt49e)* **(g) new class scenario**

| rank | depth | true | false | unknown | macro precision α=0.99 | stdev | pred. bins | macro recall | stdev | real bins | sum true | sum false | overall prec. | description |
|---|---|---|---|---|---|---|---|---|---|---|---|---|---|---|
| unassigned | 0 | 94817 | 0 | 0 | 100.0 | 0.0 | 1 | 100.0 | 0.0 | 1 | 340147 | 11467 | 96.7 | root+superkingdom |
| superkingdom | 1 | 122665 | 11467 | 0 | 93.2 | 2.8 | 2 | 45.9 | 31.8 | 2 | | | | |
| phylum | 2 | 5392 | 7208 | 0 | 25.1 | 30.7 | 19 | 6.4 | 11.6 | 20 | | | | |
| class | 3 | 0 | 4356 | 0 | 0.0 | 0.0 | 43 | 0.0 | 0.0 | 23 | 5392 | 13709 | 28.2 | phylum+class+order |
| order | 4 | 0 | 2145 | 0 | 0.0 | 0.0 | 88 | 0.0 | 0.0 | 32 | | | | |
| family | 5 | 0 | 2203 | 0 | 0.0 | 0.0 | 172 | 0.0 | 0.0 | 36 | | | | |
| genus | 6 | 0 | 4437 | 0 | 0.0 | 0.0 | 446 | 0.0 | 0.0 | 41 | 0 | 19128 | 0.0 | family+genus+species |
| species | 7 | 0 | 12488 | 0 | 0.0 | 0.0 | 657 | 0.0 | 0.0 | 49 | | | | |
| avg/sum | 1.4 | 128057 | 44304 | 0 | 16.9 | 4.8 | 203.9 | 7.5 | 6.2 | 29.0 | | | 74.3 | all but unassigned |
| avg/sum | 0.9 | 222874 | 44304 | 0 | 27.3 | 4.2 | 178.5 | 19.0 | 5.4 | 25.5 | | | 83.4 | all with unassigned |

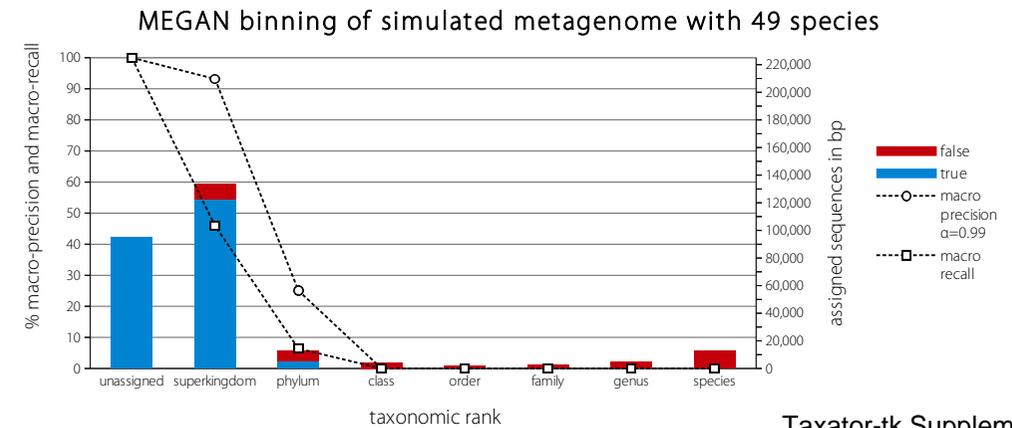

*Supplementary Figure 10: MEGAN binning of simulated metagenome with 49 species (simArt49e)* **(h) new phylum scenario**

| rank | depth | true | false | unknown | macro precision α=0.99 | stdev | pred. bins | macro recall | stdev | real bins | sum true | sum false | overall prec. | description |
|---|---|---|---|---|---|---|---|---|---|---|---|---|---|---|
| unassigned | 0 | 95948 | 0 | 0 | 100.0 | 0.0 | 1 | 100.0 | 0.0 | 1 | 334074 | 18476 | 94.8 | root+superkingdom |
| superkingdom | 1 | 119063 | 18476 | 0 | 84.6 | 8.8 | 2 | 39.5 | 33.1 | 2 | | | | |
| phylum | 2 | 0 | 9079 | 0 | 0.0 | 0.0 | 25 | 0.0 | 0.0 | 20 | | | | |
| class | 3 | 0 | 2445 | 0 | 0.0 | 0.0 | 45 | 0.0 | 0.0 | 23 | 0 | 14887 | 0.0 | phylum+class+order |
| order | 4 | 0 | 3363 | 0 | 0.0 | 0.0 | 96 | 0.0 | 0.0 | 32 | | | | |
| family | 5 | 0 | 1394 | 0 | 0.0 | 0.0 | 197 | 0.0 | 0.0 | 36 | | | | |
| genus | 6 | 0 | 3960 | 0 | 0.0 | 0.0 | 494 | 0.0 | 0.0 | 41 | 0 | 18804 | 0.0 | family+genus+species |
| species | 7 | 0 | 13450 | 0 | 0.0 | 0.0 | 814 | 0.0 | 0.0 | 49 | | | | |
| avg/sum | 1.3 | 119063 | 52167 | 0 | 12.1 | 1.3 | 239.0 | 5.6 | 4.7 | 29.0 | | | 69.5 | all but unassigned |
| avg/sum | 0.8 | 215011 | 52167 | 0 | 23.1 | 1.1 | 209.3 | 17.4 | 4.1 | 25.5 | | | 80.5 | all with unassigned |

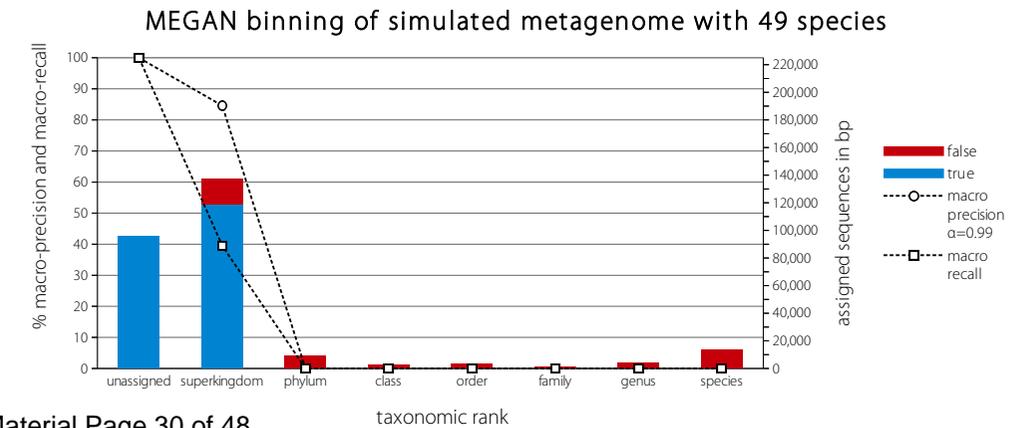

| rank | depth | true | false | unknown | macro precision α=0.99 | stdev | pred. bins | macro recall | stdev | real bins | sum true | sum false | overall prec. | description |
|---|---|---|---|---|---|---|---|---|---|---|---|---|---|---|
| unassigned | 0 | 75644.4 | 0.0 | 0 | 100.0 | 0.0 | 1 | 100.0 | 0.0 | 1 | 288633.9 | 10293.3 | 96.6 | root+superkingdom |
| superkingdom | 1 | 106494.7 | 10293.3 | 0 | 96.9 | 2.5 | 2 | 56.8 | 33.5 | 2 | | | | |
| phylum | 2 | 7691.6 | 9493.1 | 0 | 94.9 | 9.2 | 16 | 18.2 | 13.5 | 20 | 17180.7 | 13851.4 | 55.4 | phylum+class+order |
| class | 3 | 3656.6 | 2344.0 | 0 | 91.2 | 21.5 | 21 | 18.3 | 11.6 | 23 | | | | |
| order | 4 | 5832.6 | 2014.3 | 0 | 85.9 | 31.8 | 34 | 16.2 | 9.5 | 32 | | | | |
| family | 5 | 7550.6 | 1079.7 | 0 | 76.4 | 39.8 | 44 | 13.8 | 8.2 | 36 | 39774.7 | 3938.7 | 91.0 | family+genus+species |
| genus | 6 | 20271.9 | 1397.1 | 0 | 65.9 | 46.4 | 58 | 9.4 | 7.7 | 41 | | | | |
| species | 7 | 11952.3 | 1461.9 | 0 | 61.1 | 47.2 | 65 | 2.5 | 4.4 | 49 | | | | |
| avg/sum | 2.1 | 163450.1 | 28083.4 | 0 | 81.8 | 28.3 | 34.3 | 19.3 | 12.6 | 29.0 | | | 85.3 | all but unassigned |
| avg/sum | 1.5 | 239094.6 | 28083.4 | 0 | 84.0 | 24.8 | 30.1 | 29.4 | 11.0 | 25.5 | | | 89.5 | all with unassigned |

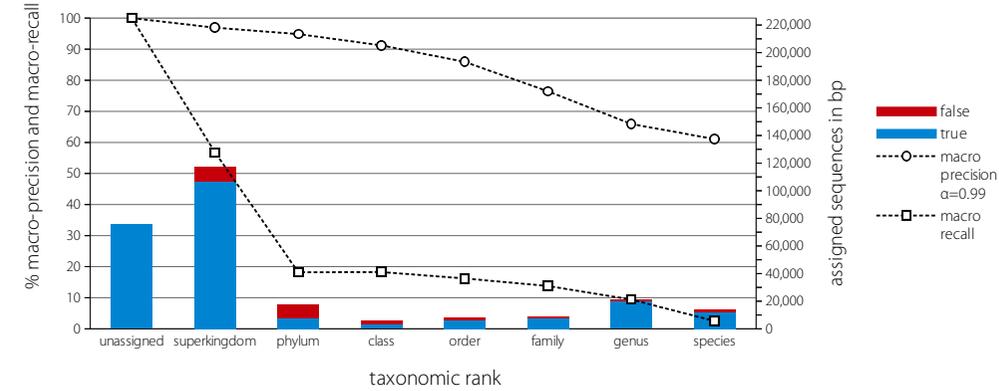

| rank | depth | true | false | unknown | macro precision α=0.99 | stdev | pred. bins | macro recall | stdev | real bins | sum true | sum false | overall prec. | description |
|---|---|---|---|---|---|---|---|---|---|---|---|---|---|---|
| unassigned | 0 | 34453 | 0 | 0 | 100.0 | 0.0 | 1 | 100.0 | 0.0 | 1 | 105775 | 0 | 100.0 | root+superkingdom |
| superkingdom | 1 | 35661 | 0 | 0 | 100.0 | 0.0 | 2 | 68.1 | 8.4 | 2 | | | | |
| phylum | 2 | 2897 | 0 | 0 | 100.0 | 0.0 | 17 | 46.5 | 28.4 | 20 | 13098 | 0 | 100.0 | phylum+class+order |
| class | 3 | 1947 | 0 | 0 | 100.0 | 0.0 | 21 | 52.5 | 29.1 | 23 | | | | |
| order | 4 | 8254 | 0 | 0 | 100.0 | 0.0 | 29 | 51.7 | 30.6 | 32 | | | | |
| family | 5 | 19632 | 0 | 0 | 100.0 | 0.0 | 32 | 52.1 | 31.5 | 36 | 183965 | 1 | 100.0 | family+genus+species |
| genus | 6 | 80667 | 0 | 0 | 100.0 | 0.0 | 34 | 43.4 | 34.0 | 41 | | | | |
| species | 7 | 83666 | 1 | 0 | 100.0 | 0.0 | 34 | 17.7 | 30.5 | 49 | | | | |
| avg/sum | 4.4 | 232724 | 1 | 0 | 100.0 | 0.0 | 24.1 | 47.4 | 27.5 | 29.0 | | | 100.0 | all but unassigned |
| avg/sum | 3.6 | 267177 | 1 | 0 | 100.0 | 0.0 | 21.3 | 54.0 | 24.1 | 25.5 | | | 100.0 | all with unassigned |

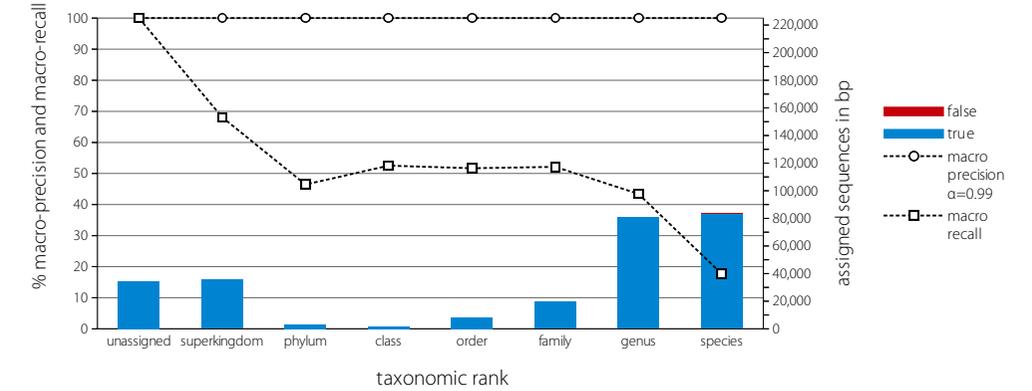

| rank | depth | true | false | unknown | macro precision α=0.99 | stdev | pred. bins | macro recall | stdev | real bins | sum true | sum false | overall prec. | description |
|---|---|---|---|---|---|---|---|---|---|---|---|---|---|---|
| unassigned | 0 | 63523 | 0 | 0 | 100.0 | 0.0 | 1 | 100.0 | 0.0 | 1 | 214561 | 4247 | 98.1 | root+superkingdom |
| superkingdom | 1 | 75519 | 4247 | 0 | 99.1 | 0.9 | 2 | 74.7 | 16.4 | 2 | | | | |
| phylum | 2 | 8526 | 1834 | 0 | 99.4 | 1.4 | 15 | 39.3 | 28.6 | 20 | 29377 | 2671 | 91.7 | phylum+class+order |
| class | 3 | 6516 | 515 | 0 | 99.7 | 0.5 | 18 | 42.8 | 27.4 | 23 | | | | |
| order | 4 | 14335 | 322 | 0 | 99.7 | 0.5 | 25 | 39.9 | 26.9 | 32 | | | | |
| family | 5 | 21470 | 246 | 0 | 99.7 | 0.8 | 28 | 34.9 | 28.6 | 36 | 82706 | 9135 | 90.1 | family+genus+species |
| genus | 6 | 61236 | 1365 | 0 | 88.7 | 30.3 | 26 | 22.6 | 28.3 | 41 | | | | |
| species | 7 | 0 | 7524 | 0 | 0.0 | 0.0 | 48 | 0.0 | 0.0 | 49 | | | | |
| avg/sum | 3.4 | 187602 | 16053 | 0 | 83.8 | 4.9 | 23.1 | 36.3 | 22.3 | 29.0 | | | 92.1 | all but unassigned |
| avg/sum | 2.6 | 251125 | 16053 | 0 | 85.8 | 4.3 | 20.4 | 44.3 | 19.5 | 25.5 | | | 94.0 | all with unassigned |

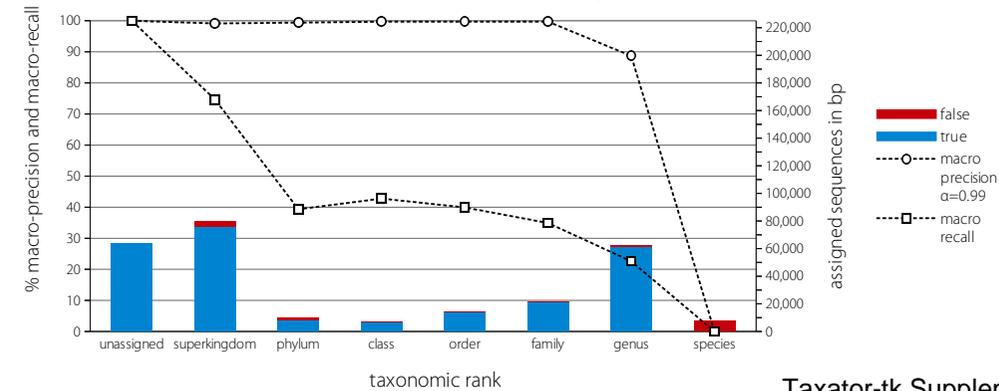

| rank | depth | true | false | unknown | macro precision α=0.99 | stdev | pred. bins | macro recall | stdev | real bins | sum true | sum false | overall prec. | description |
|---|---|---|---|---|---|---|---|---|---|---|---|---|---|---|
| unassigned | 0 | 82737 | 0 | 0 | 100.0 | 0.0 | 1 | 100.0 | 0.0 | 1 | 318017 | 8730 | 97.3 | root+superkingdom |
| superkingdom | 1 | 117640 | 8730 | 0 | 98.1 | 1.4 | 2 | 61.5 | 31.7 | 2 | | | | |
| phylum | 2 | 12404 | 4567 | 0 | 98.2 | 3.3 | 13 | 20.1 | 21.3 | 20 | 31213 | 7247 | 81.2 | phylum+class+order |
| class | 3 | 6555 | 1439 | 0 | 98.5 | 1.9 | 16 | 20.9 | 21.8 | 23 | | | | |
| order | 4 | 12254 | 1241 | 0 | 97.1 | 4.9 | 22 | 17.6 | 20.2 | 32 | | | | |
| family | 5 | 11752 | 1508 | 0 | 56.2 | 46.1 | 33 | 9.5 | 16.7 | 36 | 11752 | 7859 | 59.9 | family+genus+species |
| genus | 6 | 0 | 4633 | 0 | 0.0 | 0.0 | 52 | 0.0 | 0.0 | 41 | | | | |
| species | 7 | 0 | 1718 | 0 | 0.0 | 0.0 | 49 | 0.0 | 0.0 | 49 | | | | |
| avg/sum | 1.8 | 160605 | 23836 | 0 | 64.0 | 8.2 | 26.7 | 18.5 | 16.0 | 29.0 | | | 87.1 | all but unassigned |
| avg/sum | 1.3 | 243342 | 23836 | 0 | 68.5 | 7.2 | 23.5 | 28.7 | 14.0 | 25.5 | | | 91.1 | all with unassigned |

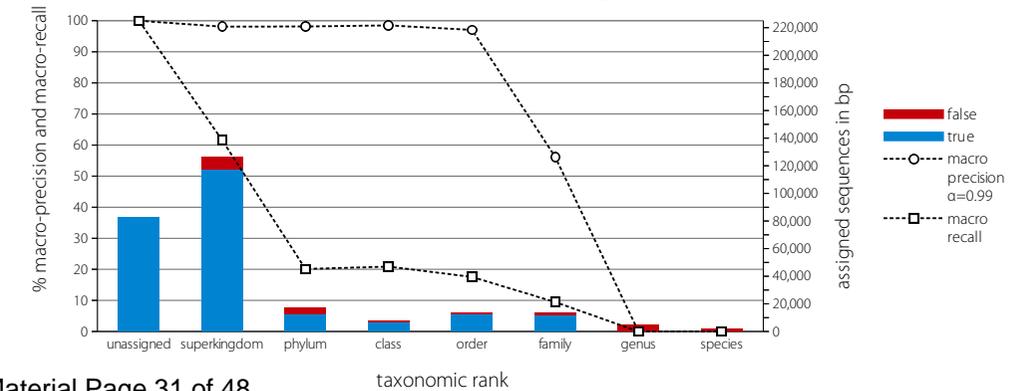

*Supplementary Figure 11: Taxator-tk binning of simulated metagenome with 49 species (simArt49e)* **(e) new family scenario**

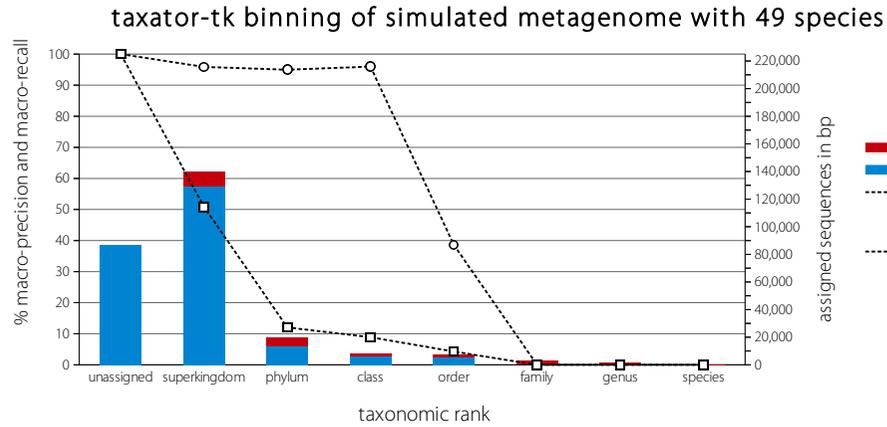

*Supplementary Figure 11: Taxator-tk binning of simulated metagenome with 49 species (simArt49e)* **(f) new order scenario**

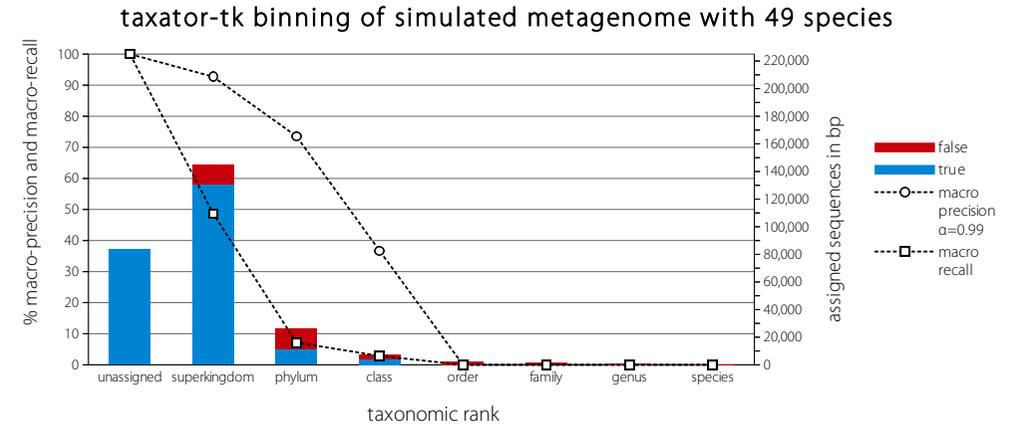

*Supplementary Figure 11: Taxator-tk binning of simulated metagenome with 49 species (simArt49e)* **(g) new class scenario**

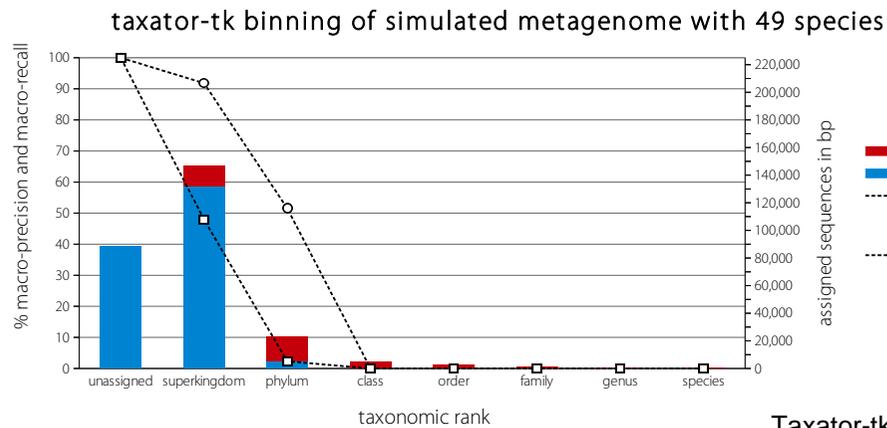

*Supplementary Figure 11: Taxator-tk binning of simulated metagenome with 49 species (simArt49e)* **(h) new phylum scenario**

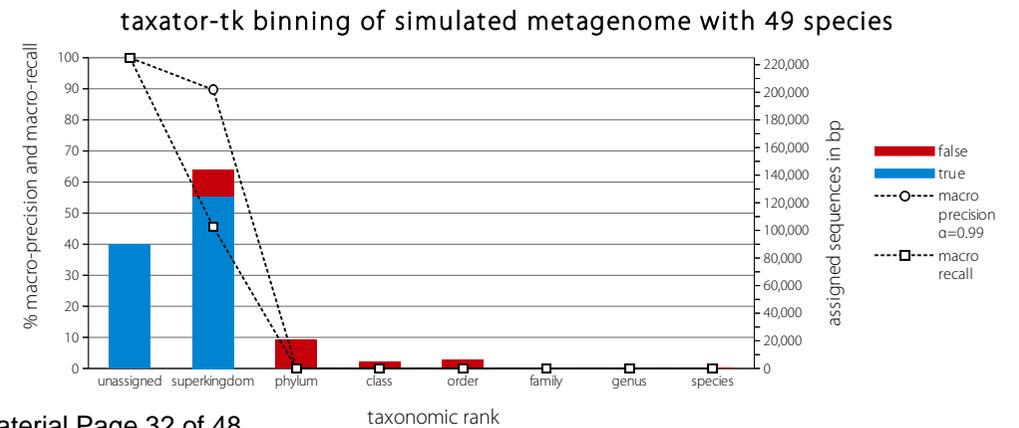

*Supplementary Figure 12 - Binning for FAMeS SimMC scenario (Nature Methods 2011)* **(a) CARMA**

| rank | depth | true | false | unknown | macro precision α=0.95 | stdev | pred. bins | macro recall | stdev | real bins | sum true | sum false | overall prec. | description |
|---|---|---|---|---|---|---|---|---|---|---|---|---|---|---|
| unassigned | 0 | 1788.99 | 0 | 0 | 100.0 | 0.0 | 1 | 100.0 | 0.0 | 1 | 2125.95 | 0 | 100.0 | root+superkingdom |
| superkingdom | 1 | 168.48 | 0 | 0 | 100.0 | 0.0 | 1 | 44.7 | 44.7 | 2 | | | | |
| phylum | 2 | 1071.86 | 49.59 | 0 | 48.1 | 40.4 | 3 | 45.9 | 36.4 | 8 | | | | |
| class | 3 | 2853.36 | 446.35 | 0 | 38.2 | 46.0 | 5 | 37.3 | 34.0 | 12 | 9749.83 | 1265.46 | 88.5 | phylum+class+order |
| order | 4 | 5824.61 | 769.52 | 0 | 23.6 | 35.1 | 19 | 39.7 | 37.0 | 22 | | | | |
| family | 5 | 1266.67 | 1107.7 | 0 | 13.7 | 28.0 | 50 | 30.3 | 41.8 | 29 | | | | |
| genus | 6 | 364.11 | 796.45 | 0 | 6.8 | 23.0 | 93 | 13.5 | 32.1 | 37 | 1719.92 | 2347.61 | 42.3 | family+genus+species |
| species | 7 | 89.11 | 443.46 | 0 | 2.5 | 15.1 | 135 | 2.2 | 14.4 | 47 | | | | |
| avg/sum | 3.9 | 11638.23 | 3613.07 | 0 | 33.3 | 26.8 | 43.7 | 30.5 | 34.4 | 22.4 | | | 76.3 | all but unassigned |
| avg/sum | 3.5 | 13427.22 | 3613.07 | 0 | 41.6 | 23.4 | 38.4 | 39.2 | 30.1 | 19.8 | | | 78.8 | all with unassigned |

*Supplementary Figure 12 - Binning for FAMeS SimMC scenario (Nature Methods 2011)* **(b) MEGAN**

| rank | depth | true | false | unknown | macro precision α=0.95 | stdev | pred. bins | macro recall | stdev | real bins | sum true | sum false | overall prec. | description |
|---|---|---|---|---|---|---|---|---|---|---|---|---|---|---|
| unassigned | 0 | 1014.03 | 0 | 0 | 100.0 | 0.0 | 1 | 100.0 | 0.0 | 1 | 4082.47 | 0 | 100.0 | root+superkingdom |
| superkingdom | 1 | 1534.22 | 0 | 0 | 100.0 | 0.0 | 1 | 47.0 | 47.0 | 2 | | | | |
| phylum | 2 | 1896.2 | 1.83 | 0 | 100.0 | 0.0 | 1 | 34.5 | 38.5 | 8 | | | | |
| class | 3 | 1021.99 | 108.21 | 0 | 65.6 | 45.6 | 4 | 29.4 | 31.7 | 12 | 4643.84 | 228.86 | 95.3 | phylum+class+order |
| order | 4 | 1725.65 | 118.82 | 0 | 29.6 | 40.6 | 11 | 30.1 | 35.2 | 22 | | | | |
| family | 5 | 935.47 | 266.12 | 0 | 13.6 | 32.0 | 17 | 15.3 | 30.8 | 29 | | | | |
| genus | 6 | 18.97 | 1684.92 | 0 | 9.2 | 25.9 | 36 | 3.8 | 14.5 | 37 | 1411.9 | 8207.44 | 14.7 | family+genus+species |
| species | 7 | 457.46 | 6256.4 | 0 | 6.1 | 21.2 | 38 | 0.2 | 1.0 | 47 | | | | |
| avg/sum | 3.5 | 7589.96 | 8436.3 | 0 | 46.3 | 23.6 | 15.3 | 22.9 | 28.4 | 22.4 | | | 47.4 | all but unassigned |
| avg/sum | 3.1 | 8603.99 | 8436.3 | 0 | 53.0 | 20.7 | 13.5 | 32.5 | 24.8 | 19.8 | | | 50.5 | all with unassigned |

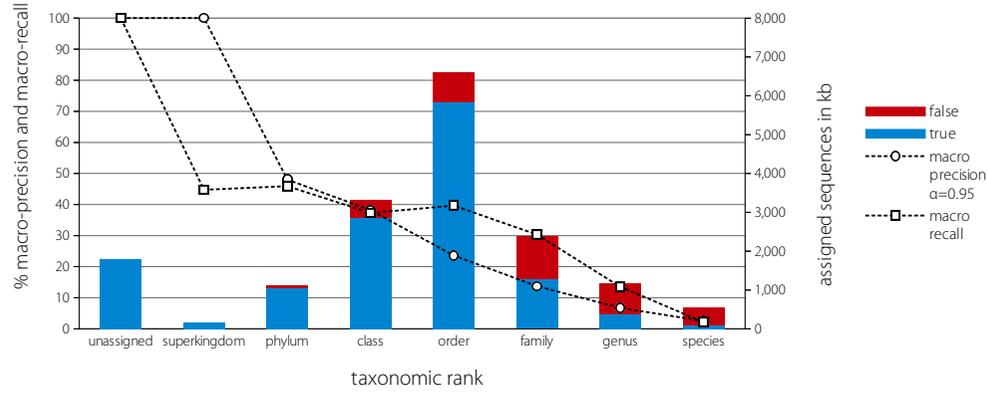

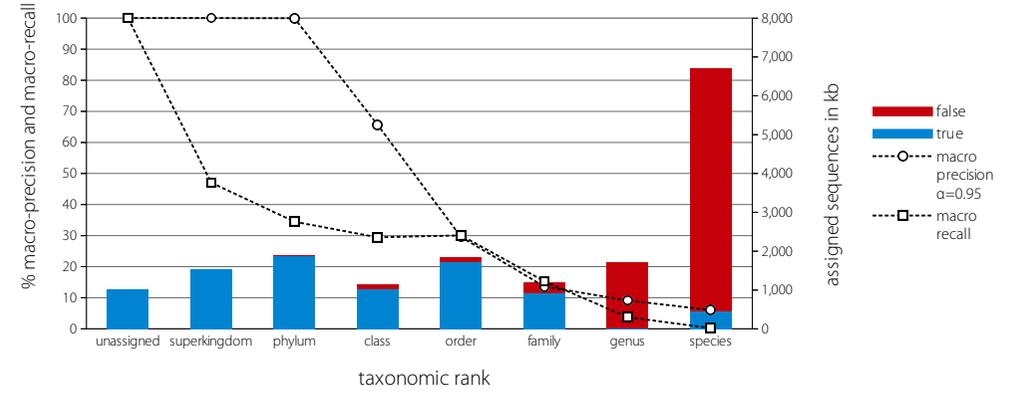

*Supplementary Figure 12 - Binning for FAMeS SimMC scenario (Nature Methods 2011)* **(c) taxator-tk**

| rank | depth | true | false | unknown | macro precision α=0.95 | stdev | pred. bins | macro recall | stdev | real bins | sum true | sum false | overall prec. | description |
|---|---|---|---|---|---|---|---|---|---|---|---|---|---|---|
| unassigned | 0 | 199.2 | 0 | 0 | 100.0 | 0.0 | 1 | 100.0 | 0.0 | 1 | 9526.04 | 2.03 | 100.0 | root+superkingdom |
| superkingdom | 1 | 4663.42 | 2.03 | 0 | 100.0 | 0.0 | 1 | 49.4 | 49.4 | 2 | | | | |
| phylum | 2 | 7936.17 | 2.11 | 0 | 100.0 | 0.0 | 1 | 29.9 | 36.1 | 8 | | | | |
| class | 3 | 2215.89 | 25.89 | 0 | 69.5 | 42.7 | 3 | 18.9 | 28.7 | 12 | 11881.62 | 56.97 | 99.5 | phylum+class+order |
| order | 4 | 1729.56 | 28.97 | 0 | 68.1 | 45.1 | 6 | 20.1 | 32.0 | 22 | | | | |
| family | 5 | 191.38 | 13.42 | 0 | 60.1 | 47.2 | 14 | 17.1 | 33.0 | 29 | | | | |
| genus | 6 | 19 | 11.53 | 0 | 50.0 | 50.0 | 10 | 9.1 | 25.3 | 37 | 212.11 | 24.95 | 89.5 | family+genus+species |
| species | 7 | 1.73 | 0 | 0 | 100.0 | 0.0 | 1 | 2.1 | 14.4 | 47 | | | | |
| avg/sum | 2.1 | 16757.15 | 83.95 | 0 | 78.2 | 26.4 | 5.1 | 20.9 | 31.3 | 22.4 | | | 99.5 | all but unassigned |
| avg/sum | 2.1 | 16956.35 | 83.95 | 0 | 81.0 | 23.1 | 4.6 | 30.8 | 27.4 | 19.8 | | | 99.5 | all with unassigned |

*Supplementary Figure 12 - Binning for FAMeS SimMC scenario (Nature Methods 2011)* **(d) original PhyloPythiaS**

| rank | depth | true | false | unknown | macro precision α=0.95 | stdev | pred. bins | macro recall | stdev | real bins | sum true | sum false | overall prec. | description |
|---|---|---|---|---|---|---|---|---|---|---|---|---|---|---|
| unassigned | 0 | 0 | 0 | 0 | 100.0 | 0.0 | 1 | 100.0 | 0.0 | 1 | 1664.72 | 0 | 100.0 | root+superkingdom |
| superkingdom | 1 | 832.36 | 0 | 0 | 100.0 | 0.0 | 1 | 49.9 | 49.9 | 2 | | | | |
| phylum | 2 | 517.19 | 25.6 | 0 | 100.0 | 0.0 | 1 | 50.8 | 41.1 | 8 | | | | |
| class | 3 | 1297.42 | 52.21 | 0 | 67.6 | 45.5 | 3 | 54.5 | 40.0 | 12 | 2474.3 | 154.19 | 94.1 | phylum+class+order |
| order | 4 | 659.69 | 76.38 | 0 | 49.6 | 44.9 | 6 | 37.4 | 34.2 | 22 | | | | |
| family | 5 | 5715.02 | 272.58 | 0 | 49.3 | 48.2 | 6 | 33.8 | 40.7 | 29 | | | | |
| genus | 6 | 7116.9 | 474.94 | 0 | 49.1 | 46.0 | 6 | 23.3 | 37.4 | 37 | 12831.92 | 747.52 | 94.5 | family+genus+species |
| species | 7 | | | 0 | | | | | | 47 | | | | |
| avg/sum | 5.0 | 16138.58 | 901.71 | 0 | 69.2 | 30.8 | 3.8 | 41.6 | 40.6 | 18.3 | | | 94.7 | all but unassigned |
| avg/sum | 5.0 | 16138.58 | 901.71 | 0 | 73.6 | 26.4 | 3.4 | 49.9 | 34.8 | 15.9 | | | 94.7 | all with unassigned |

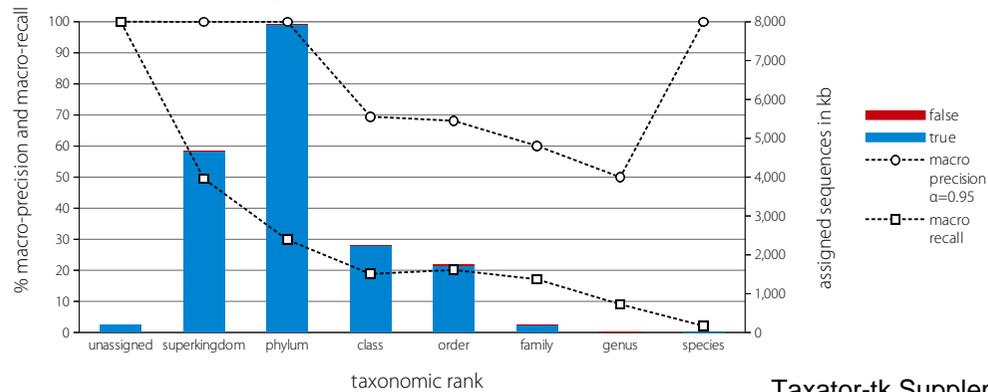

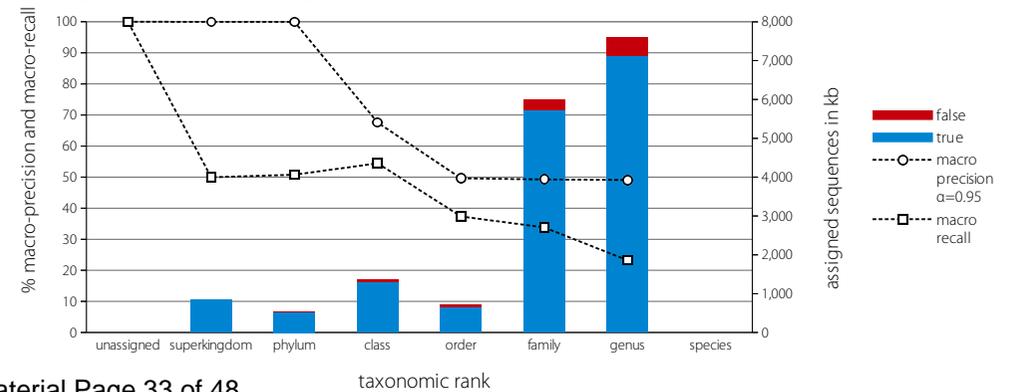

*Supplementary Figure 13 - Binning for partitioned cow rumen sample* **(a) CARMA**

| rank | depth | consistent | inconsistent | unknown | macro consistency α=0.99 | stdev | pred. bins | macro recall | stdev | cons. bins | sum true | sum false | overall consist. | description |
|---|---|---|---|---|---|---|---|---|---|---|---|---|---|---|
| unassigned | 0 | 144478 | 0 | 0 | 100.0 | 0.0 | 1 | 100.0 | 0.0 | 1 | 350414 | 154 | 100.0 | root+superkingdom |
| superkingdom | 1 | 102968 | 154 | 0 | 99.9 | 0.0 | 1 | 42.6 | 13.9 | 2 | | | | |
| phylum | 2 | 25730 | 3872 | 22 | 66.6 | 20.9 | 13 | 13.0 | 5.8 | 30 | 41974 | 7032 | 85.7 | phylum+class+order |
| class | 3 | 2256 | 1350 | 54 | 61.1 | 18.7 | 28 | 11.1 | 4.7 | 52 | | | | |
| order | 4 | 13988 | 1810 | 42 | 55.4 | 17.1 | 62 | 9.7 | 4.5 | 99 | | | | |
| family | 5 | 2400 | 964 | 104 | 52.1 | 23.2 | 167 | 9.0 | 4.8 | 198 | 7952 | 12724 | 38.5 | family+genus+species |
| genus | 6 | 5552 | 1090 | 132 | 52.6 | 36.6 | 572 | 9.2 | 5.1 | 446 | | | | |
| species | 7 | 0 | 10670 | 890 | 0.0 | 0.0 | 1254 | 0.0 | 0.0 | 926 | | | | |
| avg/sum | 1.8 | 152894 | 19910 | 1244 | 55.4 | 16.6 | 299.6 | 13.5 | 5.5 | 250.4 | | | 88.5 | all but unassigned |
| avg/sum | 1.0 | 297372 | 19910 | 1244 | 61.0 | 14.5 | 262.3 | 24.3 | 4.8 | 219.3 | | | 93.7 | all with unassigned |

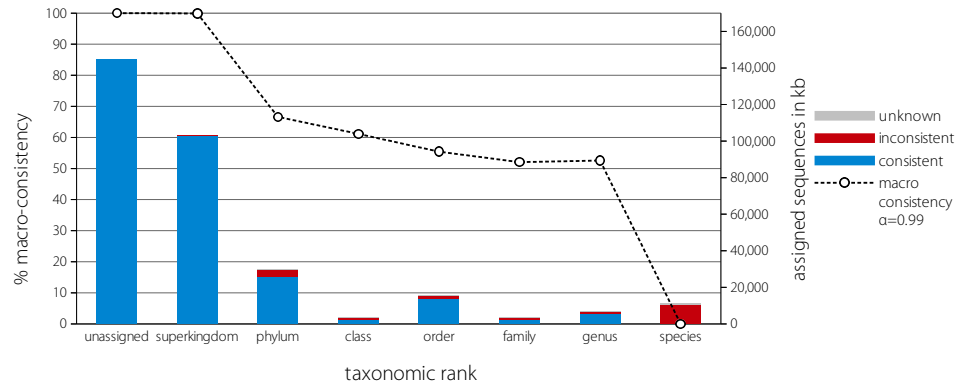

*Supplementary Figure 13 - Binning for partitioned cow rumen sample* **(b) MEGAN**

| rank | depth | consistent | inconsistent | unknown | macro consistency α=0.99 | stdev | pred. bins | macro recall | stdev | cons. bins | sum true | sum false | overall consist. | description |
|---|---|---|---|---|---|---|---|---|---|---|---|---|---|---|
| unassigned | 0 | 87760 | 0 | 0 | 100.0 | 0.0 | 1 | 100.0 | 0.0 | 1 | 218880 | 116 | 99.9 | root+superkingdom |
| superkingdom | 1 | 65560 | 116 | 0 | 99.9 | 0.0 | 1 | 43.8 | 26.2 | 3 | | | | |
| phylum | 2 | 35352 | 2802 | 34 | 67.4 | 24.6 | 12 | 24.7 | 16.9 | 27 | 79676 | 7200 | 91.7 | phylum+class+order |
| class | 3 | 2242 | 1090 | 42 | 51.9 | 27.7 | 25 | 19.7 | 13.9 | 48 | | | | |
| order | 4 | 42082 | 3308 | 66 | 39.6 | 26.3 | 51 | 15.5 | 11.9 | 88 | | | | |
| family | 5 | 2802 | 3220 | 178 | 34.9 | 21.1 | 132 | 13.6 | 9.2 | 168 | 46888 | 28304 | 62.4 | family+genus+species |
| genus | 6 | 12764 | 6726 | 436 | 33.6 | 20.4 | 264 | 12.6 | 8.0 | 295 | | | | |
| species | 7 | 31322 | 18358 | 2266 | 38.9 | 21.9 | 564 | 11.7 | 7.6 | 535 | | | | |
| avg/sum | 2.7 | 192124 | 35620 | 3022 | 52.3 | 20.3 | 149.9 | 20.2 | 13.4 | 166.3 | | | 84.4 | all but unassigned |
| avg/sum | 1.8 | 279884 | 35620 | 3022 | 58.3 | 17.8 | 131.3 | 30.2 | 11.7 | 145.6 | | | 88.7 | all with unassigned |

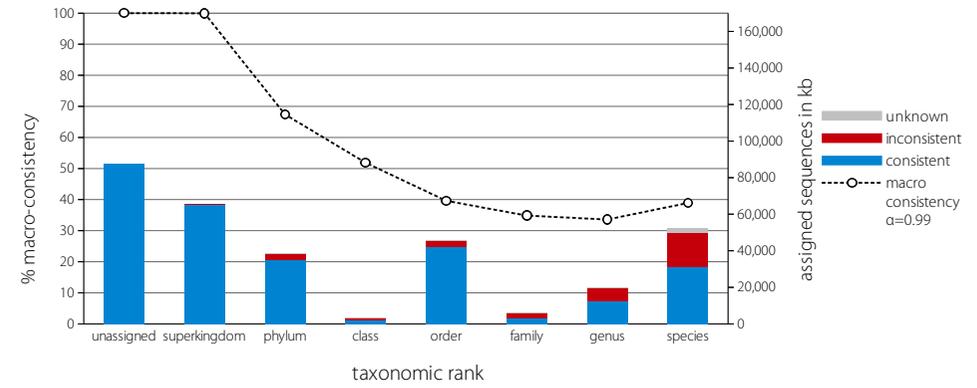

*Supplementary Figure 13 - Binning for partitioned cow rumen sample* **(c) taxator-tk**

| rank | depth | consistent | inconsistent | unknown | macro consistency α=0.99 | stdev | pred. bins | macro recall | stdev | cons. bins | sum true | sum false | overall consist. | description |
|---|---|---|---|---|---|---|---|---|---|---|---|---|---|---|
| unassigned | 0 | 122146 | 0 | 0 | 100.0 | 0.0 | 1 | 100.0 | 0.0 | 1 | 353002 | 62 | 100.0 | root+superkingdom |
| superkingdom | 1 | 115428 | 62 | 0 | 100.0 | 0.0 | 1 | 56.8 | 7.3 | 2 | | | | |
| phylum | 2 | 39828 | 1152 | 4 | 87.7 | 16.9 | 7 | 16.4 | 12.3 | 22 | 65086 | 2212 | 96.7 | phylum+class+order |
| class | 3 | 1676 | 334 | 28 | 80.2 | 17.0 | 14 | 13.7 | 11.3 | 34 | | | | |
| order | 4 | 23582 | 726 | 28 | 78.3 | 20.2 | 16 | 11.7 | 10.8 | 56 | | | | |
| family | 5 | 2524 | 198 | 100 | 79.3 | 19.8 | 50 | 10.3 | 8.8 | 84 | 12810 | 440 | 96.7 | family+genus+species |
| genus | 6 | 8938 | 198 | 94 | 76.2 | 35.9 | 110 | 9.8 | 7.8 | 94 | | | | |
| species | 7 | 1348 | 44 | 88 | 78.0 | 37.4 | 123 | 8.6 | 6.7 | 103 | | | | |
| avg/sum | 1.9 | 193324 | 2714 | 342 | 82.8 | 21.0 | 45.9 | 18.2 | 9.3 | 56.4 | | | 98.6 | all but unassigned |
| avg/sum | 1.2 | 315470 | 2714 | 342 | 85.0 | 18.4 | 40.3 | 28.4 | 8.1 | 49.5 | | | 99.1 | all with unassigned |

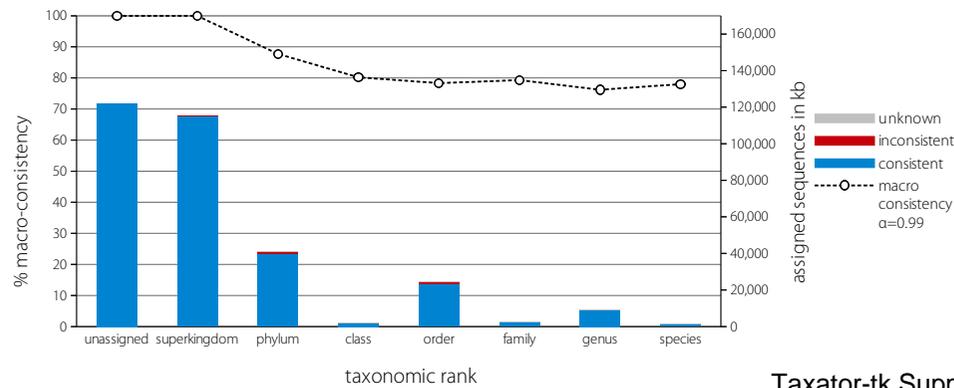

*Supplementary Figure 13 - Binning for partitioned cow rumen sample* **(d) PhyloPythiaS**

| rank | depth | consistent | inconsistent | unknown | macro consistency α=0.99 | stdev | pred. bins | macro recall | stdev | cons. bins | sum true | sum false | overall consist. | description |
|---|---|---|---|---|---|---|---|---|---|---|---|---|---|---|
| unassigned | 0 | 0 | 0 | 0 | 100.0 | 0.0 | 1 | 100.0 | 0.0 | 1 | 296276 | 2810 | 99.1 | root+superkingdom |
| superkingdom | 1 | 148138 | 2810 | 0 | 100.0 | 0.0 | 1 | 81.6 | 17.5 | 2 | | | | |
| phylum | 2 | 65136 | 24220 | 4 | 66.6 | 15.1 | 4 | 31.0 | 14.7 | 7 | 106338 | 40640 | 72.3 | phylum+class+order |
| class | 3 | 9468 | 9930 | 568 | 55.3 | 22.0 | 10 | 21.0 | 10.7 | 13 | | | | |
| order | 4 | 31734 | 6490 | 828 | 56.2 | 21.0 | 19 | 12.8 | 4.9 | 25 | | | | |
| family | 5 | 3990 | 1438 | 708 | 58.7 | 18.9 | 30 | 10.4 | 4.1 | 39 | 13842 | 3054 | 81.9 | family+genus+species |
| genus | 6 | 6144 | 1078 | 1072 | 62.8 | 18.5 | 33 | 9.2 | 4.2 | 45 | | | | |
| species | 7 | 3708 | 538 | 524 | 64.7 | 29.0 | 64 | 8.0 | 4.6 | 67 | | | | |
| avg/sum | 2.0 | 268318 | 46504 | 3704 | 66.3 | 17.8 | 23.0 | 24.9 | 8.7 | 28.3 | | | 85.2 | all but unassigned |
| avg/sum | 2.0 | 268318 | 46504 | 3704 | 70.6 | 15.6 | 20.3 | 34.3 | 7.6 | 24.9 | | | 85.2 | all with unassigned |

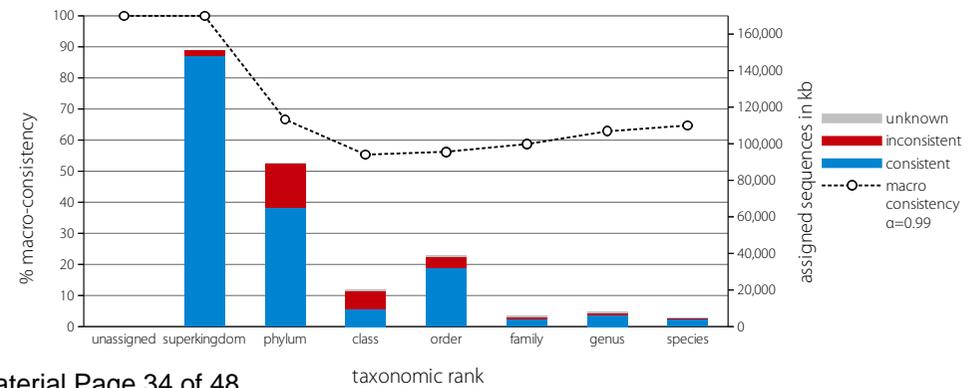

**Supplementary Figure 14:** Parallel speedup of program *taxator*

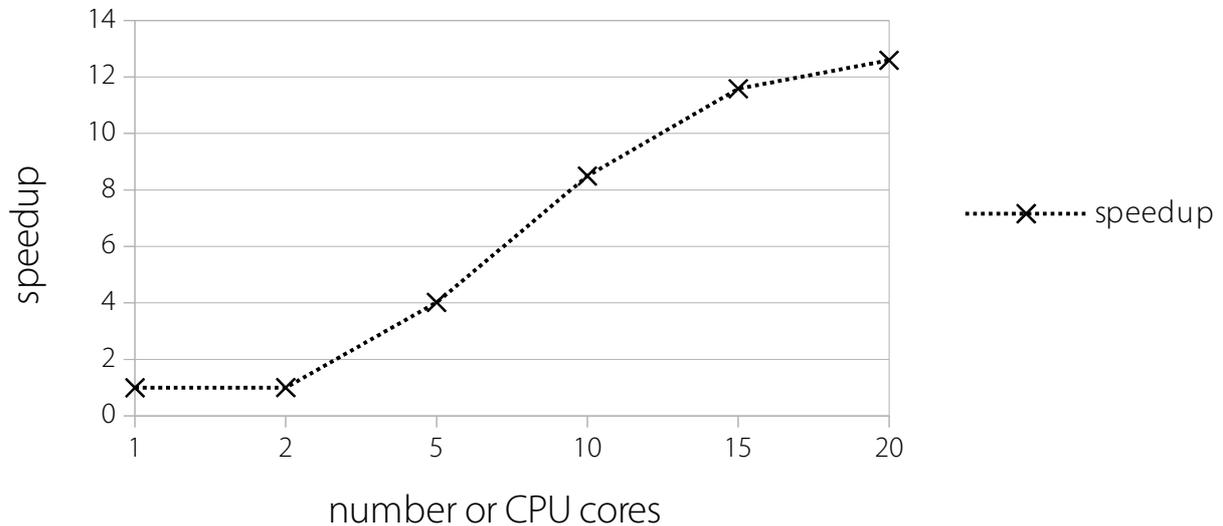

Execution time analysis with taxator for parallelized processing with multiple CPU cores. Taxonomic placement of sequence segments with taxator on input alignments for sequences of length 1000 bp (*syn1000* data set aligned against *mRefSeq47* with *LAST*). The speedup was calculated using wall clock time for a parallelized run relative to serial execution with one CPU thread. With multiple threads there is always one producer thread (consumer-producer model), thus for more than two threads, multiple consumers work on the input data in parallel. Approximate linear scale-up was observed up to 15 threads and saturation effects appear when using 20 CPU cores on our system.

$$speedup = \frac{T_1}{T_p}, with$$

$T_1$: serial execution time

$T_p$: execution time using p threads and CPU cores

**Supplementary Figure 15:** Effect of input sequence length and segmentation on *taxator-tk* processing time.

(a)

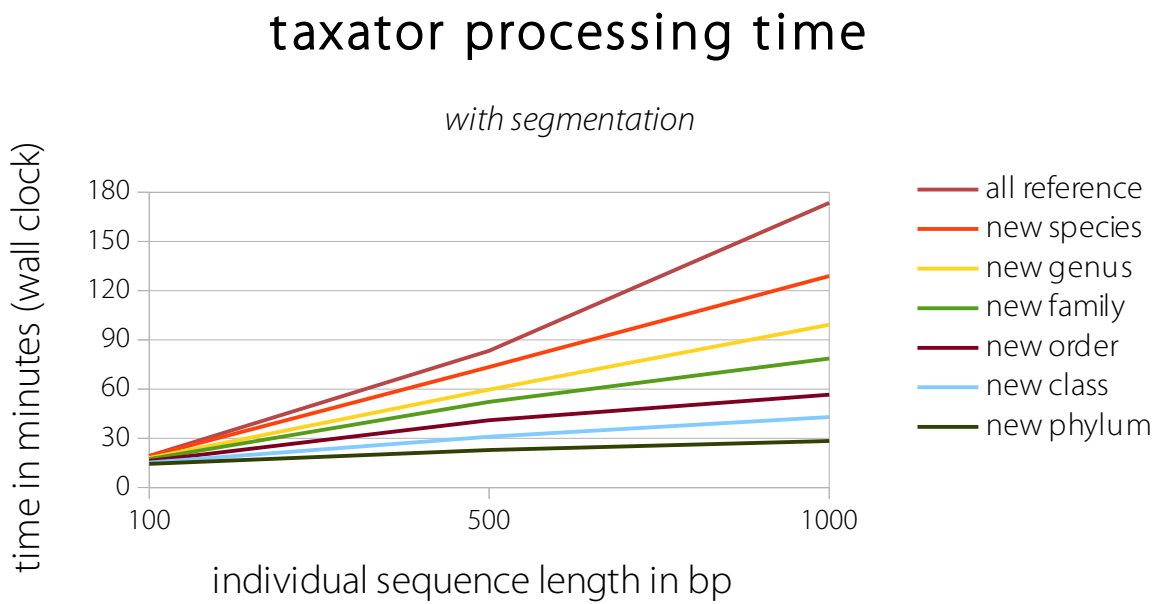

(b)

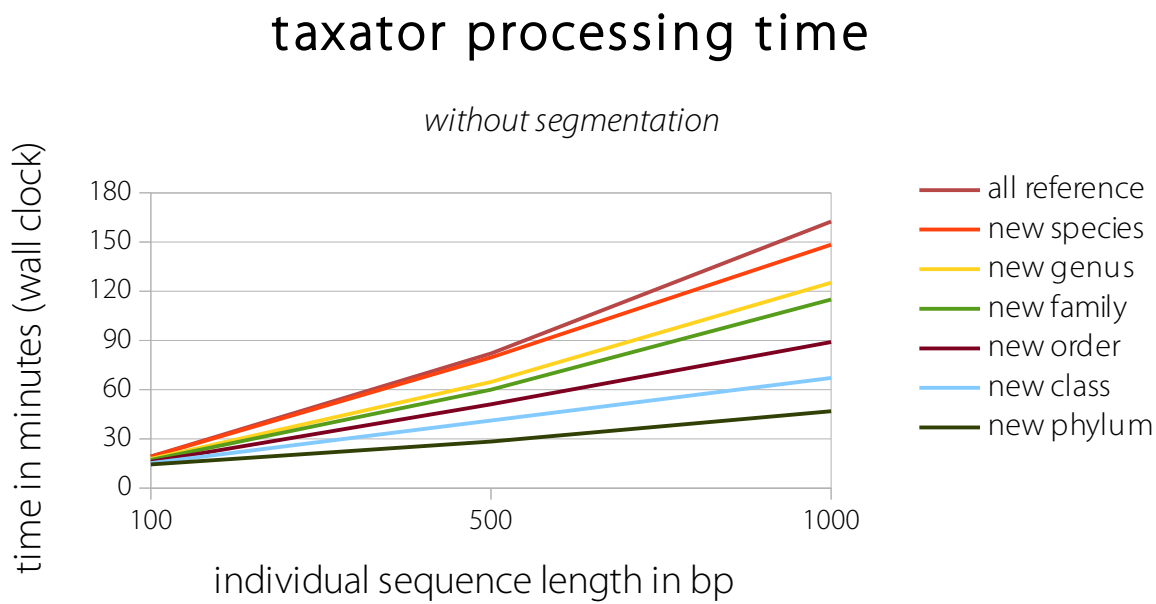

We processed approximately the same number of sequences of length 100, 500 and 1000 bp with *taxator-tk* (*syn100,syn500,syn1000*), once with the segmentation procedure being enabled (a) and once with segmentation disabled (b). The time increases for both cases approximately linear with the input length, with the slope depends on the completeness of the reference sequence data. With all reference data available, the time increases slightly more than linear, as there is no segmentation of queries during computations. For all other cases, segmentation substantially decreses the execution time.

**Supplementary Figure 16:** Taxonomic assignment of segments

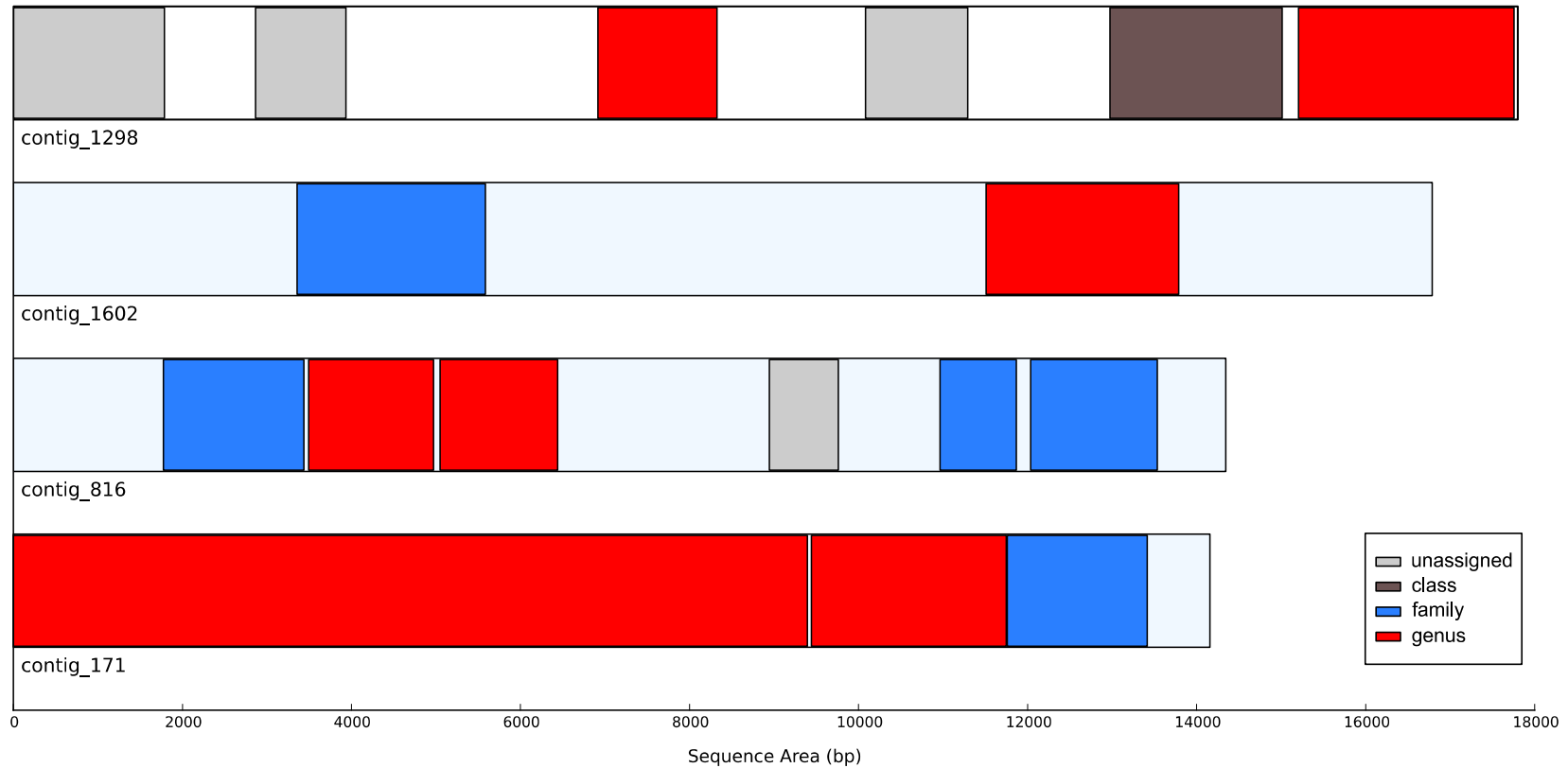

Four long contigs of the SimMC data set. Colored boxes show segments that where assigned by *taxator*, when all species reference data was removed (new species simulation). Light regions in between had no region aligned by a local alignment search and have therefore no homologs for assignment. All assigned regions in this example are consistently assigned. The segments are used by the program *binner* to derive consistent whole-sequence taxonomic assignments as used in our evaluations.

**Supplementary Figure 17:** Bin precision plots for 49 species simulated metagenomic sample (simArt49e)

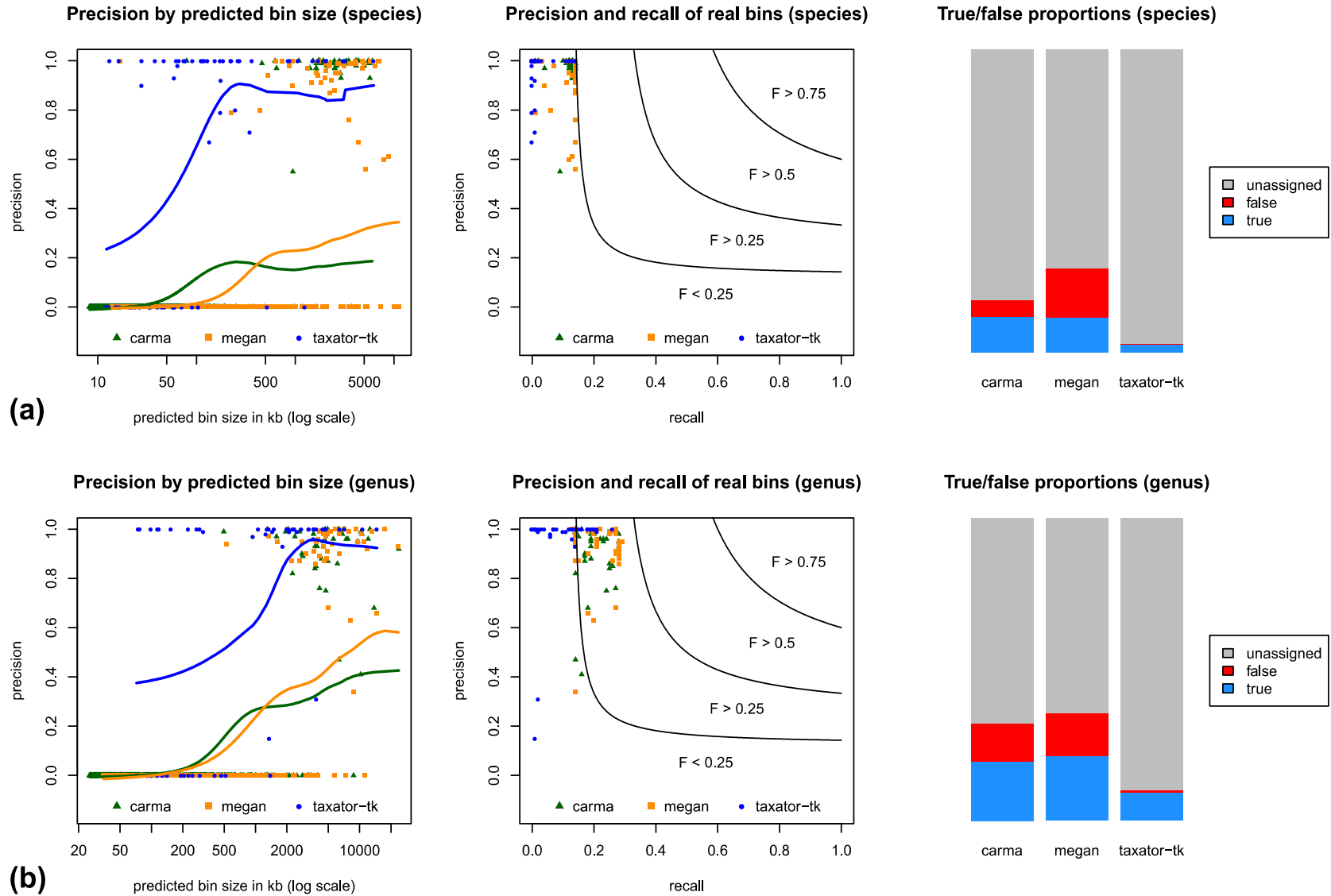

**Supplementary Figure 17:** Bin precision plots for 49 species simulated metagenomic sample (simArt49e)

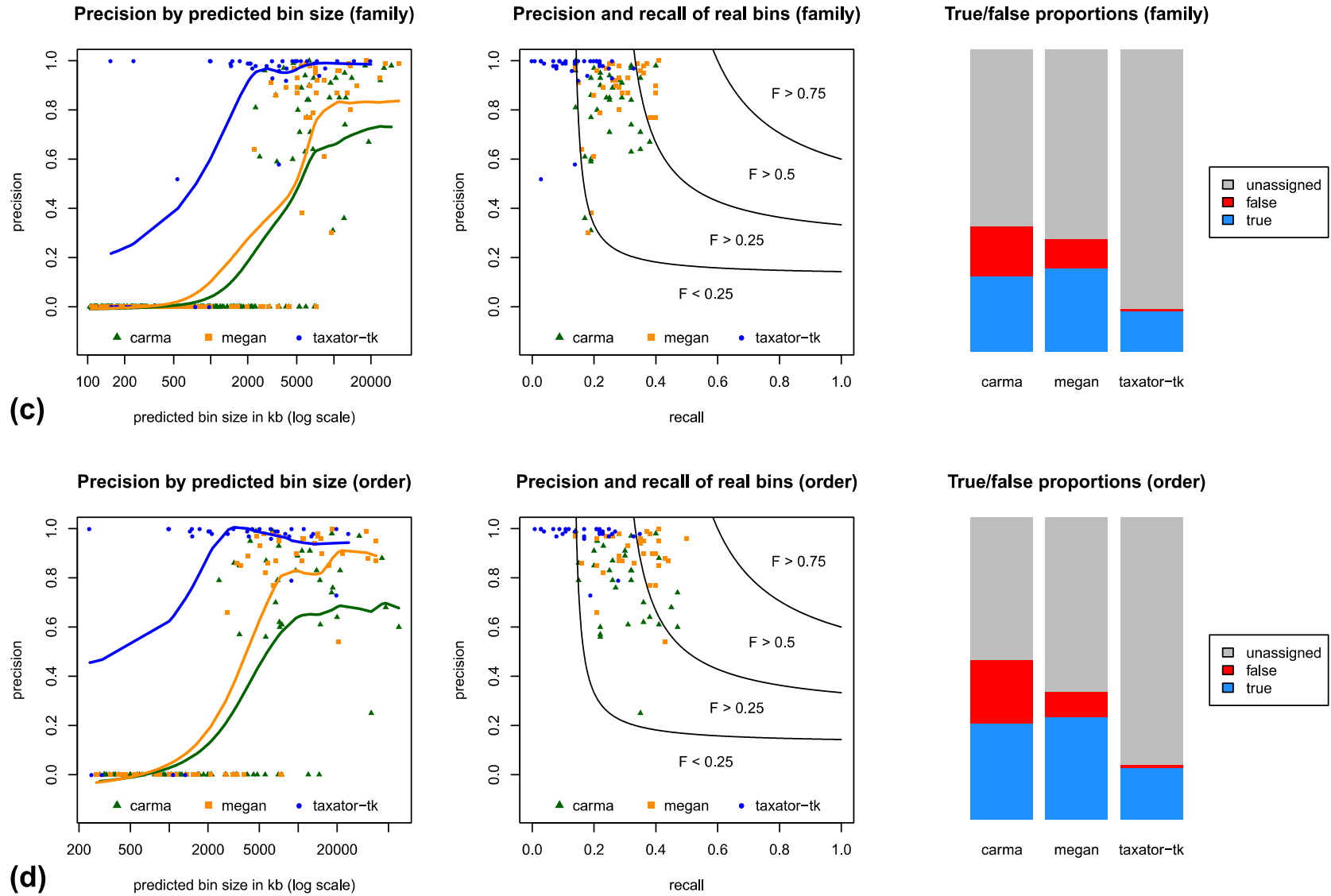

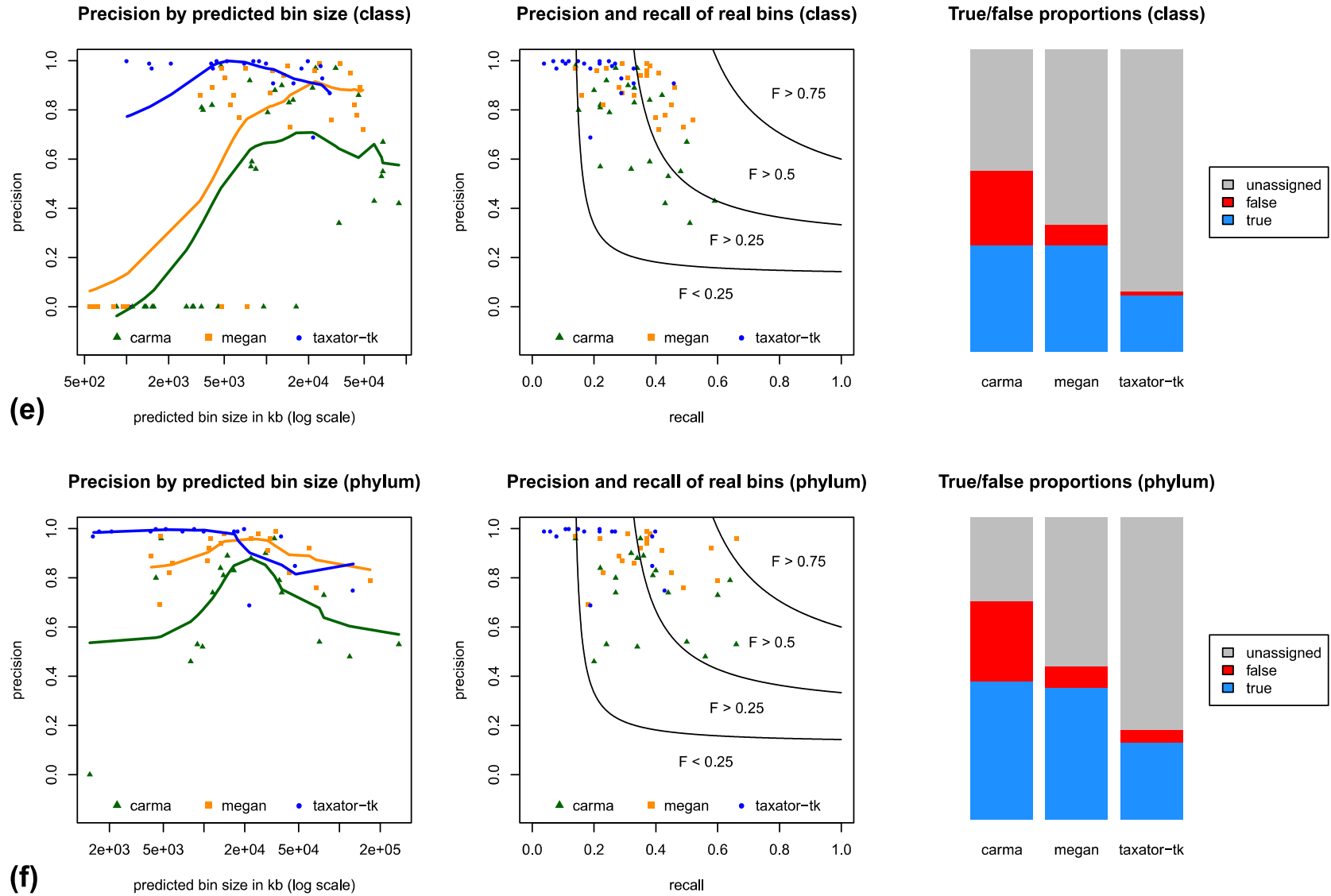

**Supplementary Figure 17:** Bin precision plots for 49 species simulated metagenomic sample (simArt49e)

**Supplementary Figure 17:** Bin precision plots for 49 species simulated metagenomic sample (simArt49e)

Comparison of assignment quality of CARMA3, MEGAN4 and taxator-tk for a simulated metagenome sample from a 49 species microbial community. Values are shown for the summary scenario (sum of all seven cross-validation scenarios), for assignments to the (a) species, (b) genus, (c) family, (d) order, (e) class and (f) phylum ranks, respectively. The first of each panels shows the precision and size for every predicted bin (after removing low abundance bins). The colored line shows a smoothed k-nearest-neighbor estimate of the mean precision as a function of predicted bin size using the R function wapply (width=0.3) followed by smooth.spline (df=10). The second panel for each rank shows bin precisions relative to recall. The F-score partitioning helps to identify similar quality bins if precision and recall are equally weighted, however we consider precision more important than recall. The third panel illustrates the total number of true (blue) and false (red) and unassigned (grey) portion of assignments at the respective ranks. Note that partially incorrect assignments are considered incorrect for the low ranking false part of the assignment and correct for the higher-ranks.

**Supplementary Figure 18:** Taxonomic composition of microbial RefSeq 47

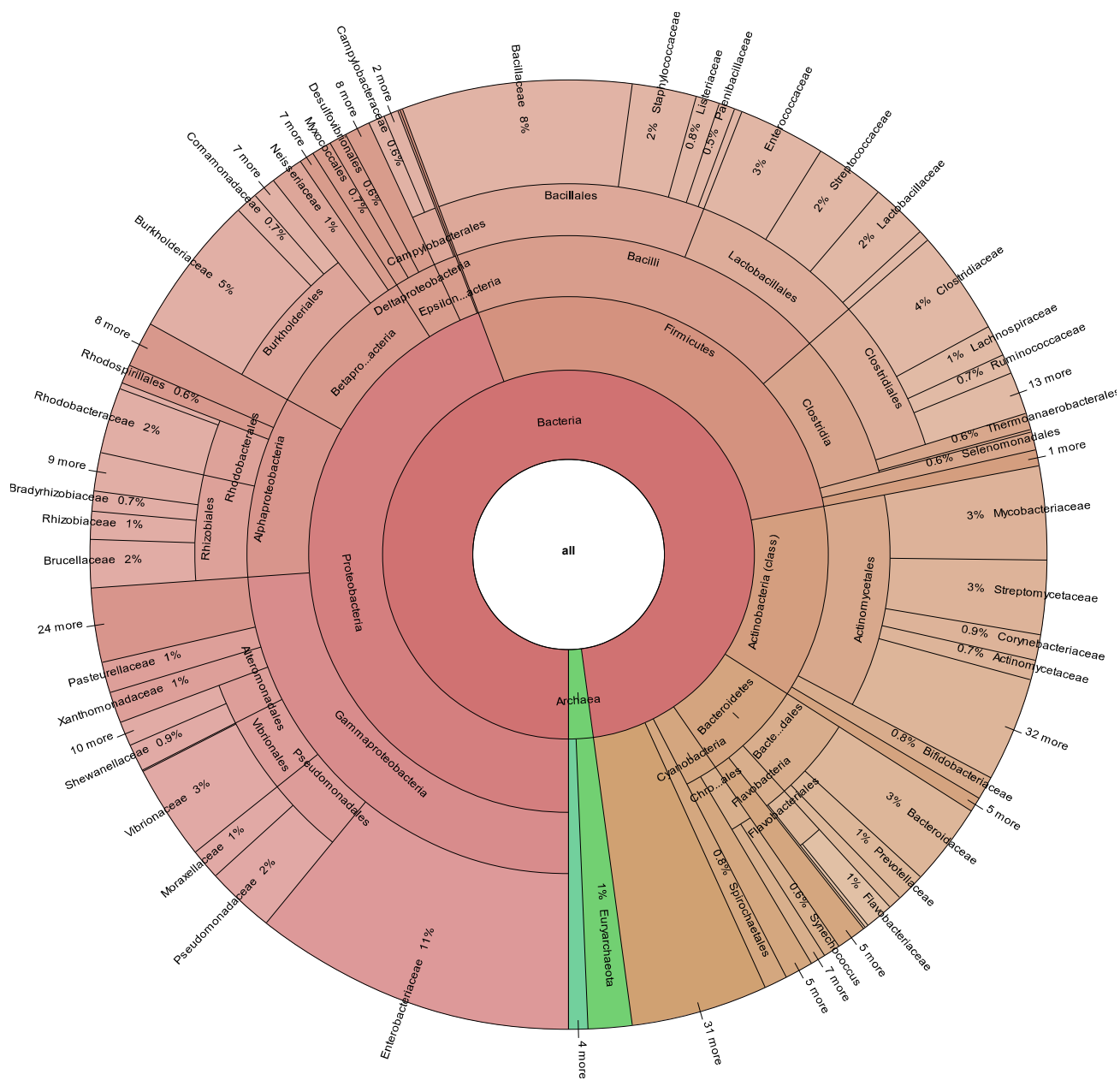

Taxonomic composition down to family level of the microbial (Bacteria, Archaea, Viruses) portion of the *RefSeq47* sequence data collection using Krona (http://krona.sourceforge.net). An interactive version can be found in the supplementary files (RefSeq47.krona.html). Abundance is measured in terms of accumulated sequence lengths per clade.

**Supplementary Figure 19:** Taxonomic composition of 16S genes extracted from *RefSeq47*

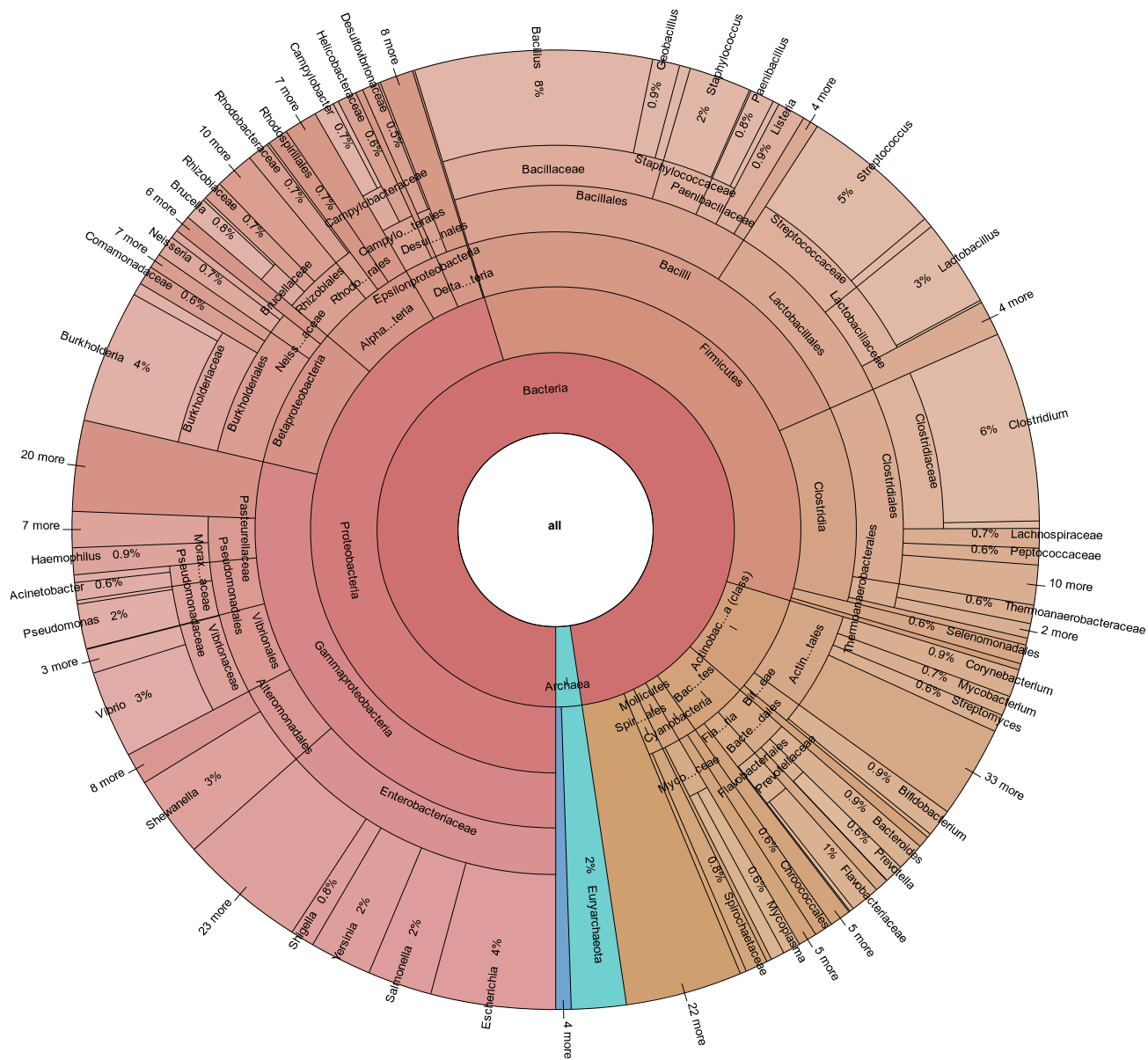

Taxonomic composition down to genus level of the 16S benchmark data set using Krona (http://krona.sourceforge.net). The data set was simulated by extracting every annotated 16S gene in *RefSeq47* which was at least 1000 bp long. An interactive version can be found in the supplementary files (refseq-16S.krona.html). Abundance is measured as the number of 16S genes.

**Supplementary Figure 20:** Taxonomic composition of microbial *RefSeq54*

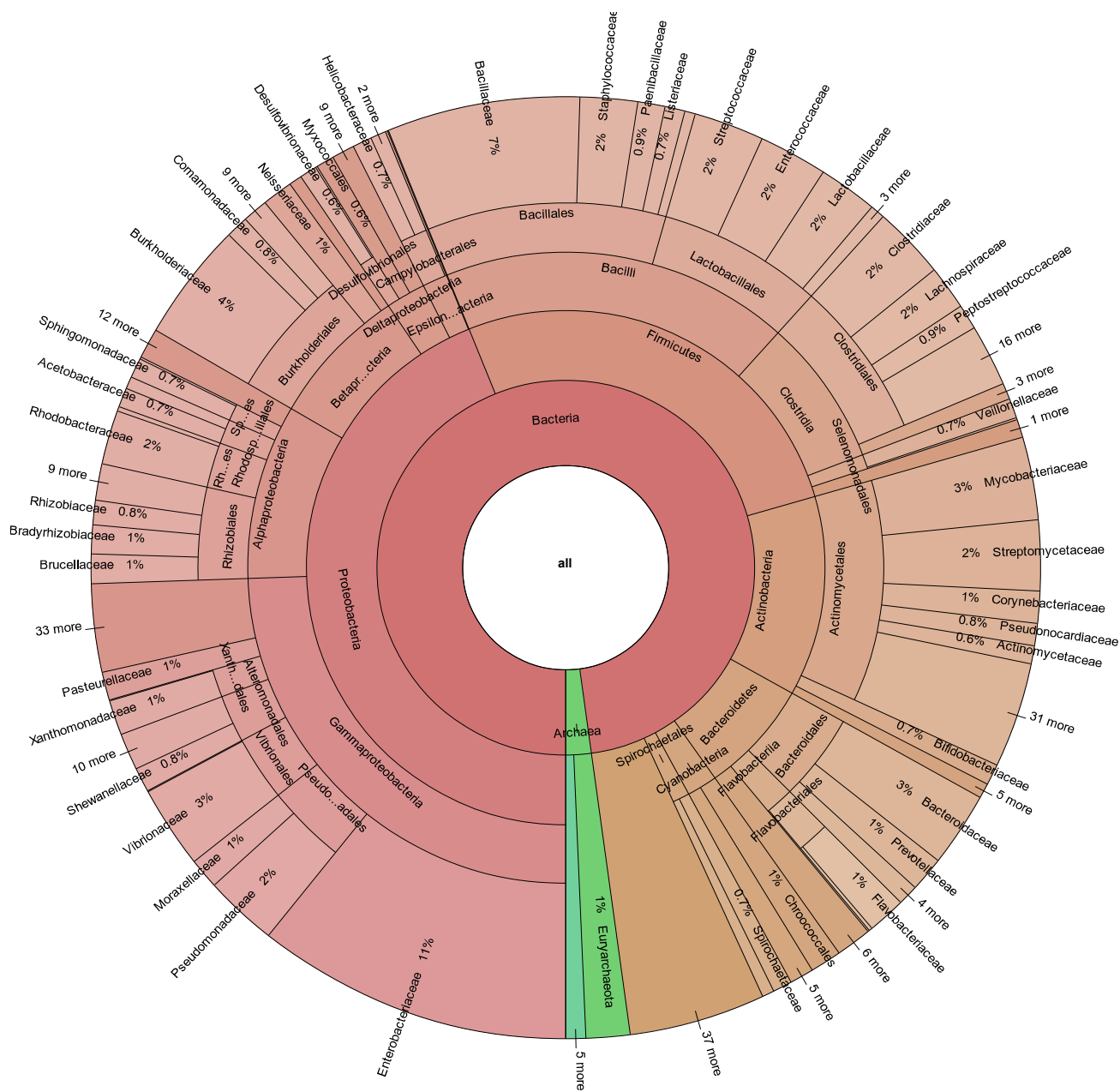

Taxonomic composition down to family level of the microbial (Bacteria, Archaea, Viruses) portion of the *RefSeq54* sequence data collection using Krona (http://krona.sourceforge.net). An interactive version can be found in the supplementary files (RefSeq54.krona.html). Abundance is measured in terms of accumulated sequence lengths per clade.

**Supplementary Figure 21:** Taxonomic composition of SimMC/AMD

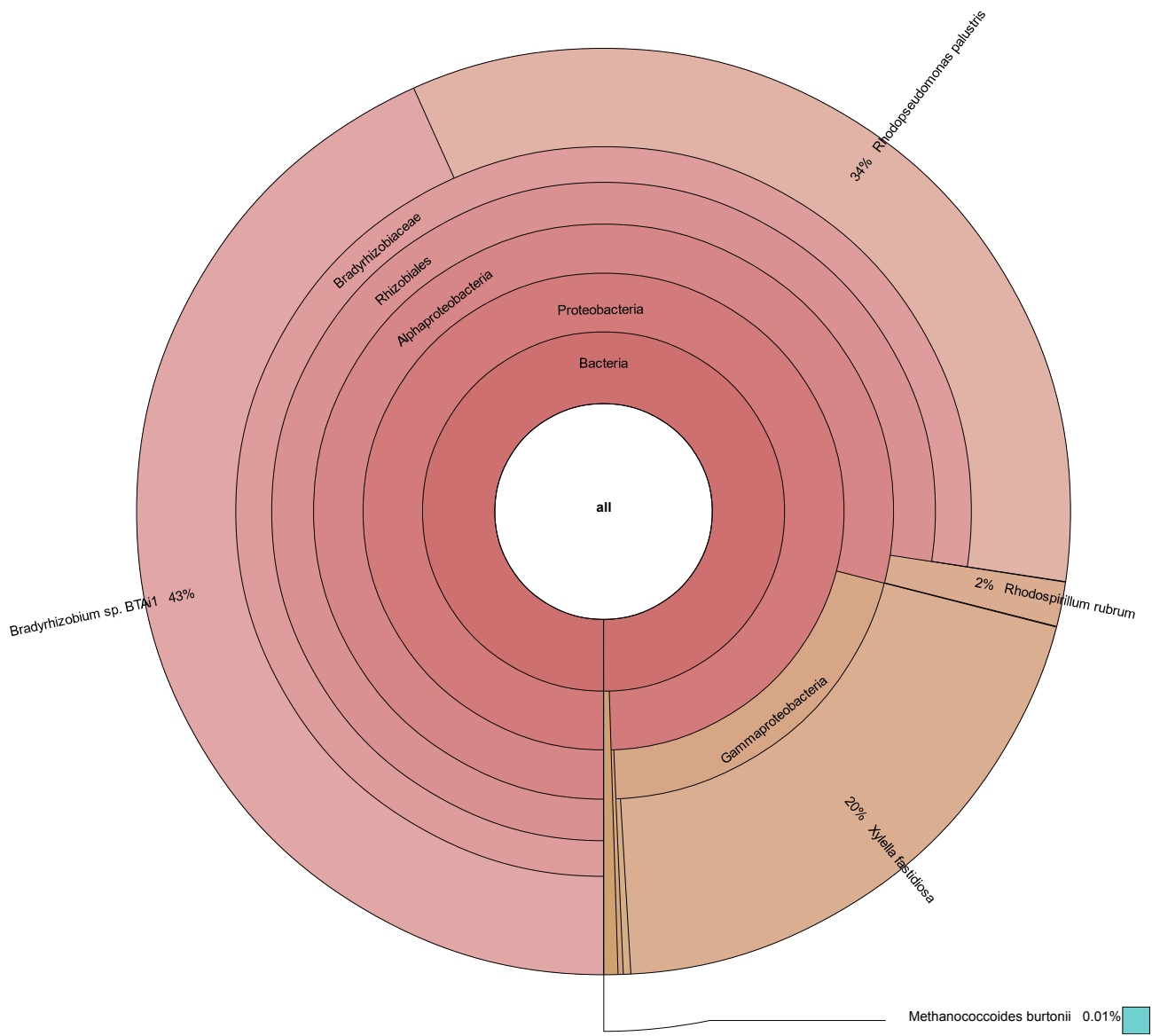

Taxonomic composition of the FAMeS simulated metagenome sample SimMC/AMD using Krona (http://krona.sourceforge.net). An interactive version can be found in the supplementary files (SimMC.krona.html). Abundance is measured in terms of accumulated contigs lengths.

**Supplementary Figure 22:** Taxonomic composition of SimHC/soil

Taxonomic composition of the FAMeS simulated metagenome sample SimHC/soil using Krona (http://krona.sourceforge.net). An interactive version can be found in the supplementary files (SimHC.krona.html). Abundance is measured in terms of accumulated contigs lengths.

**Supplementary Figure 23:** Taxonomic composition of simArt49e

Taxonomic composition of the simulated metagenome sample simArt49e using Krona (http://krona.sourceforge.net). An interactive version can be found in the supplementary files (simArt49e.krona.html). Abundance is measured in terms of accumulated contigs lengths. The data set was simulated using equal coverage for every strain, so differences in the data proportions result from a variable genome size and assembly bias.

**Supplementary Figure 24:** Query sequence segmentation and segment splicing

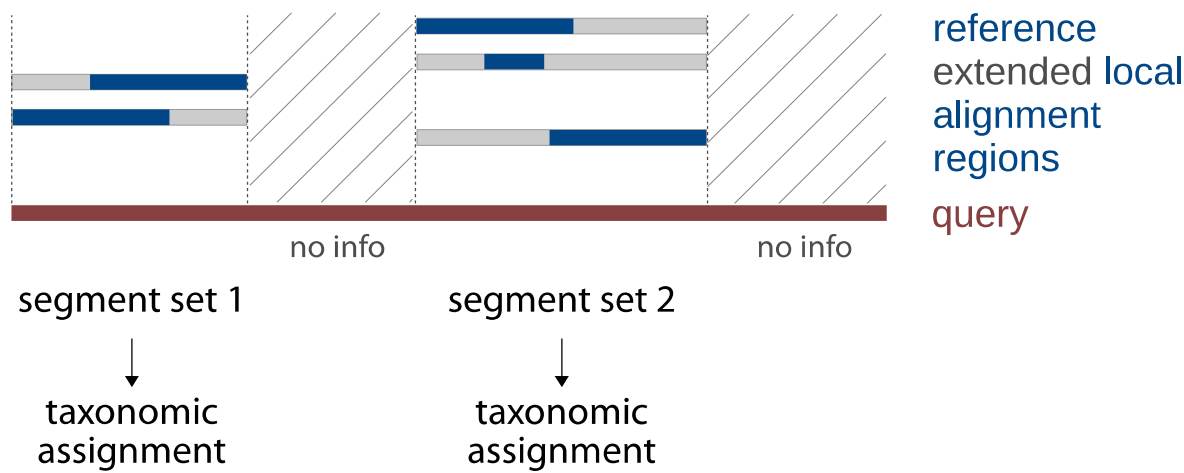

Query and corresponding reference segments from local alignment region extension and splicing. Blue bars correspond to original local alignment regions on reference nucleotide sequences which are positionally aligned to the query nucleotide sequence in red. These alignments are generated by a local (nucleotide) sequence aligner such as *BLAST* or *LAST* before running *taxator*. If alignments overlap on the query, they are joined into query segments which are flanked by regions without detected similarity to any known reference sequence. Reference segments are constructed from the original alignment reference regions (blue) by extension (grey bars) with the same number of nucleotides which are missing to match the length of the query segment. The corresponding sets of homologs are the input to the core taxonomic assignment algorithm in *taxator*.